\documentclass[a4paper]{article}
\usepackage{geometry}
\usepackage{amssymb}
\usepackage{hyperref}
\usepackage{amsmath}
\usepackage{amsthm}
\usepackage{bbold}

\usepackage[ruled]{algorithm}
\usepackage{algorithmic}

\algsetup{linenodelimiter=.}

\hypersetup{colorlinks=true}
\hypersetup{linkcolor=black,citecolor=black,urlcolor=black,filecolor=black,menucolor=black}

\newtheorem{theorem}{Theorem}
\newtheorem{definition}[theorem]{Definition}
\newtheorem{lemma}[theorem]{Lemma}
\newtheorem{corollary}[theorem]{Corollary}
\newtheorem{problem}[theorem]{Problem}
\newtheorem{example}[theorem]{Example}
\theoremstyle{remark}
\newtheorem{remark}[theorem]{\bf Remark}

\DeclareMathOperator{\linspan}{span}
\DeclareMathOperator{\coeff}{coeff}
\DeclareMathOperator{\const}{Const}

\DeclareMathOperator{\lcm}{lcm}
\DeclareMathOperator{\num}{num}
\DeclareMathOperator{\den}{den}
\DeclareMathOperator{\supp}{supp}
\DeclareMathOperator{\lc}{lc}
\DeclareMathOperator{\lm}{lm}
\DeclareMathOperator{\lt}{lt}
\DeclareMathOperator{\im}{im}

\newcommand{\Der}{\ensuremath{\partial}}
\newcommand{\param}{\theta}

\begin{document}

\title{Reduction systems and degree bounds for integration\footnote{This research was supported by the Austrian Science Fund (FWF): P 31952 and by the National Natural Science Foundation of China (NSFC): 12201065.}}

\author{Hao Du\textsuperscript{a,b}, Clemens G. Raab\textsuperscript{c,}\footnote{Corresponding author. Present address: RICAM, Austrian Academy of Sciences, Linz, Austria}}
\hypersetup{pdfauthor={Hao Du, Clemens G. Raab}}
\date{}

\maketitle

\begin{center}
\textsuperscript{a}School of Sciences, Beijing University of Posts and Telecommunications (BUPT), China\\
\textsuperscript{b}Key Laboratory of Mathematics and Information Networks (BUPT), Ministry of Education, China\\
\textsuperscript{c}Institute for Algebra, Johannes Kepler University Linz (JKU), Austria\\[\medskipamount]
Email: \href{mailto:haodu@bupt.edu.cn}{\tt haodu@bupt.edu.cn}, \href{mailto:clemensr@algebra.uni-linz.ac.at}{\tt clemensr@algebra.uni-linz.ac.at}
\end{center}

\begin{abstract}
In symbolic integration, the Risch--Norman algorithm aims to find closed forms of elementary integrals over differential fields by an ansatz for the integral, which usually is based on heuristic degree bounds.
Norman presented an approach that avoids degree bounds and only relies on the completion of reduction systems.
We give a formalization of his approach and we develop a refined completion process, which terminates in more instances.
In some situations when the completion process does not terminate, one can detect patterns allowing to still describe infinite reduction systems that are complete.
We present such infinite systems for the fields generated by Airy functions and complete elliptic integrals, respectively. 
Moreover, we show how complete reduction systems can be used to find rigorous degree bounds.
In particular, we give a general formula for weighted degree bounds and we apply it to find tight bounds in the above examples.
\end{abstract}

\paragraph{Keywords}
Symbolic integration, Risch--Norman algorithm, Completion process, Infinite reduction systems
\paragraph{MSC 2020} 33F10, 68W30, 12H05, 13P99, 15A06

\tableofcontents

\section{Introduction}

Integration in finite terms is concerned with algorithmically finding closed-form expressions for antiderivatives. Typically, \emph{elementary integrals} are considered, which can be described as being representable in terms of logarithms, exponentials, and algebraic functions applied to functions already appearing in the integrand. The main theoretical result for integration in finite terms is Liouville's Theorem, which states that if the integrand $f$ has an elementary integral, then the integral can be written so that new functions appear as constant linear combination of logarithms. 
More explicitly, there exist $u_0, u_1, \dots,u_m$ consisting of functions appearing in $f$ and constants $c_1,\dots,c_m$ such that
\[
 f = \Der{u_0} + \sum_{i=1}^mc_i\frac{\Der{u_i}}{u_i}.
\]
The first purely algebraic proof of Liouville's Theorem was given by Rosenlicht in \cite{RosenlichtLiouville}, where a precise formulation of the theorem can be found as well. Such an elementary integral of $f$ may be written as the sum of the so-called \emph{rational part} $u_0$ and the so-called \emph{logarithmic part} $\sum_{i=1}^mc_i\ln(u_i)$.

Liouville's Theorem and its various refinements on the structure of elementary integrals are the main theoretical foundation for many algorithms in symbolic integration. Risch gave the first complete algorithm \cite{RischTransc} for elementary integration of a large class of elementary integrands. His algorithm was subsequently generalized to other classes of integrands, for references see \cite{Bronstein,RaabRisch}. Since these algorithms are very involved because of their recursive structure, a simpler and more efficient approach was devised: the Risch--Norman algorithm \cite{Norman}. Even though it is not a complete algorithm for elementary integration, i.e.\ it may fail to find the elementary integral even if the given integrand has one, it is nonetheless a powerful heuristic in practice. Moreover, it is rather easy to implement and can even be generalized to many classes of integrands for which no other algorithm is available.

Several authors have given explicit examples to illustrate that standard versions of the Risch--Norman algorithm cannot find all elementary integrals. For instance, the following integrand involving the tangent function was discussed by Norman when he introduced a complementary approach based on completion of reduction systems \cite{NormanCriticalPair}.
\begin{equation}\label{eq:NormanExample}
 \int\frac{x}{\tan(x)^2+1}\,dx = \frac{x^2\tan(x)^2+2x\tan(x)+x^2+1}{4(\tan(x)^2+1)}
\end{equation}
Another example was given more recently by Boettner when he observed that the following antiderivative cannot be found by recent extensions of the Risch--Norman algorithm \cite[Ex.~8.7]{BoettnerPhD}.
\begin{equation}\label{eq:BoettnerExample}
 \int\mathrm{Ai}^\prime(x)^2\,dx = \frac{x}{3}\mathrm{Ai}^\prime(x)^2+\frac{2}{3}\mathrm{Ai}(x)\mathrm{Ai}^\prime(x)-\frac{x^2}{3}\mathrm{Ai}(x)^2
\end{equation}
The Airy function $\mathrm{Ai}(x)$ appearing in this integrand satisfies the second-order differential equation $y^{\prime\prime}(x)-xy(x)=0$ and arises in many applications, see \cite{AiryBook} for example.

In general, the Risch--Norman algorithm may fail to find the elementary integral for various reasons: it might not detect all terms necessary to form the logarithmic part, the denominator of the rational part might contain factors not predicted by the algorithm, or the numerator of the rational part might involve the constituent functions to a much larger degree than the integrand does. The latter is true for the terms $x^2\tan(x)^2$ and $\frac{x^2}{3}\mathrm{Ai}(x)^2$ in the two integrals without logarithmic part shown above.

In this paper, we propose improvements to make the heuristic Risch--Norman algorithm more powerful by addressing the problem of finding the numerator of the rational part of the integral.
To this end, we develop Norman's completion-based approach \cite{NormanCriticalPair} further.
In fact, this completion process can be understood by Gaussian elimination on infinite matrices.
For an outline of the Risch--Norman algorithm and a more detailed explanation and motivation of our aim to solve Problem~\ref{prob:MainProblem}, see Section~\ref{sec:RischNorman}.
Relation to other work is discussed also in Section~\ref{sec:discussion}.

The rest of this paper is organized as follows.  
We recall some notions in differential algebra and symbolic integration for later use and give a brief self-contained introduction to the Risch--Norman algorithm and related reduction systems in Section~\ref{sec:pre}.
A new formalization of such complete reduction systems and a refinement of Norman's completion process are presented in Section~\ref{sec:RedSys}, and also we give a proof of correctness of our refined procedure, while Norman did not indicate a correctness proof of his process in~\cite{NormanCriticalPair}.
In addition, we work out in detail one example posed as open problem by Norman and show that our refined procedure terminates in this case.
Since the completion process does not always terminate, we present three examples of infinite complete reduction systems in Section~\ref{sec:InfiniteSystems} for Airy functions and complete elliptic integrals.
Instead of applying reduction directly to given integrands to obtain the rational part of the integral, we show that complete reduction systems can also be used to find rigorous degree bounds of the numerator.
In Section~\ref{sec:degbound}, we discuss general properties of such rigorous weighted degree bounds and how they are obtained by general formulae under certain conditions, which we illustrate by examples.
Finally, we discuss additional aspects, possible improvements for implementations, and outlook of future study in Section~\ref{sec:discussion}.

\section{Prerequisites and notation}\label{sec:pre}

Throughout this paper, $(F,\Der)$ always denotes a differential field of characteristic zero. Recall that a differential field $(F,\Der)$ is a field $F$ together with a derivation $\Der$ on it, i.e.\ $\Der:F\to{F}$ is an additive map that satisfies the Leibniz rule $\Der(fg)=(\Der{f})g+f\Der{g}$ for all $f,g \in F$. The set of constant elements in $F$ forms a subfield denoted by $\const_\Der(F)=\{f\in{F}\ |\ \Der{f}=0\}$.
Moreover, we only consider the case where the field $F$ is given as a purely transcendental extension $F=C(t_1,\dots,t_n)$ of a field of constants $C\subseteq\const_\Der(F)$ by elements $t_1,\dots,t_n \in F$ that are algebraically independent over $C$.
Hence, $\Der$ is a $C$-linear map on the multivariate rational function field $C(t_1,\dots,t_n)$ and $t_1,\dots,t_n$ model algebraically independent functions.
Often, the stronger condition $C=\const_\Der(F)$ is desirable and can be satisfied in practical examples.
However, permitting constants in $F\setminus{C}$ to exist allows for more flexibility in the representation of $F$ when applying the methods of this paper, since it eliminates the need to actually prove that there are no such constants for a given choice of $\Der$ and $t_i$.

Numerators and denominators of elements of $F$ are defined by viewing those elements as rational functions in $t_1,\dots,t_n$.
Recall that a derivation on such a field is completely determined by the elements $\Der{t_1},\dots,\Der{t_n}$ via $\Der=\sum_{i=1}^n(\Der{t_i}){\cdot}\Der_i$, where $\Der_i$ is the standard partial derivation with respect to $t_i$. Conversely, any choice of $\Der{t_1},\dots,\Der{t_n} \in F$ yields a derivation on $F$ in this way.
The following definition is based on \cite[Ch.~10]{Bronstein} and \cite{BronsteinParallel}.

\begin{definition}
\label{def:denominator}
 For $(F,\Der)=(C(t_1,\dots,t_n),\Der)$ with $C\subseteq\const_\Der(F)$ such that $t_1,\dots,t_n$ are algebraically independent over $C$, we define the \emph{denominator of $\Der$} as
 \[
  \den(\Der):=\lcm(\den(\Der{t_1}),\dots,\den(\Der{t_n}))
 \]
 and to $\Der$ we associate the derivation $\tilde{\Der}:F\to{F}$ defined by $\tilde{\Der}f:=\den(\Der){\cdot}\Der{f}$.
\end{definition}

Note that, in contrast to $\Der$, the derivation $\tilde{\Der}$ necessarily maps polynomials to polynomials so that $(C[t_1,\dots,t_n],\tilde{\Der})$ is a differential subring of $(F,\tilde{\Der})$.

In any polynomial ring $R[t]=R[t_1,\dots,t_n]$ over an integral domain $R$, we denote the coefficient of a monomial $t^\alpha=t_1^{\alpha_1}{\cdot}\dots{\cdot}t_n^{\alpha_n}$ in a polynomial $p$ by $\coeff(p,t^\alpha)$.
The support of a polynomial $p \in R[t]$ is given by $\supp(p):=\{t^\alpha\ |\ \alpha\in\mathbb{N}^n,\coeff(p,t^\alpha)\neq0\}$.
For any nonzero $w \in \mathbb{R}^n$, the weighted degree of a polynomial $p \in R[t]$ w.r.t.\ $w$ can be defined by
\[
 \deg_w(p):=\sup\{w_1\alpha_1+\ldots+w_n\alpha_n\ |\ t^\alpha \in \supp(p)\}.
\]
This implies $\deg_w(0)=-\infty$ and $\deg_w(1)=0$.
Note that the weights in $w$ (and hence the degree of a polynomial) may also be negative and that these definitions and notations extend also to Laurent polynomials $R[t,t^{-1}]=R[t_1,\dots,t_n,t_1^{-1},\dots,t_n^{-1}]$ by considering all $\alpha \in \mathbb{Z}^n$.

Recall that a semigroup order is a total order on a semigroup that is compatible with multiplication, i.e.\ $x<y$ implies $xz<yz$ for all elements $x,y,z$.
In the literature, a semigroup order on (the monoid of) monomials is called a \emph{monomial order} if it satisfies $t_i>1$ for all $i$.

As usual, a semigroup order on monomials determines the leading monomial, leading coefficient, and leading term of any nonzero (Laurent) polynomial.
For the zero polynomial, we also define $\lm(0):=0$, $\lc(0):=0$, and $\lt(0):=0$ so that $\lt(p)=\lc(p)\lm(p)$ holds for all (Laurent) polynomials.
Moreover, we extend the semigroup order to include $0$ as new least element.
Consequently, $\lm(p-\lt(p))<\lm(p)$ holds for all nonzero (Laurent) polynomials $p$.

\begin{definition}
 Given a semigroup order $<$ on monomials, a (Laurent) polynomial $P=\sum_{\alpha\in\mathbb{Z}^n}c_\alpha t^\alpha$, and a (Laurent) monomial $m$, we define the \emph{truncation of $P$ at $m$} by $P_{\le m}:=\sum_{\substack{\alpha\in\mathbb{Z}^n\\t^\alpha\le m}}c_\alpha t^\alpha$ and the \emph{truncation of $P$ below $m$} by $P_{<m}:=\sum_{\substack{\alpha\in\mathbb{Z}^n\\t^\alpha<m}}c_\alpha t^\alpha$.
\end{definition}

Occasionally, we will use the Pochhammer symbol $(x)_k$ to abbreviate the product $\prod_{i=0}^{k-1}(x+i)$ in coefficients of (Laurent) polynomials.

\subsection{Risch--Norman algorithm}
\label{sec:RischNorman}

In short, given $f \in F$, the approach of the Risch--Norman algorithm for computing an explicit antiderivative of $f$ relies on making an appropriate ansatz.
This ansatz is a linear combination of expressions in $F$ or in extensions of $F$, whose derivatives lie in $F$.
Then, the derivative of the ansatz is matched to $f$ by coefficient comparison.
Both the effectiveness and the computational effort depend on the details of the ansatz, i.e.\ how accurately it predicts the structure of the antiderivative.

Choosing an ansatz is based on refinements of Liouville's Theorem.
Let $(F,\Der)$ and $\tilde{\Der}$ as in Definition~\ref{def:denominator} and consider an integrand $f \in F^*$.
Then, a structure theorem given by Bronstein (see Thm.~10.2.1 in \cite{Bronstein} resp.\ Thm.~1 in \cite{BronsteinParallel}) justifies an ansatz of the form
\begin{equation}\label{eq:ParallelAnsatz}
 f = \Der\left(\frac{u}{v}\right) + \sum_{i=1}^m\alpha_i\frac{\Der{p_i}}{p_i} + \sum_{i=1}^k\beta_i\frac{\Der{s_i}}{s_i},
\end{equation}
with $u,v \in C[t]=C[t_1,\dots,t_n]$ and $\alpha_1,\dots,\alpha_m,\beta_1,\dots,\beta_k \in \overline{C}$ for some $s_1,\dots,s_k \in \overline{C}[t]$ satisfying $s_i|\tilde{\Der}s_i$, where $p_1,\dots,p_m \in \overline{C}[t]$ are the irreducible factors of $\den(f)$ satisfying $\gcd(p_i,\tilde{\Der}p_i)=1$.
In fact, the theorem guarantees the existence of such $u,v,\alpha_i,\beta_i,s_i$, if $f$ has an elementary integral over $(F,\Der)$ and $C=\const_\Der(F)$.
In addition, the theorem also yields upper bounds for the multiplicities of $p_i$ in the polynomial $v \in C[t]$.
Based on \eqref{eq:ParallelAnsatz}, the following three main steps need to be carried out to compute an elementary integral of $f$:
\begin{enumerate}
 \item\label{step:special} Find candidates for the polynomials $v$ and $s_1,\dots,s_k$.
 \item\label{step:support} Find a finite set of monomials in $t_1,\dots,t_n$ that contains $\supp(u)$.
 \item\label{step:coeffs} Compute the constants $\alpha_1,\dots,\alpha_m,\beta_1,\dots,\beta_k$ and the constant coefficients of $u$ via linear algebra.
\end{enumerate}

In fact, steps \ref{step:support} and \ref{step:coeffs} may also be performed in an interleaving manner by iteratively determining parts of $\supp(u)$ and some of the constants involved.
In the literature and in implementations, there are various heuristics for performing steps \ref{step:special} and \ref{step:support}, see \cite{Norman,Fitch,DavenportParallel3,GeddesStefanus,Bronstein,BoettnerPhD}, for example.
No general algorithm is known that would be complete for choosing candidates for $v$ or for $s_1,\dots,s_k$ or for $\supp(u)$.
Under certain conditions on the derivation, however, some results are known for a comprehensive choice of $v$ and $s_1,\dots,s_k$, see \cite{DavenportParallel1,DavenportParallel2,BronsteinParallel}.
In particular, there is the well-known case of rational function integration corresponding to $(C(t_1),\Der)$ with $\Der{t_1}=1$, where even a comprehensive choice of candidate monomials appearing in $u$ can be given based on $f$.

We do not discuss the details of choosing $v$ and $s_1,\dots,s_k$ here.
Instead, we shall focus on the situation after step \ref{step:special}, assuming polynomials $v$ and $s_1,\dots,s_k$ are given and only $u \in C[t]$ and the constants $\alpha_1,\dots,\alpha_m$, $\beta_1,\dots,\beta_k$ remain to be found in \eqref{eq:ParallelAnsatz}.
Determining $u$ is challenging because of possible cancellations in the derivative $\Der{u}$ resp.\ $\Der\frac{u}{v}$.
In practice, usually various heuristic degree bounds have been used to determine a finite ansatz for $u$.
We mention some of these for comparison, using the notation of \eqref{eq:ParallelAnsatz}.
If only elementary monomials $t_i$ are considered, the bound
\begin{equation}\label{eq:degboundElem}
 \deg_{t_i}(u) \le 1+\max(\deg_{t_i}(\num(f)),\deg_{t_i}(\den(f)))-\min(1,\deg_{t_i}(\Der{t_i}))
\end{equation}
of partial degrees is a common choice, cf.~\cite{GeddesStefanus,Bronstein}. For the general case, the bound
\begin{equation}\label{eq:degboundBro}
 \deg(u) \le 1+\deg(\num(f))+\max(0,\deg(\den(\Der))-\max_i\deg(\tilde{\Der}t_i))
\end{equation}
for the total degree was suggested in \cite{Bronstein}. The bound on the total degree of $u$ used by the \textsc{Maple} program \texttt{pmint} \cite{pmint} amounts to
\begin{equation}\label{eq:degboundpmint}
 \deg(u) \le 1+\deg(\tfrac{v}{\gcd(\den(f),\tilde{\Der}\den(f))})+\max(\deg(\num(f)),\deg(\den(f))).
\end{equation}
The \textsc{Sage} program \texttt{parrisch} \cite{BoettnerPhD} essentially bounds the partial degrees of $u$ by 
\begin{equation}\label{eq:degboundparrisch}
 \deg_{t_i}(u) \le 1+\max(\deg_{t_i}(v),\deg_{t_i}(\tfrac{\den(\Der)}{\gcd(\den(\Der),\den(f))})+\deg_{t_i}(\num(f))).
\end{equation}
For certain derivations $\Der$, it is even possible to obtain a rigorous bound on $\deg_{t_i}(u)$ for some of the $i\in\{1,\dots,n\}$ based on Thm.~4.4.4 or Lemma~5.1.2 in \cite{Bronstein}.
Doing that simultaneously for all $i\in\{1,\dots,n\}$ to obtain a finite ansatz for $u$ is possible only if $\Der{t_i} \in C[t_i]$ holds for all $i$.

\begin{example}\label{ex:AiryBoettner}
For the integral \eqref{eq:BoettnerExample}, consider the differential field $(C(t_1,t_2,t_3),\Der)$ with $\Der{t_1}=1$, $\Der{t_2}=t_3$, and $\Der{t_3}=t_1t_2$.
The generators $t_1,t_2,t_3$ correspond to the functions $x$, $\mathrm{Ai}(x)$, and $\mathrm{Ai}^\prime(x)$, respectively.
In the notation of \eqref{eq:ParallelAnsatz}, we have $f=t_3^2$, i.e.\ $m=0$.
The integral is given by
\begin{equation}
 u=\frac{1}{3}t_1t_3^2+\frac{2}{3}t_2t_3-\frac{1}{3}t_1^2t_2^2
\end{equation}
and $v=1$.
Note that this integral violates all degree bounds \eqref{eq:degboundElem}--\eqref{eq:degboundparrisch} mentioned above.
\end{example}

In general, based on the Leibniz rule, we can reformulate \eqref{eq:ParallelAnsatz} as first-order linear differential equation
\begin{equation}
 \Der{u} - \frac{\Der{v}}{v}u = v\left(f - \sum_{i=1}^m\alpha_i\frac{\Der{p_i}}{p_i} - \sum_{i=1}^k\beta_i\frac{\Der{s_i}}{s_i}\right)
\end{equation}
for the polynomial $u$. To obtain a left hand side that is a polynomial for every $u \in C[t]$, we can multiply the equation by $\den(\Der)\den\left(\frac{\tilde{\Der}v}{v}\right)$ yielding
\begin{equation}\label{eq:PolynomialAnsatz}
 \frac{v}{\gcd(v,\tilde{\Der}{v})}\tilde{\Der}u - \frac{\tilde{\Der}v}{\gcd(v,\tilde{\Der}{v})}u = \frac{v^2}{\gcd(v,\tilde{\Der}{v})} \left(\den(\Der)f - \sum\limits_{i=1}^m\alpha_i\frac{\tilde{\Der}p_i}{p_i} - \sum\limits_{i=1}^k\beta_i\frac{\tilde{\Der}s_i}{s_i}\right)
\end{equation}
in terms of $\tilde{\Der}$.
Equations of this form will be our main concern.
We do not rely on a special way of choosing $v \in C[t]$ in terms of $f$ here.
Note that, for a solution $u \in C[t]$ to exist, the right hand side of \eqref{eq:PolynomialAnsatz} necessarily has to lie in $C[t]$, which puts restrictions on possible choices of $v$ and of $\alpha_1,\dots,\alpha_m,\beta_1,\dots,\beta_k$.
In particular, $v$ has to be divisible by $\gcd(b,\tilde{\Der}b)$ where $b:=\den(\den(\Der)f)$.
In general, we end up with the following problem.

\begin{problem}
\label{prob:MainProblem}
 Given $(F,\Der)=(C(t_1,\dots,t_n),\Der)$ with $C\subseteq\const_\Der(F)$ and $f_0,\dots,f_m,v \in C[t]$, $v\neq0$, find $u \in C[t]$ and $c_1,\dots,c_m \in C$ such that
 \begin{equation}\label{eq:MainEquation}
  \frac{v}{\gcd(v,\tilde{\Der}{v})}\tilde{\Der}u - \frac{\tilde{\Der}v}{\gcd(v,\tilde{\Der}{v})}u = f_0 + \sum_{i=1}^mc_if_i
 \end{equation}
\end{problem}

\subsection{Reduction systems}

Instead of finding $u$ via an ansatz with undetermined constant coefficients as explained in Section~\ref{sec:RischNorman}, Norman \cite{NormanCriticalPair} discussed a reduction-based approach to this problem.
In short, for fixed $v$, the approach uses derivatives $\Der\left(\frac{q}{v}\right)$ with different $q \in C[t]$ chosen algorithmically to reduce given integrands to zero.
Then, all $q$ used during reduction are the contributions to the solution $u$.
To find such $q$, Norman's idea is to systematically generate a suitable reduction system via a completion process, see also Section~\ref{sec:NormanCompletion}.
In fact, analogous to how \eqref{eq:PolynomialAnsatz} was obtained above, he considers the problem over a common denominator $d \in C[t]$, which amounts to considering an inhomogeneous equation with differential operator $L(u):=d\Der\!\left(\frac{u}{v}\right)$.
His reduction rules are based on identities relating certain polynomials involving parameters in their coefficients and exponents.
These parameterized polynomials are image and preimage of each other under $L$, see also Example~\ref{ex:NormanMain} below.
In our paper, we choose $d:=\frac{\den(\Der)v^2}{\gcd(v,\tilde{\Der}{v})}$, yielding $L(u)$ equal to the left hand side of \eqref{eq:MainEquation}.
Polynomials in the image of $L$ obtained from instantiating these parameters are then used to reduce the given right hand side of \eqref{eq:MainEquation} w.r.t.\ some monomial order, collecting the contributions to the solution $u \in C[t]$ in the process.
Note that, here, we reduce modulo the $C$-vector space $\im(L)$, in contrast to polynomial reduction in the context of Gr\"obner bases, where one reduces modulo an ideal.
Consequently, to reduce a given term, a polynomial used for reducing can only be multiplied by a constant coefficient to match its leading term with the given term.

\begin{example}\label{ex:NormanMain}
For the integral \eqref{eq:NormanExample}, consider the differential field $(C(t_1,t_2),\Der)$ with $\Der{t_1}=1$ and $\Der{t_2}=t_2^2+1$.
The generators $t_1,t_2$ correspond to the functions $x$ and $\tan(x)$, respectively, so the integrand is given by $f=\frac{t_1}{t_2^2+1}$.
Using the setting of \eqref{eq:ParallelAnsatz}, we hence have $m=0$ and, to compute an integral, we may choose $v=t_2^2+1$ and $k=0$.
This leads to $\Der{u}-2t_2u=t_1$ in \eqref{eq:MainEquation} and we abbreviate the left hand side by $L(u):=\Der{u}-2t_2u$.
For a monomial order with $t_1>t_2$, Norman \cite[p.~203]{NormanCriticalPair} computed four reduction rules.
The first and the last of them rely on the two identities
\begin{align}
 (j-3)t_1^it_2^j+(j-1)t_1^it_2^{j-2}+it_1^{i-1}t_2^{j-1} &= L(t_1^it_2^{j-1}) \quad\text{and}\label{eq:NormanR1}\\
 (2i+2)t_1^i+(i^2+i)t_1^{i-1}t_2 &= L(t_1^{i+1}t_2^2+t_1^{i+1}+(i+1)t_1^it_2)\label{eq:NormanR4}
\end{align}
respectively.
To solve $L(u)=t_1$ by reducing the right hand side to zero, we can set $i=1$ in \eqref{eq:NormanR4} and use its left hand side to reduce $t_1$ to $-\frac{1}{2}t_2$ first, obtaining contribution $\frac{1}{4}t_1^2t_2^2+\frac{1}{4}t_1^2+\frac{1}{2}t_1t_2$ to the solution.
Then, we can set $i=0$ and $j=1$ in \eqref{eq:NormanR1} to reduce $-\frac{1}{2}t_2$ to zero, yielding the contribution $\frac{1}{4}$ to the solution.
Altogether, we obtain the solution
\begin{equation}
 u=\frac{1}{4}t_1^2t_2^2+\frac{1}{2}t_1t_2+\frac{1}{4}t_1^2+\frac{1}{4}
\end{equation}
and we see that this integral violates the total degree bounds \eqref{eq:degboundBro} and \eqref{eq:degboundpmint}, but satisfies the bounds on partial degrees given by \eqref{eq:degboundElem} and \eqref{eq:degboundparrisch}.
\end{example}

A formal treatment of such reduction rules and reduction relations induced by them is presented in the next section.
Throughout the paper, we use some terminology from rewriting theory \cite{BaaderNipkow}, where reduction is seen as an abstract relation of objects such that $f$ is in relation to $g$ if and only if there is a reduction step from $f$ to $g$.
Recall that such a reduction relation is called \emph{normalizing} if for every reducible element $f$ there is an irreducible element $g$, also called \emph{normal form}, such that $f$ can be reduced to $g$ in finitely many steps.
It is called \emph{terminating} if, for every element $f$, every chain of reduction steps starting in $f$ is finite.

\section{Complete reduction systems}
\label{sec:RedSys}

In this section, we are going to formalize the reduction-based approach given in~\cite{NormanCriticalPair} so that there is a framework for handling the identities used for reduction.
First, we discuss general properties of such reduction systems.
Then, using our formalization, we represent Norman's completion process.
Moreover, we also exhibit a refinement and prove its correctness.
In addition, we work out one example posed as open problem in~\cite{NormanCriticalPair} precisely and show that our refined procedure terminates in this case.

The left hand side of \eqref{eq:MainEquation} leads to a differential operator $L$ acting on polynomials $u$.
In fact, to apply reduction systems, $L$ is considered as a linear map only.
Throughout this section, we consider the ring of polynomials with coefficients in a field $C$ of characteristic zero in the indeterminates $t_1,\dots,t_n$ and a $C$-linear map $L:C[t] \to C[t]$ defined on the monomial basis by some Laurent polynomial $p \in C[\param][t,t^{-1}]$ via
\begin{equation}\label{eq:defL}
 L(t^\alpha):=p(\alpha,t)t^\alpha
\end{equation}
for all $\alpha \in \mathbb{N}^n$.
Note that the coefficients of $p$ are polynomials in indeterminates $\param_1,\dots,\param_n$ and are restricted by the fact that $p(\alpha,t)t^\alpha$ has to lie in $C[t]$ for all $\alpha \in \mathbb{N}^n$.
So, for each Laurent monomial $m \in \supp_t(p)$ and $i\in\{1,\dots,n\}$, $k:=-\deg_{t_i}(m)>0$ implies that $\coeff(p,m)$ is divisible by $(\param_i-k+1)_k$ in $C[\param]$.

In the context of the Risch--Norman algorithm, for fixed nonzero $v \in C[t]$, the left hand side of \eqref{eq:MainEquation} amounts to $L(u)$, where $L$ can be given in the form \eqref{eq:defL} by
\begin{equation}\label{eq:defP}
 p:=\frac{v}{\gcd(v,\tilde{\Der}{v})}\sum_{i=1}^n\param_i\frac{\tilde{\Der}t_i}{t_i}-\frac{\tilde{\Der}v}{\gcd(v,\tilde{\Der}{v})} \in C[\param][t,t^{-1}].
\end{equation}
For $L$ defined this way, we always have $v \in \ker(L)$.

For understanding the completion process presented later, it may be helpful to consider the linear algebra view.
Whenever doing so, to simplify notation, we assume that an order of monomials is chosen such that for every monomial there are only finitely many monomials that are smaller.
This assumption allows us to enumerate all monomials $m_0,m_1,\ldots$ in $\{t^\alpha\ |\ \alpha\in\mathbb{N}^n\}$ in ascending order.
This gives an ordered basis of $C[t]$ and, in terms of this basis, we can represent any polynomial $q \in C[t]$ by an infinite vector (resp.\ sequence) of coefficients
\[\coeff(q):=(\coeff(q,m_i))_{i\in\mathbb{N}}.\]
More generally, using an arbitrary monomial order, the set of monomials is no longer order isomorphic to the natural numbers in general.
This would result in more complicated notation for the linearly ordered index set, which is needed to enumerate monomials in the given monomial order, but the basic linear algebra view would still apply mutatis mutandis.

Similarly, the $C$-linear map $L$ yields the infinite matrix $M \in C^{\mathbb{N}\times\mathbb{N}}$ whose columns are given by the infinite vectors $\coeff(L(m_0)),\coeff(L(m_1)),\dots$.
We will use this notation again in Sections \ref{sec:NormanCompletion} and \ref{sec:RefinedCompletion} for discussing the linear algebra view.
Since $L$ maps polynomials to polynomials, there are only finitely many nonzero entries in each column of $M$ and, since $L$ is given by \eqref{eq:defL}, also each row has only finitely many nonzero entries.

For two polynomials $f,g \in C[t]$, the identity $L(g)=f$ is equivalent to
\begin{equation}\label{eq:linearredrule}
 \coeff(f)=M\cdot\coeff(g).
\end{equation}
Since $\im(L)$ is infinite dimensional by definition \eqref{eq:defL} whenever $p\neq0$, there are infinitely many linearly independent identities of the form $L(g)=f$.
In order to deal with them in a finite way, we exploit the patterns of coefficients implied by \eqref{eq:defL} to represent identities in a parametrized way as follows.

\begin{definition}
\label{def:conditionalidentity}
 Let $P,Q \in C[\param][t,t^{-1}]$ and let $B$ be a logical combination (i.e.\ using $\wedge$, $\vee$, and $\neg$) of equations, inequations, and inequalities in $C[\param]$. We say that $(P,Q,B)$ encodes a \emph{conditional identity} for $L$, if
 \begin{equation}\label{eq:conditionalidentity}
  B|_{\param=\alpha} \Longrightarrow L(Q(\alpha,t)t^\alpha)=P(\alpha,t)t^\alpha
 \end{equation}
 holds for all $\alpha\in\mathbb{N}^n$.
 If $(P,Q,B)$ encodes a conditional identity for $L$ with $P\neq0$ and $<$ is a semigroup order on monomials, we call the quotient $\delta(P,Q,B):=\lm_t(Q)/\lm_t(P)$ the \emph{offset} of $(P,Q,B)$ w.r.t.\ $<$.
\end{definition}

The conditional identity \eqref{eq:conditionalidentity} determined by the triple $(P,Q,B)$ generalizes the identity \eqref{eq:defL} determined by the polynomial $p$ resp.\ by $(P,Q,B)=(p,1,true)$.
Fixing a semigroup order $<$ on monomials, we can define when and how a conditional identity can be used for reduction.

\begin{definition}\label{DEF:reductionrule}
 Let $(P,Q,B)$ encode a conditional identity for $L$.
 We call $(P,Q,B)$ a \emph{reduction rule} for $L$ w.r.t.~$<$ if and only if $\lm_t(P)=1$ and for all $\alpha \in \mathbb{N}^n$ satisfying $B|_{\param=\alpha}$ we have $\lc_t(P)|_{\param=\alpha}\neq0$.
 If in addition $\lc_t(Q)|_{\param=\alpha}\neq0$ holds for all $\alpha \in \mathbb{N}^n$ satisfying $B|_{\param=\alpha}$, we say that the reduction rule $(P,Q,B)$ has \emph{exact offset}.
 We call a set of reduction rules a \emph{reduction system}.
\end{definition}

To illustrate the notions, the algorithm, and procedures in this section, we use \eqref{eq:NormanExample}, which is the same example as was used in \cite{NormanCriticalPair}.
\begin{example}
\label{ex:NormanFCI}
 As in Example~\ref{ex:NormanMain}, let $(C(t_1,t_2),\Der)$ be defined by $\Der{t_1}=1$ and $\Der{t_2}=t_2^2+1$ and let $v=t_2^2+1$.
 Then, we have $\den(\Der)=1$ and hence $\tilde\Der=\Der$.
 So, the definition~\eqref{eq:defP} yields $p=(\param_2-2)t_2+\param_2t_2^{-1}+\param_1t_1^{-1}$.
 Then, the $C$-linear map $L$ defined by \eqref{eq:defL} satisfies $L(u)=v\Der(u/v)$ for all $u \in C[t_1, t_2]$.
 This agrees with $L(u)=\Der{u}-2t_2u$ in Example~\ref{ex:NormanMain}.
 Trivially, $(P,Q,B)=(p,1,true)$ forms a conditional identity, since it satisfies \eqref{eq:conditionalidentity} for all $\alpha\in\mathbb{N}^2$ by definition.
 Moreover, to order monomials, we use the lexicographic order with $t_2<t_1$.
 Then, $(p,1,true)$ has offset $t_2^{-1}$ w.r.t.\ the order $<$.
 However, $(p,1,true)$ cannot be a reduction rule since $\lm_t(p)=t_2\neq1$.
\par
 To obtain a reduction rule for $L$, we simply perform a shift by $\beta=(0,1)$ to get $(P,Q,B)=(p(\param-\beta,t)/t^\beta,1/t^\beta,\param_2\ge1\wedge\param_2\neq3)$ with $\lm_t(P)=1$.
 Explicitly, $P=(\param_2-3)+(\param_2-1)t_2^{-2}+\param_1t_1^{-1}t_2^{-1}$ and $Q=t_2^{-1}$.
 With this choice of $B$, we check that \eqref{eq:conditionalidentity} still holds for all $\alpha\in\mathbb{N}^2$:
\[
 \alpha_2\ge1\wedge\alpha_2\neq3 \Longrightarrow L(\underbrace{t^{\alpha+(0,-1)}}_{\in C[t_1,t_2]}) = \underbrace{p(\alpha+(0,-1),t)t^{\alpha+(0,-1)}}_{v\Der (t^{\alpha+(0,-1)}/v)}.
\]
 Since $\lc_t(P)|_{\param=\alpha}=\alpha_2-3\neq0$ for all $\alpha\in\mathbb{N}^2$ with $B|_{\param=\alpha}$, $r_1:=(P,Q,B)$ is a reduction rule with offset $\delta(r_1)=t_2^{-1}$.
 Moreover, $\lc_t(Q)=1$ implies that $r_1$ has exact offset.
 Condition $B$ implies that every monomial $t^\alpha$ with $\alpha_2\ge1\wedge\alpha_2\neq3$ is reducible by $r_1$.
\end{example}

In conditional identities and in reduction rules, $\param$ stands for the exponent vectors of monomials and hence ranges only over $\mathbb{N}^n$.
Note that, in Definitions \ref{def:conditionalidentity} and \ref{DEF:reductionrule}, it is not required that there is an $\alpha\in\mathbb{N}^n$ such that $B|_{\param=\alpha}$ is true.
Moreover, the offset of a reduction rule $(P,Q,B)$ is given by $\lm_t(Q)$ and, for all $\alpha \in \mathbb{N}^n$ satisfying $B|_{\param=\alpha}$, it is at least as large as the quotient $\lm_t(Q(\alpha,t)t^\alpha)/\lm_t(P(\alpha,t)t^\alpha)$ in terms of $<$.
If the reduction rule has exact offset, then this quotient is equal to the offset for all $\alpha \in \mathbb{N}^n$ satisfying $B|_{\param=\alpha}$.

If $(P,Q,B)$ is a reduction rule, then for any $\alpha \in \mathbb{N}^n$ satisfying $B|_{\param=\alpha}$, $\lm_t(P)=1$ and $\lc_t(P)|_{\param=\alpha}\neq0$ together imply $\lm_t(P(\alpha,t)t^\alpha)=t^\alpha$. Hence, we can rewrite the monomial $t^\alpha$ as the sum of an element in the image of $L$ and a polynomial that is either zero or has a leading monomial smaller than $t^\alpha$:
\begin{equation}\label{eq:rewriterule}
 t^\alpha = L\left(\frac{Q(\alpha,t)t^\alpha}{\lc_t(P)|_{\param=\alpha}}\right)-\frac{P_{<1}(\alpha,t)t^\alpha}{\lc_t(P)|_{\param=\alpha}}.
\end{equation}
Since the rewrite rule \eqref{eq:rewriterule} allows to reduce $t^\alpha$ modulo the image of $L$ to a polynomial with leading monomial smaller than $t^\alpha$, we have the following definition.

\begin{definition}
 Let $r=(P,Q,B)$ be a reduction rule.
 We say that a monomial $t^\alpha$ is \emph{reducible by $r$}, if $B|_{\param=\alpha}$ holds.
 More generally, $f \in C[t]$ is \emph{reducible by $r$}, if some element of $\supp(f)$ is reducible by $r$.
 One \emph{reduction step} using $r$ replaces $f \in C[t]$ by $f-\frac{\coeff(f,t^\alpha)}{\lc_t(P)|_{\param=\alpha}}P(\alpha,t)t^\alpha$, if $t^\alpha\in\supp(f)$ and $B|_{\param=\alpha}$.
\end{definition}

As illustrated in Example~\ref{ex:NormanMain}, part of the preimage can be computed in a reduction step using \eqref{eq:rewriterule}.
\begin{example}
\label{ex:NormanReduction}
 Continuing Example~\ref{ex:NormanFCI}, we have $v=t_2^2+1$ and
 \[
  r_1=(\underbrace{(\param_2-3)+(\param_2-1)t_2^{-2}+\param_1t_1^{-1}t_2^{-1}}_{P},\underbrace{t_2^{-1}}_{Q},\underbrace{\param_2\ge1\wedge{\param_2}\neq3}_{B}).
 \]
 Condition $B$ implies that every monomial $t^\alpha$ with $\alpha_2\ge1\wedge\alpha_2\neq3$ is reducible by $r_1$.
 To illustrate the use of $r_1$, we present the reduction of $t_1t_2^2$ step by step.
 First, with $\alpha=(1,2)$, we obtain
 \[
  t_1t_2^2 = L\left(\frac{Q((1,2),t){\cdot}t_1t_2^2}{-1}\right)-\frac{P_{<1}((1,2),t){\cdot}t_1t_2^2}{-1} = L\left(-t_1t_2\right)+t_1+t_2,
 \]
 i.e.\ we reduced $t_1t_2^2$ to $t_1+t_2$.
 Next, to further reduce $t_1+t_2$ by $r_1$, we can only choose $\alpha=(0,1)$ and obtain
 \[
  t_2 = L\left(\frac{Q((0,1),t){\cdot}t_2}{-2}\right)-\frac{P_{<1}((0,1),t){\cdot}t_2}{-2} = L\left(-\tfrac{1}{2}\right),
 \]
 i.e.\ $t_1+t_2$ is reduced to $t_1$ and cannot be reduced further by $r_1$.
 Altogether, this implies that $t_1t_2^2 = L\left(-t_1t_2-\frac{1}{2}\right)+t_1$.
 In other words, the integration of $t_1t_2^2/v$, which can be viewed as $\frac{x\tan(x)^2}{\tan(x)^2+1}$, has been simplified to the integration of $t_1/v$, which corresponds to \eqref{eq:NormanExample}, by applying the reduction rule $r_1$ twice.
\end{example}

Note that, while the conditional identity in Example~\ref{ex:NormanFCI} contained information about $L(t^\alpha)$ for $\alpha_2=2$, the reduction rule $r_1$ does not cover such cases because this would mean $\lc_t(P)|_{\param=\alpha}=0$ when setting $\param=\alpha$ with $\alpha_2=3$ in $r_1$.
In order not to lose any information when turning conditional identities into reduction rules, one may need to construct several rules from a single conditional identity.
This is the purpose of Algorithm~\ref{alg:CItoRR} in the following section.
More generally, we introduce the notion of precompleteness in the following definition to formally express that all necessary information about $L$ is there.

\begin{definition}\label{DEF:complete}
 Let $S$ be a set of triples $(P,Q,B)$ encoding conditional identities for $L$.
 We say that $S$ is \emph{precomplete} for $L$, if
 \[
  \linspan_C\{P(\alpha,t)t^\alpha\ |\ (P,Q,B)\in S,\alpha\in\mathbb{N}^n,B|_{\param=\alpha}\}=\im(L).
 \]
 If $S$ is a reduction system for $L$ w.r.t.\ $<$, we call it \emph{complete} for $L$ w.r.t.\ $<$, if it is precomplete for $L$ and the leading monomial of every nonzero $f\in\im(L)$ is reducible by $S$.
\end{definition}

Evidently, a set $S$ of triples $(P,Q,B)$ encoding conditional identities for $L$ is precomplete for $L$ if and only if
\begin{equation}\label{eq:weakcompleteness}
 \linspan_C\{Q(\alpha,t)t^\alpha\ |\ (P,Q,B)\in{S},\alpha\in\mathbb{N}^n,B|_{\param=\alpha}\} + \ker(L) = C[t].
\end{equation}
Later, we will use the following straightforward sufficient criterion to show that a given precomplete reduction system is in fact complete.
\begin{lemma}\label{lem:NoRedundancy}
 If $S$ is a precomplete reduction system for $L$ w.r.t.\ $<$ such that every monomial can be reduced by at most one element of $S$, then $S$ is complete for $L$.
\end{lemma}
\begin{proof}
 Let $f \in \im(L)$ be nonzero.
 By precompleteness, there are nonzero $c_1,\dots,c_m \in C$ and pairwise different $f_1,\dots,f_m \in C[t]$ such that $f=\sum_{i=1}^mc_if_i$ and for every $i$ there are $(P,Q,B) \in S$ and $\alpha\in\mathbb{N}^n$ with $B|_{\param=\alpha}\wedge{f_i=P(\alpha,t)t^\alpha}$.
 By construction, each $\lm(f_i)$ is reducible by $S$.
 Since $f_1,\dots,f_m$ are pairwise different, $\lm(f_1),\dots,\lm(f_m)$ are pairwise different by assumption on $S$.
 Therefore, we have that $\lm(f)=\max_i\lm(f_i)$ and hence $\lm(f)$ is reducible by $S$.
\end{proof}
If $S$ is a complete reduction system for $L$ w.r.t.\ $<$ and induces a normalizing reduction relation, then the set of monomials that cannot be reduced by $S$ spans a direct complement of $\im(L)$.

Note that the requirement of being precomplete is not redundant in the definition of completeness, since it is not implied by reducibility of leading monomials of every nonzero $f\in\im(L)$ in general.
If the induced reduction relation is terminating, however, we easily obtain the following lemma characterizing completeness of the reduction system.

\begin{lemma}
\label{lem:completeness}
 Let $S$ be a reduction system for $L$ w.r.t.\ $<$.
 If the induced reduction relation on $C[t]$ is terminating, then the following are equivalent.
 \begin{enumerate}
  \item\label{item:lmreducible} $S$ is complete for $L$.
  \item\label{item:allreducible} Every nonzero $f\in\im(L)$ is reducible by $S$.
  \item\label{item:tozero} Every $f\in\im(L)$ is eventually reduced to zero by $S$.
 \end{enumerate}
\end{lemma}
\begin{proof}
 Property \ref{item:lmreducible} trivially implies property \ref{item:allreducible}.
 To show property \ref{item:tozero} from \ref{item:allreducible}, we let $f\in\im(L)$.
 Since the reduction relation is terminating, iteratively reducing $f$ by $S$ eventually yields some $g \in C[t]$ which cannot be reduced further.
 By $f\in\im(L)$, it follows that $g\in\im(L)$, which implies $g=0$ by property \ref{item:allreducible}.
\par
 To show property \ref{item:lmreducible} from \ref{item:tozero}, we let $f\in\im(L)\setminus\{0\}$.
 Since $f$ can be reduced to zero by $S$, it is a $C$-linear combination of some $P(\alpha,t)t^\alpha$ with $(P,Q,B)\in{S},\alpha\in\mathbb{N}^n,B|_{\param=\alpha}$.
 Moreover, $\lm(f)\neq0$ has to be reducible by $S$, since $f$ could not be reduced to zero by $S$ otherwise.
\end{proof}

In general, however, a complete reduction system for $L$ may not be able to reduce every $f\in\im(L)$ to zero in finitely many steps.
In fact, a reduction system $S$ for $L$ can reduce every $f\in\im(L)$ to zero in finitely many steps if and only if $S$ is both complete for $L$ and induces a normalizing reduction relation on $\im(L)$.
Moreover, if every $f\in\im(L)$ can be reduced to zero by $S$ in finitely many steps, then $S$ induces a confluent reduction relation on $C[t]$, since any two polynomials that differ by an element from $\im(L)$ can be reduced to the same polynomial in finitely many steps in that case.

Based on property \ref{item:tozero} listed in Lemma~\ref{lem:completeness}, a complete reduction system for $L$ that induces a terminating reduction relation enables to decide straightforwardly if
\begin{equation}\label{eq:ODE}
 L(u)=f
\end{equation}
has a solution $u \in C[t]$ for given $f \in C[t]$.
At the same time, it allows to compute such a solution if it exists, as illustrated in Example~\ref{ex:NormanMain}.
In fact, for any polynomial $f$, we can compute an additive decomposition $f=L(u)+r$ in this way, where $r$ is the normal form of $f$.
Since the reduction system is complete, the normal form lies in a direct complement of $\im(L)$ and hence is unique.
Moreover, if the right hand side of \eqref{eq:ODE} has the form $f_0+\sum_{i=1}^mc_if_i$ with $f_i \in C[t]$ and undetermined $c_i \in C$ as in \eqref{eq:MainEquation}, then, using such a reduction system to reduce every $f_i$ to its normal form, we can obtain an explicit representation of all $c \in C^m$ that permit a solution $u \in C[t]$ along with a representation of a corresponding $u$.

If $p \in C[\param][t,t^{-1}]$ is given such that \eqref{eq:defL} holds, a precomplete set of \emph{basic reduction rules} $(P,Q,B)$ where $Q$ is just a (Laurent) monomial can be created easily.
However, these basic rules do not form a complete reduction system in general.
So, given $L$ and $<$, the general outline to construct a complete reduction system naturally is as follows, see also~\cite{NormanCriticalPair}.
\begin{enumerate}
 \item Create basic rules $r_1,\dots,r_m$ from $(p,1,true)$.
 \item Compute a complete reduction system $S$ from $\{r_1,\dots,r_m\}$.
\end{enumerate}
How this can be done in detail will be discussed for the rest of this section.
In particular, Algorithm~\ref{alg:CItoRR} can produce the basic rules, which are the input for the completion process given by Procedure~\ref{proc:NormanCompletion} or by Procedure~\ref{proc:RefinedCompletion}.
Before entering this discussion, we still need to point out an important algorithmic aspect.

When reduction rules are given, it is straightforward to decide if any given polynomial is reducible by simply plugging exponent vectors into conditions.
In contrast, the definitions and procedures in the rest of this section also involve more difficult problems.
In particular, these are equivalent to asking whether a certain formula $B$ formed from the constituents of reduction rules can be satisfied by an element of $\mathbb{N}^n$.
Even if these conditions $B$ were restricted to conjunctions of equations, it is known from the Davis--Putnam--Robinson--Matiyasevich Theorem that $\exists \,\alpha\in\mathbb{N}^n:B|_{\param=\alpha}$ is undecidable in general, see e.g.\ \cite{JonesMatiyasevich}.
To deal with the undecidability of existential statements over $\mathbb{N}$ in the formalized steps of the completion process that follow, we adopt the following paradigm, since we do not want to restrict applicability of those procedures to instances that can be decided in practice.

\begin{remark}\label{rem:relaxation}
 In order to enable evaluating each instance of existential statements over $\mathbb{N}$ arising during an algorithm or procedure in finite time, we allow also false positive answers (but no false negative answers) to all statements quantified by $\exists \, \alpha\in\mathbb{N}^n$ at any step of the algorithm and procedures stated in the rest of this section.
 We will always use this viewpoint when showing correctness and other properties of those algorithm and procedures.
 For easier readability, we do not introduce extra notation to signify this tolerance for false positive evaluations of existential statements.
 In other words, when existence of an element of $\mathbb{N}^n$ satisfying a given condition is asked in an algorithm or procedure, one may safely proceed assuming that such an element exists if one cannot detect easily that such an element does not exist.
 \qed
\end{remark}

In theory, this would even permit to replace all statements quantified by $\exists \, \alpha\in\mathbb{N}^n$ with the truth value $true$ in the formalization of the completion process.
However, since false positives in general will cause unnecessary computations and can even lead to non-termination of computations that otherwise would terminate, at least some effort should be made to detect nonexistence.
In practice, there is a wide range of options from simple Boolean manipulations that exhibit contradictions to more sophisticated manipulations of the polynomials involved.
For instance, one can relax the existential statement into a decidable one such that every witness $\alpha\in\mathbb{N}^n$ of the original statement is also a witness of the relaxed statement.
One way is to remove all restrictions in the condition that are inequalities so that only a logical combination of equations and inequations in $C[\param]$ remains and no solutions are lost.
Existence of solutions in $\overline{C}^n$ can then be decided via Gr\"obner bases.
If $C$ is presented as a field extension of $\mathbb{Q}$, another way is to replace every equation (resp.\ inequation) in $C[\param]$ by an equivalent conjunction of equations (resp.\ disjunction of inequations) in $\mathbb{Q}[\param]$ that has the same solutions in $\overline{\mathbb{Q}}^n$.
Existence of solutions in $(\overline{\mathbb{Q}}\cap[0,\infty[)^n$ can then be decided via cylindrical algebraic decomposition and related methods.
Moreover, techniques from satisfiability modulo theories (SMT) can be applied to detect that solutions of conditions in $\mathbb{Q}[\param]$ do or do not exist.

\subsection{From conditional identities to reduction rules}
\label{sec:CItoRR}

Let $(P,Q,B)$ encode a conditional identity for $L$ such that $P(\alpha,t)\neq0$ for at least one $\alpha\in\mathbb{N}^n$ that satisfies $B|_{\param=\alpha}$.
Such a conditional identity gives rise to one or more reduction rules.
In particular, if there exists $\alpha\in\mathbb{N}^n$ satisfying $B|_{\param=\alpha}$ such that $\lc_t(P)|_{\param=\alpha}$ is nonzero, then, with $\beta$ being the exponent vector of $\lm_t(P)$, the main rule is given by $r_1=(P_1,Q_1,B_1)$ where $P_1:=P(\param-\beta,t)/t^\beta$, $Q_1:=Q(\param-\beta,t)/t^\beta$ and $B_1:=(B \wedge \lc_t(P)\neq0)|_{\param=\param-\beta}$, as in Example~\ref{ex:NormanFCI}.
Further reduction rules arise from degenerate cases when $\lc_t(P)|_{\param=\alpha}=0$.
In full generality, we follow the algorithm below to generate reduction rules from a triple $(P,Q,B)$ encoding a conditional identity for $L$.

\begin{algorithm}[ht]
\caption{Convert a conditional identity into reduction rules}
\label{alg:CItoRR}
\begin{algorithmic}[1]
\REQUIRE a semigroup order $<$ on monomials and a triple $(P,Q,B)$ encoding a conditional identity for $L$
\ENSURE reduction rules $r_1,\dots,r_m$ for $L$ w.r.t.\ $<$ s.t.\ for every $\alpha\in\mathbb{N}^n$ satisfying $B|_{\param=\alpha}$ either $P(\alpha,t)=0$ or there is $r_i=(P_i,Q_i,B_i)$ among $r_1,\dots,r_m$ such that the exponent vector $\gamma$ of $\lm_t(P(\alpha,t)t^\alpha)$ satisfies $B_i|_{\param=\gamma}$, $P(\alpha,t)t^\alpha=P_i(\gamma,t)t^\gamma$, $Q(\alpha,t)t^\alpha=Q_i(\gamma,t)t^\gamma$, and $\lm_t(Q)t^\alpha=\lm_t(Q_i)t^\gamma$
\STATE $m:=0$
\WHILE{$P\neq0\wedge\exists \, {\alpha\in\mathbb{N}^n}: B|_{\param=\alpha} \wedge P(\alpha,t)\neq0$}\label{line:checknonzero}
\IF{$\exists \, {\alpha\in\mathbb{N}^n}: B|_{\param=\alpha} \wedge \lc_t(P)|_{\param=\alpha}\neq0$}\label{line:checkleadingcoeff}
\STATE\label{line:findshift} $m:=m+1,\beta:=\text{exponent vector of }\lm_t(P)$
\STATE\label{line:newrule} $P_m:=P(\param-\beta,t)/t^\beta$, $Q_m:=Q(\param-\beta,t)/t^\beta$,
\newline
$B_m:=\left(B|_{\param=\param-\beta} \wedge \lc_t(P)|_{\param=\param-\beta}\neq0 \wedge \bigwedge\limits_{\substack{i=1\\\beta_i>0}}^n\param_i\ge\beta_i\right)$
\STATE $r_m:=(P_m,Q_m,B_m)$
\STATE\label{line:updateB} $B:=(B \wedge \lc_t(P)=0)$
\ENDIF
\STATE\label{line:reduceP} $P:=P-\lt_t(P)$
\ENDWHILE
\RETURN $r_1,\dots,r_m$
\end{algorithmic}
\end{algorithm}

\begin{lemma}\label{lem:invariant}
 Let $(P_0,Q_0,B_0)$ be the input $(P,Q,B)$ of Algorithm~\ref{alg:CItoRR}.
 At the beginning of each iteration of the loop in Algorithm~\ref{alg:CItoRR}, the modified triple $(P,Q,B)$ encodes a conditional identity for $L$ such that $P=(P_0)_{\le\lm_t(P)}$ is nonzero, $Q=Q_0$, and $B$ is equivalent to $B_0\wedge\bigwedge\{\coeff(P_0,t^\gamma)=0\ |\ t^\gamma\in\supp_t(P_0),t^\gamma>\lm_t(P)\}$ over $\mathbb{N}$.
\end{lemma}
\begin{proof}
 For the first iteration of the loop, this holds trivially.
 For the inductive proof in subsequent iterations, we denote $(P,Q,B)$ at the beginning of the current iteration by $(P_\mathrm{old},Q_\mathrm{old},B_\mathrm{old})$ and assume it satisfies the claim.
 At the end of the current iteration, we distinguish two cases.
 If $B$ was modified, we have that $B=(B_\mathrm{old}\wedge\lc_t(P_\mathrm{old})=0)$.
 If $B$ was not modified, we have by Remark~\ref{rem:relaxation} that the condition in line~\ref{line:checkleadingcoeff} was indeed false, i.e.\ $B_\mathrm{old}$ implies $\lc_t(P_\mathrm{old})=0$ over $\mathbb{N}$.
 In both cases, $B$ is equivalent to $B_\mathrm{old}\wedge\coeff(P_\mathrm{old},\lm_t(P_\mathrm{old}))=0$ over $\mathbb{N}$.
 Unless the current iteration is the last, we explicitly have $P\neq0$ at the beginning of the next iteration, since that part of line~\ref{line:checknonzero} is not affected by Remark~\ref{rem:relaxation}.
 Altogether, these properties imply that $(P,Q,B)=(P_\mathrm{old}-\lt_t(P_\mathrm{old}),Q_\mathrm{old},B)$ satisfies the claim since $(P_\mathrm{old},Q_\mathrm{old},B_\mathrm{old})$ does and since $\lm_t(P_\mathrm{old})$ is the only monomial of $P_\mathrm{old}$ larger than $\lm_t(P)$.
\end{proof}

\begin{theorem}\label{thm:correctness}
 Algorithm~\ref{alg:CItoRR} terminates and is correct.
\end{theorem}
\begin{proof}
 In each iteration of the loop, line~\ref{line:reduceP} removes one term from $P$.
 Since the inequation $P\neq0$ in line~\ref{line:checknonzero} therefore would eventually be violated even if the existential statement in line~\ref{line:checknonzero} is always considered true, the algorithm terminates.
\par
 Let $(P_0,Q_0,B_0)$ be the input of Algorithm~\ref{alg:CItoRR} and let $r_1,\dots,r_m$ be its output.
 At the time when $r_i$, $i\in\{1,\dots,m\}$, is created, we have $P\neq0$ and $t^\beta=\lm_t(P)$ by lines \ref{line:checknonzero} and \ref{line:findshift}, which implies $\lm_t(P_i)=\lm_t(P)/t^\beta=1$ by line~\ref{line:newrule}.
 Moreover, for all $\alpha\in\mathbb{N}^n$ with $B_i|_{\param=\alpha}$, the definitions in line~\ref{line:newrule} imply that $\alpha-\beta\in\mathbb{N}^n$, $B|_{\param=\alpha-\beta}$, and $\lc_t(P_i)|_{\param=\alpha} = \lc_t(P)|_{\param=\alpha-\beta} \neq 0$.
 By Lemma~\ref{lem:invariant}, it follows that $L(Q_i(\alpha,t)t^\alpha) = L(Q(\alpha-\beta,t)t^{\alpha-\beta}) = P(\alpha-\beta,t)t^{\alpha-\beta} = P_i(\alpha,t)t^\alpha$ for all such $\alpha$.
 These properties of $r_i$ hold regardless whether the existential statements in lines \ref{line:checknonzero} and \ref{line:checkleadingcoeff} were wrongly or correctly assumed to be true.
 Altogether, all $r_1,\dots,r_m$ are reduction rules for $L$ w.r.t.\ $<$.
\par
 Now, let $\alpha\in\mathbb{N}^n$ satisfy $B_0|_{\param=\alpha}$ and assume $P_0(\alpha,t)\neq0$.
 We let $t^{\tilde{\beta}}:=\lm(P_0(\alpha,t))$ and $\gamma:=\alpha+\tilde{\beta}$.
 As long as the monomial $t^{\tilde{\beta}}:=\lm(P_0(\alpha,t))$ is not yet removed from $P$ in Algorithm~\ref{alg:CItoRR}, the condition in line~\ref{line:checknonzero} is always fulfilled and the loop continues.
 Eventually, there is an iteration in which $\lm_t(P)=t^{\tilde{\beta}}$ holds at the beginning.
 In this particular iteration, $B|_{\param=\alpha}$ and $\lc_t(P)|_{\param=\alpha} = \coeff(P_0,t^{\tilde{\beta}})|_{\param=\alpha} = \lc(P_0(\alpha,t))\neq0$ hold by Lemma~\ref{lem:invariant}.
 So, the condition in line~\ref{line:checkleadingcoeff} is true and the exponent vector $\tilde{\beta}$ is chosen in line~\ref{line:findshift}.
 The rule $r_i=(P_i,Q_i,B_i)$ created afterwards satisfies $P_i(\gamma,t)t^\gamma=P_0(\alpha,t)t^\alpha$, $Q_i(\gamma,t)t^\gamma=Q_0(\alpha,t)t^\alpha$, and $\lm_t(Q_i)t^\gamma=\lm_t(Q_0)t^\alpha$ by line~\ref{line:newrule} and by Lemma~\ref{lem:invariant}.
 For the same reason, we also have $B_i|_{\param=\gamma}$ since $\gamma\ge\tilde{\beta}$ holds componentwise.
\end{proof}

From the definition of $L$, we immediately get the conditional identity \eqref{eq:conditionalidentity} where $P:=p$, $Q:=1$, and $B:=true$.
We can apply Algorithm~\ref{alg:CItoRR} to $(p,1,true)$ to obtain basic reduction rules $r_1,\dots,r_m$.
The first rule $r_1$ is called the \emph{generic rule}.
In particular, $r_1=(P_1,Q_1,B_1)$ is given by $P_1:=p(\param-\beta,t)/\lm_t(p)$, $Q:=1/\lm_t(p)$, and $B_1:=(\lc_t(p)\neq0)|_{\param=\param-\beta}$, where $\beta$ is the exponent vector of $\lm_t(p)$, cf.\ Example~\ref{ex:NormanFCI}.
\begin{lemma}\label{lem:basicrules}
 The output of applying Algorithm~\ref{alg:CItoRR} to $(p,1,true)$ forms a precomplete reduction system for $L$ w.r.t.\ $<$.
 Moreover, each element $r_i=(P_i,Q_i,B_i)$ of this reduction system has exact offset and is such that $Q_i$ is a Laurent monomial.
\end{lemma}
\begin{proof}
 Correctness of the algorithm implies that the output forms a reduction system for $L$ w.r.t.\ $<$ and that any $t^\alpha$, $\alpha\in\mathbb{N}^n$, either lies in $\ker(L)$ or is of the form $Q_i(\gamma,t)t^\gamma$ for some $i\in\{1,\dots,m\}$ and $\gamma\in\mathbb{N}^n$ s.t.\ $B_i|_{\param=\gamma}$.
 Therefore, the reduction system satisfies \eqref{eq:weakcompleteness}.
 Any $Q_i$ created in line~\ref{line:newrule} is of the form $t^{-\beta}$ by Lemma~\ref{lem:invariant}.
 In particular, $\lc_t(Q_i)=1$ implies exact offset.
\end{proof}

\begin{example}
\label{ex:NormanBasicRules}
For $L$ and $<$ as defined in Example~\ref{ex:NormanFCI}, we compute the basic rules by applying Algorithm~\ref{alg:CItoRR} to $(P,Q,B)=((\param_2-2)t_2+\param_2t_2^{-1}+\param_1t_1^{-1},1,true)$.
First, we obtain $\beta=(0,1)$ and the reduction rule
\[
 r_1=(\underbrace{(\param_2-3)+(\param_2-1)t_2^{-2}+\param_1t_1^{-1}t_2^{-1}}_{P_1},\underbrace{t_2^{-1}}_{Q_1},\underbrace{\param_2\ge1\wedge\param_2\neq3}_{B_1})
\]
exactly as in Example~\ref{ex:NormanFCI}.
Then, $B$ and $P$ are updated to $B=(\param_2=2)$ and $P=\param_2t_2^{-1}+\param_1t_1^{-1}$.
So, in the next iteration of the loop, we have $\beta=(0,-1)$.
Based on line~\ref{line:newrule}, this yields the reduction rule
\[
 r_2=(\underbrace{(\param_2+1)+\param_1t_1^{-1}t_2}_{P_2},\underbrace{t_2}_{Q_2},B_2)
\]
with condition $B_2=(\param_2=1\wedge\param_2\neq-1)$, which can be simplified to the equivalent condition $B_2=(\param_2=1)$.
Next, $B$ and $P$ are updated to $B=(\param_2=2\wedge\param_2=0)$ and $P=\param_1t_1^{-1}$.
Now, the condition in line~\ref{line:checknonzero} is violated since the new condition $B$ cannot be satisfied by any element of $\mathbb{N}^2$.
Hence the algorithm stops and returns the basic rules $r_1,r_2$.
\end{example}

It is easy to see that the reduction rules computed by Algorithm~\ref{alg:CItoRR} have pairwise distinct offset, if the input satisfies $Q\neq0$.
However, if the input satisfies $Q=0$, we have $P(\alpha,t)=0$ whenever $B|_{\param=\alpha}$ by \eqref{eq:conditionalidentity} and consequently the output of Algorithm~\ref{alg:CItoRR} will be empty or, by Remark~\ref{rem:relaxation}, consist entirely of rules whose condition $B_i$ is inconsistent over $\mathbb{N}$ by construction.
Since we will use it in Section~\ref{sec:RefinedCompletion}, we also show the following property of the output of Algorithm~\ref{alg:CItoRR}.

\begin{lemma}\label{lem:CItoRR}
 Let $r_1,\dots,r_m$ be the result of Algorithm~\ref{alg:CItoRR} with input $(P_0,Q_0,B_0)$.
 For all $i\in\{1,\dots,m\}$ and $\gamma\in\mathbb{N}^n$ with $B_i|_{\param=\gamma}$ there exists $\alpha\in\mathbb{N}^n$ such that $B_0|_{\param=\alpha}$ and $\lm_t(Q_i)t^\gamma=\lm_t(Q_0)t^\alpha$.
 Conversely, for every $\alpha\in\mathbb{N}^n$ with $B_0|_{\param=\alpha}$, there exist at most one $i\in\{1,\dots,m\}$ and $\gamma\in\mathbb{N}^n$ with $B_i|_{\param=\gamma}$ and $\lm_t(Q_i)t^\gamma=\lm_t(Q_0)t^\alpha$.
 Moreover, if $\lc_t(Q_0)|_{\param=\alpha}\neq0$ holds for all $\alpha\in\mathbb{N}^n$ with $B_0|_{\param=\alpha}$, then $r_1,\dots,r_m$ have exact offset.
\end{lemma}
\begin{proof}
 Let $\beta^{(1)},\dots,\beta^{(m)}\in\mathbb{N}^n$ be the vectors $\beta$ used in line~\ref{line:newrule} for constructing $r_1,\dots,r_m$, respectively.
 By this construction and Lemma~\ref{lem:invariant}, it follows for all $i \in \{1,\dots,m\}$ that $Q_i=Q_0(\param-\beta^{(i)},t)/t^{\beta^{(i)}}$ and that $B_i|_{\param=\gamma}$ implies $B_0|_{\param=\gamma-\beta^{(i)}}$, $t^{\beta^{(i)}}=\lm(P_0(\gamma-\beta^{(i)},t))$, and $\gamma-\beta^{(i)}\in\mathbb{N}^n$ for all $\gamma\in\mathbb{N}^n$.
\par
 Consequently, for $\gamma\in\mathbb{N}^n$ with $B_i|_{\param=\gamma}$, it follows that $\alpha:=\gamma-\beta^{(i)}$ satisfies $\alpha\in\mathbb{N}^n$, $B_0|_{\param=\alpha}$, and $\lm_t(Q_i)t^\gamma=\lm_t(Q_0)t^\alpha$.
 Moreover, together with $Q_i(\gamma,t)t^\gamma=Q_0(\alpha,t)t^\alpha$, this implies $\lc_t(Q_i)|_{\param=\gamma}=\lc_t(Q_0)|_{\param=\alpha}$.
 Conversely, let $\alpha\in\mathbb{N}^n$ satisfy $B_0|_{\param=\alpha}$, then requiring $\lm_t(Q_i)t^\gamma=\lm_t(Q_0)t^\alpha$ from $\gamma\in\mathbb{N}^n$ implies that $\gamma=\alpha+\beta^{(i)}$.
 Additionally imposing $B_i|_{\param=\gamma}$ implies $t^{\beta^{(i)}}=\lm(P_0(\alpha,t))$, which uniquely fixes this monomial independent of $i$.
 Therefore at most one $i$ can be used, since $\beta^{(1)},\dots,\beta^{(m)}$ are pairwise different as $\lm_t(P)$ decreases in terms of $<$ in each iteration of the loop.
\end{proof}

\subsection{Norman's completion process}
\label{sec:NormanCompletion}
\floatname{algorithm}{Procedure}

For formalizing the completion process given in~\cite{NormanCriticalPair}, we formalize two concepts that are needed to create new reduction rules from existing ones.
So, in addition to reduction of monomials and polynomials, we also define what it means to reduce a conditional identity.

\begin{definition}
\label{defn:reducible}
 Let $(P,Q,B)$ with $P\neq0$ encode a conditional identity for $L$, let $\beta$ be the exponent vector of $\lm_t(P)$ and let $r_1=(P_1,Q_1,B_1)$ be a reduction rule for $L$. We say that $(P,Q,B)$ is \emph{reducible by $r_1$}, if $B|_{\param=\alpha} \Longrightarrow (B_1|_{\param=\alpha+\beta} \wedge \lc_t(P_1)|_{\param=\alpha+\beta}\neq0)$ holds for all $\alpha\in\mathbb{N}^n$. In that case, the reduction $(\tilde{P},\tilde{Q},B)$ of $(P,Q,B)$ by $r_1$ is given by
 \[
 \tilde{P}:=\frac{\lc_t(P_1)|_{\param=\param+\beta}}{g}P-\frac{\lc_t(P)}{g}P_1(\param+\beta,t)t^\beta, \]
 \[\tilde{Q}:=\frac{\lc_t(P_1)|_{\param=\param+\beta}}{g}Q-\frac{\lc_t(P)}{g}Q_1(\param+\beta,t)t^\beta
 \]
 where $g:=\gcd(\lc_t(P),\lc_t(P_1)|_{\param=\param+\beta}) \in C[\param]$.
\end{definition}

Note that, reduction changes only the (Laurent) polynomials $P,Q$ and leaves the condition $B$ untouched.
If, in Definition~\ref{defn:reducible}, $(P,Q,B)$ is such that $\lm_t(P)$ is a monomial without negative exponents, then we have $\alpha+\beta\in\mathbb{N}^n$ for all $\alpha\in\mathbb{N}^n$ and the condition $B|_{\param=\alpha} \Longrightarrow (B_1|_{\param=\alpha+\beta} \wedge \lc_t(P_1)|_{\param=\alpha+\beta}\neq0)$ is equivalent to $B|_{\param=\alpha} \Longrightarrow B_1|_{\param=\alpha+\beta}$.
In particular, if $(P,Q,B)$ has $\lm_t(P)=1$ and its condition is of the form $B=\tilde{B}\wedge{B_1}$ for some $\tilde{B}$, then it is trivially reducible by $r_1$.
Therefore, no checks for reducibility will be needed in Procedure~\ref{proc:NormanCompletion} and they will appear only in line~\ref{line:checkreduction} of Procedure~\ref{proc:RefinedCompletion}.
The definition immediately implies the following properties.

\begin{lemma}
\label{lem:reduction}
 Let $(P,Q,B)$, $\beta$, $r_1=(P_1,Q_1,B_1)$, and $\tilde{P},\tilde{Q},g$ be as in Definition~\ref{defn:reducible}.
 Then, $(\tilde{P},\tilde{Q},B)$ encodes a conditional identity for $L$ and we have $\lm_t(\tilde{P})<\lm_t(P)$.
 In particular, for all $\alpha\in\mathbb{N}^n$ satisfying $B|_{\param=\alpha}$, we have $\tilde{P}(\alpha,t)=\frac{\lc_t(P_1)|_{\param=\alpha+\beta}}{g(\alpha)}P(\alpha,t)-\frac{\lc_t(P)|_{\param=\alpha}}{g(\alpha)}P_1(\alpha+\beta,t)t^\beta$ with $\lc_t(P_1)|_{\param=\alpha+\beta}$ and $g(\alpha)$ being nonzero and $\alpha+\beta\not\in\mathbb{N}^n \Longrightarrow \lc_t(P)|_{\param=\alpha}=0$.
 If $\delta(r_1)\neq\delta(P,Q,B)$, then $\delta(\tilde{P},\tilde{Q},B)>\delta(P,Q,B)$.
\end{lemma}
\begin{proof}
 By construction of $\tilde{P}$, the leading terms of $P$ and $P_1(\param+\beta,t)t^\beta$ cancel each other so that $\lm_t(\tilde{P})<\lm_t(P)$.
 Let $\alpha\in\mathbb{N}^n$ satisfy $B|_{\param=\alpha}$.
 By Definition~\ref{defn:reducible}, both $B_1|_{\param=\alpha+\beta}$ and $\lc_t(P_1)|_{\param=\alpha+\beta} \neq 0$ hold.
 Hence, $g(\alpha)$ is nonzero because $g$ is a factor of $\lc_t(P_1)|_{\param=\param+\beta}$.
 If $\alpha+\beta\not\in\mathbb{N}^n$, then $\lc_t(P)|_{\param=\alpha} \neq 0$ would imply that the leading monomial of $P(\alpha,t)t^\alpha$ involves negative exponents.
 By \eqref{eq:conditionalidentity}, this would be in contradiction with the fact that $L$ is a map of $C[t]$ into itself.
 Thus, $\alpha+\beta\not\in\mathbb{N}^n \Longrightarrow \lc_t(P)|_{\param=\alpha}=0$.
 Since $r_1$ is a reduction rule, we altogether have that $\frac{\lc_t(P)|_{\param=\alpha}}{g(\alpha)}L(Q_1(\alpha+\beta,t)t^{\alpha+\beta}) = \frac{\lc_t(P)|_{\param=\alpha}}{g(\alpha)}P_1(\alpha+\beta,t)t^{\alpha+\beta}$, regardless of $\alpha+\beta$.
 Together with the construction of $\tilde{P},\tilde{Q}$, linearity of $L$, and $(P,Q,B)$ being a conditional identity, we obtain
 \begin{align*}
  L(\tilde Q(\alpha,t)t^\alpha) &= \frac{\lc_t(P_1)|_{\param=\alpha+\beta}}{g(\alpha)}L(Q(\alpha,t)t^\alpha)-\frac{\lc_t(P)|_{\param=\alpha}}{g(\alpha)}L(Q_1(\alpha+\beta,t)t^{\alpha+\beta})\\
  &=\frac{\lc_t(P_1)|_{\param=\alpha+\beta}}{g(\alpha)}P(\alpha,t) \cdot t^\alpha-\frac{\lc_t(P)|_{\param=\alpha}}{g(\alpha)} P_1(\alpha+\beta,t)t^\beta \cdot t^\alpha\\
  &=\tilde P(\alpha,t) \cdot t^\alpha,
 \end{align*}
 which implies that $(\tilde{P},\tilde{Q},B)$ is a conditional identity for $L$.
 Moreover, if $\delta(r_1)\neq\delta(P,Q,B)$, then the leading terms of $Q$ and $Q_1(\param+\beta,t)t^\beta$ cannot be canceled in the linear combination $\tilde Q$, which implies $\lm_t(\tilde Q)=\max(\lm_t(Q), \lm_t(Q_1)t^\beta)\ge \lm_t(Q)$.
 Together with $\lm_t(\tilde P)<\lm_t(P)$, we have that $\delta(\tilde{P},\tilde{Q},B)>\delta(P,Q,B)$.
\end{proof}

\begin{definition}
 Let $r_1=(P_1,Q_1,B_1),r_2=(P_2,Q_2,B_2)$ be two distinct reduction rules for $L$ w.r.t.\ $<$. We say that they form a \emph{critical pair}, if there is $\alpha\in\mathbb{N}^n$ such that $(B_1 \wedge B_2)|_{\param=\alpha}$ holds.
\end{definition}

By Remark~\ref{rem:relaxation}, checking in Procedure \ref{proc:NormanCompletion} or \ref{proc:RefinedCompletion} if two reduction rules form a critical pair may never miss an actual critical pair.
However, two rules may be assumed to form a critical pair even if they do not, which then results in additional iterations of the main loop of the procedures.

Norman's completion process can be formalized as Procedure~\ref{proc:NormanCompletion}.
There, for simplicity of notation, we use the convention that the three components of any reduction rule $r_i$ are always denoted by $P_i,Q_i,B_i$ using the same index.

\begin{algorithm}[ht]
\caption{Formalization of Norman's completion process}
\label{proc:NormanCompletion}
\begin{algorithmic}[1]
\REQUIRE a semigroup order $<$ on monomials and a precomplete reduction system $\{r_1,\dots,r_m\}$ for $L$ w.r.t.\ $<$
\ENSURE complete reduction system $S$ for $L$ w.r.t.\ $<$
\STATE $A:=\{(i,j) \in \{1,\dots,m\}^2\ |\ i<j\text{ and $r_i,r_j$ form a critical pair}\}$
\WHILE{$A\neq\emptyset$}
\STATE choose $(i,j) \in A$
\STATE $A:=A\setminus\{(i,j)\}$
\STATE reduce $(P_j,Q_j,B_i \wedge B_j)$ by $r_i$ to obtain $(P,Q,B)$
\IF{$P\neq0$}
\STATE create new reduction rules $r_{m+1},\dots,r_{m+k}$ from $(P,Q,B)$ by Algorithm~\ref{alg:CItoRR}
\STATE $m:=m+k$
\STATE $A:=A\cup\{(i,j) \in \{1,\dots,m\}^2\ |\ i<j,m-k<j,\linebreak[0]\text{ and $r_i,r_j$}\linebreak[0]\text{ form a}\linebreak[0]\text{ critical pair}\}$
\ENDIF
\ENDWHILE
\RETURN $\{r_1,\dots,r_m\}$
\end{algorithmic}
\end{algorithm}

\begin{example}
\label{ex:NormanCompletion}
Continuing Example~\ref{ex:NormanFCI}, we apply Procedure~\ref{proc:NormanCompletion} to the basic rules computed in Example~\ref{ex:NormanBasicRules}.
At the beginning, we have $A=\{(1,2)\}$, so there is only the choice $i=1,j=2$.
We reduce $((\param_2+1)+\param_1t_1^{-1}t_2,t_2,\param_2\ge1\wedge\param_2\neq3\wedge\param_2=1)$ by $r_1$ to obtain
\begin{align*}
 P &= (\param_2-3)P_2-(\param_2+1)P_1 = (1-\param_2^2)t_2^{-2}+\param_1(\param_2-3)t_1^{-1}t_2-\param_1(\param_2+1)t_1^{-1}t_2^{-1},\\
 Q &= (\param_2-3)Q_2-(\param_2+1)Q_1 = (\param_2-3)t_2-(\param_2+1)t_2^{-1},
\end{align*}
and same condition $B$, which can be simplified to the equivalent condition $B=(\param_2=1)$.
Since $\lc_t(P)|_{\param=\alpha}=1-\alpha_2^2$ vanishes for all $\alpha\in\mathbb{N}^2$ satisfying $B|_{\param=\alpha}$, applying Algorithm~\ref{alg:CItoRR} to $(P,Q,B)$ yields the reduction rule
\[
 r_3=(\underbrace{(\param_1+1)(\param_2-4)-(\param_1+1)\param_2t_2^{-2}}_{P_3},\underbrace{(\param_2-4)t_1-\param_2t_1t_2^{-2}}_{Q_3},\underbrace{\param_2=2}_{B_3}).
\]
This results in $A=\{(1,3)\}$, leaving only the choice $i=1,j=3$ for the next iteration of the loop.
Performing the computations, we obtain the new reduction rule
\begin{multline*}
 r_4=(\overbrace{-2(\param_1+1)(\param_2^2-2)-\param_1(\param_1+1)(\param_2-2)t_1^{-1}t_2}^{P_4},\\
 \underbrace{(\param_2-1)(\param_2-2)t_1t_2^2-(\param_2-1)(\param_2+2)t_1-(\param_1+1)(\param_2-2)t_2}_{Q_4},\underbrace{\param_2=0}_{B_4}),
\end{multline*}
which does not give rise to critical pairs with previous rules.
Hence, the procedure stops and returns the complete reduction system $\{r_1,r_2,r_3,r_4\}$.
\end{example}

Finally, we mention the subtle differences between the presentation of Norman~\cite{NormanCriticalPair} and our formalization of it.
Since termination of Procedure~\ref{proc:NormanCompletion} is not guaranteed even if $<$ is a monomial order, we do not require $<$ to be Noetherian.
In contrast to Definition~\ref{DEF:reductionrule}, Norman does not require $\lc_t(P)|_{\param=\alpha}\neq0$ explicitly in reduction rules, apart from using different notation.
Related to Algorithm~\ref{alg:CItoRR}, he just exemplifies the substitutions $\param=\param-\beta$ made in line~\ref{line:newrule} and states the necessity
{\lq\lq}to solve equations (which can sometimes be nonlinear and multivariate) to determine what values for index variables can cause a reduction to degenerate{\rq\rq}
on p.~203 in his paper.
However, this is undecidable in general, as we pointed out immediately before Remark~\ref{rem:relaxation}, and he does not mention an analogue of Remark~\ref{rem:relaxation}.
Furthermore, Norman seems to allow only systems of equations as conditions for reduction rules.
So, in comparison, our Algorithm~\ref{alg:CItoRR} generally will produce more restrictive conditions for the same input.
For reduction of monomials, he compensates this by allowing reduction of $t^\alpha$ only if $\lc_t(P)|_{\param=\alpha}\neq0$ and all (Laurent) monomials appearing in $P(\alpha,t)t^\alpha$ and $Q(\alpha,t)t^\alpha$ have nonnegative exponents.
In fact, the possible reduction steps by each of $r_1,\dots,r_4$ computed in Examples \ref{ex:NormanBasicRules} and \ref{ex:NormanCompletion} therefore agree exactly with those by each of (r1)--(r4) computed in Norman's paper.

Reviewing the linear algebra view, cf.\ \eqref{eq:linearredrule}, a reduction system in our sense can be understood as representation of two infinite matrices $V$ and $W$, satisfying
\begin{equation}\label{eq:RedSysMat}
 V = M \cdot W.
\end{equation}
The columns of $V$ resp.\ $W$ are given by all $\coeff(P(\alpha,t)t^\alpha)$ resp.\ $\coeff(Q(\alpha,t)t^\alpha)$ with $B|_{\param=\alpha}$ for any rule $(P,Q,B)$ of the system.

For a vector $\mathbf{v}=(v_0,v_1,\dots) \in C^{\mathbb{N}}$, we also define $\ell(\mathbf{v}):=\max\{i\in\mathbb{N}\ |\ v_i\neq0\}$ if $\mathbf{v}$ is nonzero.
If some polynomial $q \in C[t]$ is nonzero, then $\lm(q) = m_{\ell(\coeff(q))}$.
The standard unit vectors $\mathbf{e}_0,\mathbf{e}_1,\dots$ satisfy $\mathbf{e}_i=\coeff(m_i)$ for all $i \in \mathbb{N}$.

We start with the initial precomplete reduction system obtained from $(p,1,true)$, which consists of basic reduction rules (i.e.\ preimages are chosen to be monomials) and gives rise to
\begin{equation}\label{eq:initialQ}
 V_0=M\cdot{W_0} \quad\text{with}\quad W_0=(\mathbf{e}_0,\mathbf{e}_1,\mathbf{e}_2,\dots)
\end{equation}
up to permutation of columns and up to dropping columns that would be zero in $V_0$.
Further rules are introduced, if there exist columns $\mathbf{v}_j$ and $\mathbf{v}_k$ of $V=(\mathbf{v}_0,\mathbf{v}_1,\mathbf{v}_2,\dots)$ in \eqref{eq:RedSysMat} such that $\ell(\mathbf{v}_j)=\ell(\mathbf{v}_k)$.
Actually, two reduction rules related to these vectors form a critical pair and can give rise to several (possibly infinitely many) other such pairs of columns at the same time.
The new rule comes from the linear combinations $\mathbf{v}=c_1\mathbf{v}_j+c_2\mathbf{v}_k$ such that $\ell(\mathbf{v})<\ell(\mathbf{v}_j)$.
The corresponding new columns $\mathbf{w}$ in $W$ at the same position as $\mathbf{v}$ in $V$ are just the linear combinations $\mathbf{w}=c_1\mathbf{w}_j+c_2\mathbf{w}_k$ with the same $c_1,c_2,j,k$.
Therefore, introducing a new rule gives rise to new columns of the matrices $V$ and $W$ such that $V_\mathrm{new}=\left(V_\mathrm{old}\ \ \big|\ \ V_\mathrm{old}\cdot{T}\right)$ and $W_\mathrm{new}=\left(W_\mathrm{old}\ \ \big|\ \ W_\mathrm{old}\cdot{T}\right)$ with the same $T$, which has exactly two nonzero entries in each column.
Finally, we obtain a complete reduction system as soon as for any linear combination $\mathbf{v}$ of columns in $V$, there is some existing column $\mathbf{v}_k$ such that $\ell(\mathbf{v}) = \ell(\mathbf{v}_k)$.

\subsection{Refined completion process}
\label{sec:RefinedCompletion}

One sees that the completion process is closely related to Gaussian elimination on columns in the linear algebra view.
Gaussian elimination on rows of infinite matrices was already considered in \cite{Koetteritzsch}.
If, instead of introducing new columns as in the previous paragraph, the column operation $\mathbf{v}=c_1\mathbf{v}_j+c_2\mathbf{v}_k$ replaces $\mathbf{v}_j$ when $j > k$, then this corresponds to one elimination step.
Eventually, we obtain a matrix $V$ which is an upper column echelon form of $V_0=M\cdot{W_0}$ up to permutation of columns, since we started the elimination with \eqref{eq:initialQ}.
The property that each column and each row has only finitely many nonzero entries is maintained in each elimination step.
The simplifying assumption imposed on the order of monomials in the linear algebra view implies that the order is Noetherian.
By Noetherianity, each column is affected only by finitely many elimination steps.
So, $V=V_0\cdot{T}$ can be expressed by an upper triangular transformation matrix $T$.
Since $T$ has only finitely many nonzero entries in each of its columns, every column of $V$ and $W:=W_0\cdot{T}$ indeed corresponds to a polynomial in $C[t]$.
In other words, we found a factorization $V\cdot T^{-1}$ of $V_0$ such that $T^{-1}$ is upper triangular and $V$ is an upper column echelon form up to permutation of columns.

Allowing arbitrary semigroup orders on monomials, we give a refinement of Norman's completion process in the following.
Namely, instead of just adding new rules and keeping track of pairs of rules that have not been dealt with yet, we modify existing rules so that critical pairs no longer arise after they were dealt with.
To this end, when handling critical pairs in Procedure~\ref{proc:RefinedCompletion}, we split the reduction rule with larger $\lm_t(Q)$ into two parts according to the condition $B$ of the other rule.
In the part where this condition $B$ is satisfied, we eliminate $\lt_t(P)$ in line~\ref{line:spoly2} as before, while the part where $B$ is not satisfied is used to replace the original rule.
Forming the latter part in line~\ref{line:remainingcase} makes it necessary to deal with more general conditions than conjunctions of equations only, allowing logical combinations of equations and inequations at least.
In addition, in order to delay creating new rules for as long as possible, we do further reductions before creating new reduction rules from $(P,Q,B)$ computed in line~\ref{line:spoly2}.
Altogether, this reduces the number of critical pairs being considered.
This even causes the refined version to terminate in some cases where Norman's original completion process does not terminate.
In Section~\ref{sec:NormanTanLogExample}, we look at such an example in detail.

In short, the main differences between Procedures \ref{proc:NormanCompletion} and \ref{proc:RefinedCompletion} are that our refined version removes reduction rules in line~\ref{line:removerule}, which never happens in the original version, and that an inner loop reduces $(P,Q,B)$ further before adding new rules to the reduction system.
Moreover, we reduce conditional identities during the procedure only by rules that have smaller offset so that $\lm_t(Q)$ does not change during reduction.
This preserves exact offset of rules, as Lemma~\ref{lem:exactoffset} will show.

\begin{algorithm}[ht]
\caption{Refinement of Norman's completion process}
\label{proc:RefinedCompletion}
\begin{algorithmic}[1]
\REQUIRE a semigroup order $<$ on monomials and a precomplete reduction system $\{r_1,\dots,r_m\}$ for $L$ w.r.t.\ $<$ such that for any $\gamma \in \mathbb{N}^n$ there is at most one $r_i=(P_i,Q_i,B_i)$ with $\exists \, \alpha\in\mathbb{N}^n:B_i|_{\param=\alpha}\wedge\lm_t(Q_i)t^\alpha=t^\gamma$
\ENSURE complete reduction system $S$ for $L$ w.r.t.\ $<$
\STATE $S:=\{r_1,\dots,r_m\}$
\WHILE{$S$ has two elements with distinct offset that form a critical pair}
\STATE\label{line:selectCP2} choose some $r_i=(P_i,Q_i,B_i),r_j=(P_j,Q_j,B_j) \in S$ with $\lm_t(Q_i)<\lm_t(Q_j)$ that form a critical pair
\STATE\label{line:removerule} $S:=S\setminus\{r_j\}$
\IF{$\exists \, {\alpha\in\mathbb{N}^n}:B_j|_{\param=\alpha} \wedge\neg B_i|_{\param=\alpha}$}\label{line:checkremaining}
\STATE\label{line:remainingcase} $r_{m+1}:=(P_j,Q_j,B_j \wedge\neg B_i)$
\STATE $S:=S\cup\{r_{m+1}\}$
\STATE $m:=m+1$
\ENDIF
\STATE\label{line:spoly2} reduce $(P_j,Q_j,B_i \wedge B_j)$ by $r_i$ to obtain $(P,Q,B)$
\WHILE{$P\neq0$ and $(P,Q,B)$ can be reduced by some $r \in S$ with $\delta(r)<\delta(P,Q,B)$}\label{line:checkreduction}
\STATE\label{line:selectR} choose some $r \in S$ with $\delta(r)<\delta(P,Q,B)$ that can reduce $(P,Q,B)$
\STATE replace $(P,Q,B)$ with its reduction by $r$
\ENDWHILE
\STATE\label{line:createrules2} create new reduction rules $r_{m+1},\dots,r_{m+k}$ from $(P,Q,B)$ by Algorithm~\ref{alg:CItoRR}
\STATE $S:=S\cup\{r_{m+1},\dots,r_{m+k}\}$
\STATE $m:=m+k$
\ENDWHILE
\RETURN $S$
\end{algorithmic}
\end{algorithm}

\begin{example}
Continuing Example~\ref{ex:NormanFCI}, we apply Procedure~\ref{proc:RefinedCompletion} to the basic rules $\{r_1,r_2\}$ computed in Example~\ref{ex:NormanBasicRules}.
With this input, the computation is still very similar to that of Procedure~\ref{proc:NormanCompletion} shown in Example~\ref{ex:NormanCompletion}.
In the first iteration of the main loop, the only critical pairs arise from $r_1$ and $r_2$.
Since $\lm_t(Q_1)=t_2^{-1}<t_2=\lm_t(Q_2)$, we remove $r_2$ from $S$ in line~\ref{line:removerule}.
In the next line, the condition $B_2\wedge\neg{B_1}=(\param_2=1\wedge\neg(\param_2\ge1\wedge\param_2\neq3))$ cannot be satisfied, so we do not need to include a replacement for $r_2$ into $S$.
The subsequent reduction yields the same $(P,Q,B)$ as in Example~\ref{ex:NormanCompletion}.
In line~\ref{line:checkreduction}, we determine that this $(P,Q,B)$ is not reducible by any element of $S=\{r_1\}$ since condition $\param_2=1$ does not imply $(\param_2\ge3\wedge\param_2\neq5)\wedge\param_2\neq5$ over $\mathbb{N}$.
So, we proceed with creating $r_3$ as in Example~\ref{ex:NormanCompletion} and including it into $S$.
In the next iteration of the main loop, similarly to the first one, we can only deal with $r_1$ and $r_3$ and we remove $r_3$ from $S$ since $\lm_t(Q_1)=t_2^{-1}<t_1=\lm_t(Q_3)$.
Again, due to the condition in line~\ref{line:checkremaining}, there is no immediate replacement for $r_3$ and we proceed with line~\ref{line:spoly2}.
The result $(P,Q,B)$ with $\lm_t(P)=t_2^{-2}$ and $B=(\param_2=2)$ is the same as in Example~\ref{ex:NormanCompletion}.
Since $\param_2=2$ does not imply $(\param_2\ge3\wedge\param_2\neq5)\wedge\param_2\neq5$ over $\mathbb{N}$, the inner loop cannot be entered at line~\ref{line:checkreduction} and we create $r_4$ as in Example~\ref{ex:NormanCompletion}.
Now, $S=\{r_1,r_4\}$ does not give rise to any critical pairs, so the procedure returns this complete reduction system $S$.

Then, we apply the new system $S$ to $f=t_1$, which could not be reduced in Example~\ref{ex:NormanReduction} by $r_1$ alone.
Since $r_1$ and $r_4$ correspond to identities \eqref{eq:NormanR1} and \eqref{eq:NormanR4}, respectively, this leads exactly to the reduction steps shown in Example~\ref{ex:NormanMain}.
First, since the exponent vector $\alpha=(1,0)$ of $f$ satisfies the condition $B_4$ of $r_4$, we can reduce $f$ by $r_4$.
By $P_4(\alpha,t)=8+4t_1^{-1}t_2$ and $Q_4(\alpha,t)=2t_1t_2^2+2t_1+4t_2$, we obtain $f = L\left(\frac{1}{8}Q_4(\alpha,t)t^\alpha\right)-\frac{1}{2}t_2$.
So, $f$ has been reduced to $-\frac{1}{2}t_2$, which has exponent vector $\alpha=(0,1)$ satisfying the condition $B_1$ of $r_1$.
Together with $P_1(\alpha,t)=-2$ and $Q_1(\alpha,t)=t_2^{-1}$, we have $-\frac{1}{2}t_2 = L\left(\frac{1}{4}\right)$, i.e., the remainder is reduced to zero.
Thus, $f$ belongs to $\im(L)$ with $f=L\left(\frac{1}{4}t_1^2t_2^2+\frac{1}{4}t_1^2+\frac{1}{2}t_2+\frac{1}{4}\right)$, which leads to~\eqref{eq:NormanExample} finally.
\end{example}

Keep in mind that Remark~\ref{rem:relaxation} not only affects line~\ref{line:checkremaining} of Procedure~\ref{proc:RefinedCompletion} but also allows to choose any $r_i,r_j$ with $\lm_t(Q_i)<\lm_t(Q_j)$ in line~\ref{line:selectCP2} even if they do not form a critical pair.
As a result, the triples created in lines \ref{line:remainingcase} and \ref{line:spoly2} may involve conditions that cannot be satisfied in $\mathbb{N}^n$.
In view of Remark~\ref{rem:relaxation}, checking reducibility of $(P,Q,B)$ by a reduction rule $r_1=(P_1,Q_1,B_1)$ has to be done based on the condition $\neg\exists\alpha\in\mathbb{N}^n: B|_{\param=\alpha} \wedge (\neg B_1|_{\param=\alpha+\beta} \vee \lc_t(P_1)|_{\param=\alpha+\beta}=0)$ involving the existential quantifier, which is equivalent to the condition given in Definition~\ref{defn:reducible}.
Consequently, when checking reducibility in lines \ref{line:checkreduction} and \ref{line:selectR}, an actual possibility to reduce a given $(P,Q,B)$ may be missed, but is never wrongly detected.

\begin{remark}\label{rem:termination}
 Regarding termination of Procedure~\ref{proc:RefinedCompletion}, neither of the two loops is guaranteed to terminate after finitely many iterations.
 By the discussion in the previous paragraph, the inner loop can be artificially terminated within the principles of Remark~\ref{rem:relaxation}, even if an infinite sequence of reductions would be possible.
 Artificially terminating the main loop by considering the condition of the loop to be false when it actually is fulfilled, however, is in violation of Remark~\ref{rem:relaxation}.
 In such a case, Lemma~\ref{lem:invariant2} shows that the output $S$ nevertheless is a precomplete reduction system, just completenes of $S$ is not guaranteed.
\par
 Regarding executability of lines \ref{line:selectCP2} and \ref{line:selectR}, note that the conditions of both loops imply directly that these choices are always possible, as long as Remark~\ref{rem:relaxation} is applied deterministically.
 If evaluating $\exists{\alpha\in\mathbb{N}^n}:B|_{\param=\alpha}$ is allowed to yield different results for the same $B$ within the principles of Remark~\ref{rem:relaxation}, then the choice in lines \ref{line:selectCP2} or \ref{line:selectR} may become impossible when executing those lines.
 To prevent such a case, one can combine verification of each loop condition with making the corresponding choice of reduction rules, so that no existence checks in $\mathbb{N}^n$ are performed separately for lines \ref{line:selectCP2} and \ref{line:selectR}.
 In any case, we can assume that lines \ref{line:selectCP2} and \ref{line:selectR} succeed whenever they are executed.
\qed
\end{remark}

The remainder of this section is devoted to proving correctness of Procedure~\ref{proc:RefinedCompletion} and showing that it preserves exact offset.
Recall that the reduction in line~\ref{line:spoly2} is always possible, since $r_i,r_j$ are reduction rules and reducibility $\forall\,{\alpha\in\mathbb{N}^n}:(B_i \wedge B_j)|_{\param=\alpha} \Longrightarrow (B_i|_{\param=\alpha} \wedge \lc_t(P_i)|_{\param=\alpha}\neq0)$ trivially holds in this case.
Since Procedure~\ref{proc:RefinedCompletion} may not terminate (even when $<$ is Noetherian), we prove its correctness in Theorem~\ref{thm:correctness2} only if it terminates.
Before doing so, we show the invariant properties in Lemma~\ref{lem:invariant2}, which hold independent of termination.
To this end, we need the following simple result about reduction of conditional identities by reduction rules with smaller offset.

\begin{lemma}\label{lem:ReductionWithSmallOffset}
 Let $(P,Q,B)$, $r_1$, $\tilde{P},\tilde{Q}$ as in Definition~\ref{defn:reducible}.
 If $\delta(r_1)<\delta(P,Q,B)$, then
 \begin{enumerate}
  \item $Q\neq0$ and $\lm_t(Q)=\lm_t(\tilde{Q})$;
  \item for all $\alpha\in\mathbb{N}^n$ with $B|_{\param=\alpha}$, we have $\lc_t(Q)|_{\param=\alpha}\neq0$ if and only if $\lc_t(\tilde{Q})|_{\param=\alpha}\neq0$.
 \end{enumerate}
\end{lemma}
\begin{proof}
 By $\delta(r_1)<\delta(P,Q,B)$, we have $Q\neq0$.
 With $r_1=(P_1,Q_1,B_1)$ and $\lm_t(P)=t^\beta$, we obtain 
 $$\lm_t(Q_1)t^\beta=\delta(r_1)t^\beta<\delta(P,Q,B)t^\beta=\lm_t(Q).$$
 So, with $g:=\gcd(\lc_t(P),\lc_t(P_1)|_{\param=\param+\beta}) \in C[\param]$, we have $\lt_t(\tilde{Q})=\frac{\lc_t(P_1)|_{\param=\param+\beta}}{g}\lt_t(Q)$.
 Hence, $\lm_t(\tilde{Q})=\lm_t(Q)$ and $\lc_t(\tilde{Q})=\frac{\lc_t(P_1)|_{\param=\param+\beta}}{g}\lc_t(Q)$.
 Since $\lc_t(P_1)|_{\param=\alpha+\beta}\neq0$ for all $\alpha\in\mathbb{N}^n$ with $B|_{\param=\alpha}$, we have for all such $\alpha$ that $\lc_t(Q)|_{\param=\alpha}\neq0$ if and only if $\lc_t(\tilde{Q})|_{\param=\alpha}\neq0$.
\end{proof}

\begin{lemma}\label{lem:invariant2}
 At the end of each iteration of the main loop in Procedure~\ref{proc:RefinedCompletion}, the set $S$ is again a reduction system for $L$ w.r.t.\ $<$ that satisfies all properties imposed on the input.
\end{lemma}
\begin{proof}
 Considering one iteration of the main loop, let $S_\mathrm{old}$ be the set $S$ at the beginning of the current iteration, let $S_\mathrm{mid}$ be the set $S$ immediately before executing line~\ref{line:spoly2}, and let $S_\mathrm{new}$ be the set $S$ at the end of the current iteration.
 We show that $S_\mathrm{new}$ satisfies the desired properties whenever $S_\mathrm{old}$ does.
\par
 First, it is easy to see that $S_\mathrm{new}$ is a reduction system for $L$ w.r.t.\ $<$, because the operations performed on any triples encoding a conditional identity for $L$ during the procedure again yield triples encoding a conditional identity for $L$ and any triples added into $S$ are even reduction rules w.r.t.\ $<$ by construction.
 Triples $(P,Q,B)$ whose condition $B$ cannot be satisfied in $\mathbb{N}^n$ trivially encode conditional identities for $L$ anyway.
 For any $r=(P,Q,B)$, we abbreviate $U(r):=\{\lm_t(Q)t^\alpha\ |\ \alpha\in\mathbb{N}^n,B|_{\param=\alpha}\}$ and $V(r):=\linspan_C\{P(\alpha,t)t^\alpha\ |\ \alpha\in\mathbb{N}^n,B|_{\param=\alpha}\}$ for shorter notation.
\par
 Next, we show that the sets $U(r)$ associated to all $r \in S_\mathrm{new}$ are pairwise disjoint (i.e.\ for any $\gamma \in \mathbb{N}^n$ there is at most one $(P,Q,B) \in S_\mathrm{new}$ with $\exists \, \alpha\in\mathbb{N}^n:B|_{\param=\alpha}\wedge\lm_t(Q)t^\alpha=t^\gamma$).
 By assumption, all sets $U(r)$ with $r \in S_\mathrm{old}$ are pairwise disjoint.
 In line~\ref{line:removerule}, the set $U(r_j)$ gets removed, so it suffices to show that all $U(r)$ with $r \in S_\mathrm{new}\setminus{S_\mathrm{old}}$ are pairwise disjoint subsets of $U(r_j)$.
 To this end, we split $U(r_j)$ into two disjoint sets $U(P_j,Q_j,B_j \wedge\neg B_i)$ and $U(P_j,Q_j,B_j \wedge B_i)$.
 If nonempty, the first of them is introduced by the only element of $S_\mathrm{mid}\setminus{S_\mathrm{old}}$ based on line~\ref{line:remainingcase}.
 Otherwise, any potential $r \in S_\mathrm{mid}\setminus{S_\mathrm{old}}$ has empty $U(r)$ anyway.
 For the second one, we note that $U(P_j,Q_j,B_j \wedge B_i)$ agrees with $U(P,Q,B)$ throughout all reductions in line \ref{line:spoly2} and the following small loop, because reduction by rules with smaller offset does not change $B$ or $\lm_t(Q)$ at all by Lemma~\ref{lem:ReductionWithSmallOffset}.
 Finally, by Lemma~\ref{lem:CItoRR}, the new sets $U(r)$ with $r \in S_\mathrm{new}\setminus{S_\mathrm{mid}}$ are pairwise disjoint subsets of $U(P,Q,B)$.
\par
 In order to show $S_\mathrm{new}$ is precomplete, it suffices to show that $V(r_j)$ is contained in the sum of all $V(r)$ with $r \in S_\mathrm{new}$.
 We consider $V(r_j)$ as the sum of $V(P_j,Q_j,B_j \wedge\neg B_i)$ and $V(P_j,Q_j,B_j \wedge B_i)$.
 If non-trivial, the first summand is already covered by $S_\mathrm{mid}$ based on line~\ref{line:remainingcase}.
 For the second summand, we note that it is contained in the sum of all $V(r)$ with $r \in S_\mathrm{mid}\cup\{(P,Q,B)\}$ throughout all reductions in line \ref{line:spoly2} and the following small loop.
 This is because the sum of the space $V(P,Q,B)$ generated after a reduction step and the space generated by the rule used for reduction contains the space generated by the triple being reduced, which can be shown straightforwardly using Lemma~\ref{lem:reduction}.
 Afterwards, in line~\ref{line:createrules2}, $V(P,Q,B)$ is contained in the sum of all $V(r)$ with $r \in S_\mathrm{new}\setminus{S_\mathrm{mid}}$ by correctness of Algorithm~\ref{alg:CItoRR}, which concludes the proof.
\end{proof}

\begin{theorem}\label{thm:correctness2}
 If Procedure~\ref{proc:RefinedCompletion} terminates for a given input, the output $S$ is a complete reduction system for $L$ w.r.t.\ $<$.
\end{theorem}
\begin{proof}
 By Lemma~\ref{lem:invariant2}, $S$ is a precomplete reduction system for $L$ w.r.t.\ $<$.
 If the procedure terminates, there are no critical pairs arising from $S$, since Remark~\ref{rem:relaxation} does not allow critical pairs being overlooked and Lemma~\ref{lem:invariant2} implies that rules with the same offset cannot form critical pairs.
 Consequently, $S$ is complete by Lemma~\ref{lem:NoRedundancy}.
\end{proof}

In practice, we will use Procedure~\ref{proc:RefinedCompletion} with the set of basic rules for $L$ as input, which are obtained from $(p,1,true)$ by Algorithm~\ref{alg:CItoRR}.
In addition to being a precomplete reduction system, the basic rules satisfy the conditions for the input of Procedure~\ref{proc:RefinedCompletion} by Lemma~\ref{lem:CItoRR}.
These rules always have exact offset by Lemma~\ref{lem:basicrules} and this property is preserved throughout the procedure.

\begin{lemma}
\label{lem:exactoffset}
 If each reduction rule in the input of Procedure~\ref{proc:RefinedCompletion} has exact offset, then each reduction rule created during the procedure has exact offset as well.
\end{lemma}
\begin{proof}
 It suffices to show this for one iteration of the main loop.
 The reduction rule created in line~\ref{line:remainingcase} trivially has exact offset, since it uses the same $Q_j$ as the rule $r_j$ and the condition $B_j\wedge\neg{B_i}$ implies $B_j$.
 Similarly, $(P_j,Q_j,B_i\wedge{B_j})$ in line~\ref{line:spoly2} has exact offset.
 Then, Lemma~\ref{lem:ReductionWithSmallOffset} shows that $\lc_t(Q)|_{\param=\alpha}\neq0$ holds for all $\alpha\in\mathbb{N}^n$ satisfying $B|_{\param=\alpha}$ after reduction in line~\ref{line:spoly2} and also after the subsequent loop, since $B=B_j\wedge{B_i}$.
 Therefore, the reduction rules created immediately after this inner loop from $(P,Q,B)$ will have exact offset by Lemma~\ref{lem:CItoRR}.
\end{proof}

\subsection{Norman's example involving $\tan(\ln(x))$}
\label{sec:NormanTanLogExample}

Norman reports on p.~204 of \cite{NormanCriticalPair} that his completion process does not seem to terminate when using the denominator $\tan(\ln(x))^2+1$ for the integral and he poses as open problem to characterize how the reduction system evolves in such cases.
To model $\tan(\ln(x))$, we consider the differential field $(\mathbb{Q}(t_1,t_2,t_3),\Der)$ with $\Der{t_1}=1$, $\Der{t_2}=\frac{1}{t_1}$, and $\Der{t_3}=\frac{t_3^2+1}{t_1}$ so that $t_1,t_2,t_3$ correspond to $x,\ln(x),\tan(\ln(x))$, respectively.
Then, the denominator of the integral is chosen as $v=t_3^2+1$.
This example was mentioned by Norman as being particularly problematic when he considered a monomial order with $t_3>t_2>t_1$.
He did not fully specify the monomial order he had used beyond the comparison of individual variables, however.
In what follows, we will use the lexicographic order, since the point can be made most easily with this monomial order.
In Section~\ref{sec:NormanTanLogInfinite}, we exhibit one infinite pattern that arises during  Procedure~\ref{proc:NormanCompletion}, showing that Norman's completion process indeed does not terminate in this example.
In Section~\ref{sec:NormanTanLogRefined}, we display the finitely many iterations our Procedure~\ref{proc:RefinedCompletion} takes for constructing a complete reduction system for the same input.
In fact, the refined version terminates in this example for any semigroup order $<$ that satisfies $t_2>1$ and $t_3>1$.

From the derivatives $\Der{t_i}$, we see that $\den(\Der)=t_1$. Then, $\tilde{\Der}v=2t_3v$ yields
\[
 p:=\param_1+\param_2\frac{1}{t_2}+\param_3\frac{t_3^2+1}{t_3}-2t_3
\]
by \eqref{eq:defP}.
Therefore, the basic rules $r_1,r_2,r_3,r_4$ as computed by Algorithm~\ref{alg:CItoRR} from $(p,1,true)$ are given as $r_i=(P_i,Q_i,B_i)$, where the generic rule has
\begin{equation}\label{eq:NormanTanLogBasic1}
 P_1 = (\param_3-3)+\frac{\param_1}{t_3}+\frac{\param_2}{t_2t_3}+\frac{\param_3-1}{t_3^2}, \quad Q_1 = \frac{1}{t_3}, \quad B_1=(\param_3\neq3 \wedge \param_3\ge1)
\end{equation}
and the degenerate cases are given by
\begin{align}
 P_2 &= \param_1+\param_2t_2^{-1}+\param_3t_3^{-1}, & Q_2 &= 1, & B_2 &= (\param_3=2 \wedge \param_1\neq0),\label{eq:NormanTanLogBasic2}\\
 P_3 &= (\param_2+1)+\param_3t_2t_3^{-1}, & Q_3 &= t_2, & B_3 &= (\param_3=2 \wedge \param_1=0 \wedge \param_2+1\neq0),\label{NormanTanLogBasic3}\\
 P_4 &= \param_3+1, & Q_4 &= t_3, & B_4 &= (\param_3=1 \wedge \param_1=0 \wedge \param_2=0 \wedge \param_3+1\neq0).\label{eq:NormanTanLogBasic4}
\end{align}
Critical pairs are formed by the following pairs of basic rules: $(r_1,r_2)$, $(r_1,r_3)$, $(r_1,r_4)$.

The formulae for conditions $B_i$ of new rules computed during the completion process can get quite involved.
So, in the following, we replace those conditions by simpler ones that have the same solutions in $\mathbb{N}^3$.

\subsubsection{Norman's completion process}
\label{sec:NormanTanLogInfinite}

When applying Procedure~\ref{proc:NormanCompletion} to the basic rules $r_1,r_2,r_3,r_4$ computed in \eqref{eq:NormanTanLogBasic1}--\eqref{eq:NormanTanLogBasic4}, at the beginning of each iteration of the main loop an element of $A$ needs to be chosen.
The concrete choice does not affect whether the loop is iterated infinitely often or a finite number of times.
The following choices are made to show in a straightforward way that Procedure~\ref{proc:NormanCompletion} does not terminate in this example.

In the first iteration of the loop, we start by choosing $i=1,j=2$.
We obtain the condition $B = B_1 \wedge B_2 = (\param_3\neq3 \wedge \param_3\ge1 \wedge \param_3=2 \wedge \param_1\neq0)$ and reduction yields the polynomials
\[
 P = (\param_3-3)P_2-\param_1P_1 = \frac{\param_2(\param_3-3)}{t_2}+\frac{\param_3^2-3\param_3-\param_1^2}{t_3}-\frac{\param_1\param_2}{t_2t_3}-\frac{\param_1(\param_3-1)}{t_3^2}
\]
and $Q = (\param_3-3)Q_2-\param_1Q_1 = (\param_3-3)-\frac{\param_1}{t_3}$.
Creating reduction rules from $(P,Q,B)$ by Algorithm~\ref{alg:CItoRR}, we obtain the rule $r_5$ with
\begin{align*}
 P_5 &= (\param_2+1)(\param_3-3)+(\param_3^2-3\param_3-\param_1^2)\frac{t_2}{t_3}-\frac{\param_1(\param_2+1)}{t_3}-\param_1(\param_3-1)\frac{t_2}{t_3^2},\\
 Q_5 &= (\param_3-3)t_2-\param_1\frac{t_2}{t_3},
\end{align*}
and condition $B_5=(\param_3\neq3 \wedge \param_3\ge1 \wedge \param_3=2 \wedge \param_1\neq0 \wedge (\param_2+1)(\param_3-3)\neq0)$ that can be simplified to $B_5=(\param_3=2 \wedge \param_1\neq0)$.
We also obtain $r_6$ from the degenerate case of $(P,Q,B)$.
These new rules give rise to additional critical pairs, one of the three elements added into $A$ is $(2,5)$.

For the second iteration of the loop, we choose $i=2,j=5$.
The condition $B = B_2 \wedge B_5$ can be simplified to $B = (\param_3=2 \wedge \param_1\neq0)$.
Reduction by $r_2$ yields
\begin{multline*}
 P = \param_1P_5-(\param_2+1)(\param_3-3)P_2 = - \frac{\param_2(\param_2+1)(\param_3-3)}{t_2}+\param_1(\param_3^2-3\param_3-\param_1^2)\frac{t_2}{t_3}\\
 -\frac{(\param_2+1)(\param_3^2-3\param_3+\param_1^2)}{t_3}-\param_1^2(\param_3-1)\frac{t_2}{t_3^2}
\end{multline*}
and $Q = \param_1Q_5-(\param_2+1)(\param_3-3)Q_2 = \param_1(\param_3-3)t_2-(\param_2+1)(\param_3-3)-\param_1^2\frac{t_2}{t_3}$. Then, $(P,Q,B)$ gives rise to the new rules $r_7$ and $r_8$, where the main rule has
\begin{align*}
 P_7 &= -(\param_2+1)(\param_2+2)(\param_3-3)+\param_1(\param_3^2-3\param_3-\param_1^2)\frac{t_2^2}{t_3}-(\param_2+2)(\param_3^2-3\param_3+\param_1^2)\frac{t_2}{t_3}-\param_1^2(\param_3-1)\frac{t_2^2}{t_3^2}\\
 Q_7 &= \param_1(\param_3-3)t_2^2-(\param_2+2)(\param_3-3)t_2-\param_1^2\frac{t_2^2}{t_3}
\end{align*}
and simplified condition $B_7=(\param_3=2 \wedge \param_1\neq0)$.
Among the four new elements added into $A$, we find $(2,7)$.

In subsequent iterations, inductively for $n=1,2,\dots$, we choose $i=2$ and $j=2n+5$ based on $A$.
With this choice, $r_j$ is given by
\begin{align*}
 P_{2n+5} = {}&(-1)^n(\param_2+1)_{n+1}(\param_3-3)+\param_1^n(\param_3^2-3\param_3-\param_1^2)\frac{t_2^{n+1}}{t_3}-\param_1^{n-1}(\param_2+n+1)(\param_3^2-3\param_3+\param_1^2)\frac{t_2^n}{t_3}\\
  &+\sum_{l=1}^{n-1}(-1)^{n-l+1}\param_1^{l-1}(\param_2+l+1)_{n-l+1}\param_3(\param_3-3)\frac{t_2^l}{t_3}-\param_1^{n+1}(\param_3-1)\frac{t_2^{n+1}}{t_3^2}\\
  Q_{2n+5} = {}&\sum_{l=1}^{n+1}(-1)^{n-l+1}\param_1^{l-1}(\param_2+l+1)_{n-l+1}(\param_3-3)t_2^l-\param_1^{n+1}\frac{t_2^{n+1}}{t_3}
\end{align*}
and $B_{2n+5}=(\param_3=2 \wedge \param_1\neq0)$. The case $n=1$ agrees with $r_7$ computed just before. Reduction by $r_2$ gives $P=\param_1P_{2n+5}-(-1)^n(\param_2+1)_{n+1}(\param_3-3)P_2$, $Q=\param_1Q_{2n+5}-(-1)^n(\param_2+1)_{n+1}(\param_3-3)Q_2$, and $B=(\param_3=2 \wedge \param_1\neq0)$. Collecting terms in $P$ and $Q$ we obtain
\begin{align*}
 P = &-(-1)^n\param_2(\param_2+1)_{n+1}(\param_3-3)\frac{1}{t_2}+\param_1^{n+1}(\param_3^2-3\param_3-\param_1^2)\frac{t_2^{n+1}}{t_3}-\param_1^n(\param_2+n+1)(\param_3^2-3\param_3+\param_1^2)\frac{t_2^n}{t_3}\\
 &+\sum_{l=0}^{n-1}(-1)^{n-l+1}\param_1^l(\param_2+l+1)_{n-l+1}\param_3(\param_3-3)\frac{t_2^l}{t_3}-\param_1^{n+2}(\param_3-1)\frac{t_2^{n+1}}{t_3^2}
\end{align*}
and $Q = \sum_{l=0}^{n+1}(-1)^{n-l+1}\param_1^l(\param_2+l+1)_{n-l+1}(\param_3-3)t_2^l-\param_1^{n+2}\frac{t_2^{n+1}}{t_3}$.
Converting $(P,Q,B)$ into reduction rules, we obtain $r_{2n+7}$ and $r_{2n+8}$, where $r_{2n+7}$ is given by $P_{2n+7}=P|_{\param_2=\param_2+1}t_2$, $Q_{2n+7}=Q|_{\param_2=\param_2+1}t_2$, and $B_{2n+7}=(B|_{\param_2=\param_2+1} \wedge -(-1)^n(\param_2+1)_{n+2}(\param_3-3)\neq0)$.
We can check that $P_{2n+7}$ and $Q_{2n+7}$ obtained this way agree with replacing $n$ by $n+1$ in the formulae for $P_{2n+5}$ and $Q_{2n+5}$ above.
Furthermore, we can check that $B_{2n+7}$ is equivalent to $\param_3=2 \wedge \param_1\neq0$.
Based on critical pairs involving the two new rules, $(2,2n+7)$ is among the $n+4$ elements added into $A$.
So, the next iteration of the loop is exactly as the current one, except $n$ being replaced by $n+1$.
Hence, Procedure~\ref{proc:NormanCompletion} continues indefinitely.

\subsubsection{Refinement of Norman's completion process}
\label{sec:NormanTanLogRefined}

We apply Procedure~\ref{proc:RefinedCompletion} to the basic rules $S=\{r_1,r_2,r_3,r_4\}$ given by \eqref{eq:NormanTanLogBasic1}--\eqref{eq:NormanTanLogBasic4}.

In the first iteration of the main loop, we choose $i=1$ and $j=2$ satisfying $\lm_t(Q_i)=\frac{1}{t_3}<1=\lm_t(Q_j)$.
We remove $r_2$ from the reduction system $S$ and we check that $B_2 \wedge \neg B_1$ has no solution in $\mathbb{N}^3$.
Reduction by $r_i$ yields
\begin{align*}
 P &= (\param_3-3)P_2-\param_1P_1 = \frac{\param_2(\param_3-3)}{t_2}+\frac{\param_3^2-3\param_3-\param_1^2}{t_3}-\frac{\param_1\param_2}{t_2t_3}-\frac{\param_1(\param_3-1)}{t_3^2}\\
 Q &= (\param_3-3)Q_2-\param_1Q_1 = (\param_3-3)-\frac{\param_1}{t_3}
\end{align*}
and simplified condition $B = (\param_3=2 \wedge \param_1\neq0)$, just as before when we applied Procedure~\ref{proc:NormanCompletion}.
The current offset is $\delta(P,Q,B)=t_2$.
Now, instead of immediately creating new rules from $(P,Q,B)$, we can enter the inner while loop to reduce $(P,Q,B)$ further by $r_1 \in S$ having offset $\delta(r_1)=\frac{1}{t_3}<\delta(P,Q,B)$.
This can be done four times.
Each time, there is only one choice for reduction and the current offset $\delta(P,Q,B)$ increases.
In most cases, $Q$ is only modified by introducing a new smallest term without changing the existing terms.
Eventually, we arrive at
\begin{align*}
 P &= \frac{\param_1(\param_1^2-4+8\param_3-2\param_3^2)}{t_3^2}+\frac{\param_2(3\param_1^2-4+8\param_3-2\param_3^2)}{t_2t_3^2}+\text{lower terms}\\
 Q &= (\param_3-3)(\param_3-4)-\frac{\param_1(\param_3-4)}{t_3}-\frac{\param_2(\param_3-4)}{t_2t_3}+\frac{\param_1^2+3\param_3-\param_3^2}{t_3^2}+\frac{2\param_1\param_2}{t_2t_3^2}+\frac{\param_2(\param_2-1)}{t_2^2t_3^2}.
\end{align*}
Even though all elements of $S=\{r_1,r_3,r_4\}$ have smaller offset than $\delta(P,Q,B)=t_3^2$, no further reduction is possible.
Creating new reduction rules from $(P,Q,B)$ by Algorithm~\ref{alg:CItoRR} yields only one new rule $r_5$, which we add into $S$.
\begin{align*}
 P_5 = {}&\param_1(\param_1^2+4-2\param_3^2)+\frac{\param_2(3\param_1^2+4-2\param_3^2)}{t_2}+\frac{3\param_1\param_2(\param_2-1)}{t_2^2}+\frac{\param_2(\param_2-1)(\param_2-2)}{t_2^3}\\
  &+\param_3\left(\frac{\param_1^2+2-\param_3-\param_3^2}{t_3}+\frac{2\param_1\param_2}{t_2t_3}+\frac{\param_2(\param_2-1)}{t_2^2t_3}\right)\\
 Q_5 = {}&(\param_3-1)(\param_3-2)t_3^2-\param_1(\param_3-2)t_3-\param_2(\param_3-2)\frac{t_3}{t_2}+(\param_1^2+2-\param_3-\param_3^2)+\frac{2\param_1\param_2}{t_2}+\frac{\param_2(\param_2-1)}{t_2^2}\\
 B_5 = {}&(\param_3=0 \wedge \param_1\neq0)
\end{align*}
No new critical pairs can be formed with $r_5$.
Note that condition $B_5$ implies that the last (i.e.\ smallest) three terms of $P_5$ vanish in $P_5(\alpha,t)$ for every $\alpha\in\mathbb{N}^3$ satisfying $B_5|_{\param=\alpha}$.

In the second iteration of the main loop, we choose $i=1$ and $j=3$ satisfying $\lm_t(Q_i)=\frac{1}{t_3}<t_2=\lm_t(Q_j)$.
We remove $r_3$ from $S$ and we check that $B_3 \wedge \neg B_1$ has no solution in $\mathbb{N}^3$.
Reduction by $r_i$ yields $(P,Q,B)$ with polynomials $P=(\param_3-3)P_3-(\param_2+1)P_1$, $Q=(\param_3-3)Q_3-(\param_2+1)Q_1$, and simplified condition $B=(\param_3=2 \wedge \param_1=0)$.
In the inner while loop, there is only one choice for reduction, namely by $r_1$.
The first time, we obtain $\lt_t(P)=-\frac{\param_1(\param_2+1)(\param_3-4)}{t_3}$ and $\delta(P,Q,B)=t_2t_3$.
Note that, even though $\lc_t(P)|_{\param=\alpha}=0$ for all $\alpha\in\mathbb{N}^3$ with $B|_{\param=\alpha}$, we can reduce the new $(P,Q,B)$ again and $r_1$ is the only option to do so.
After this second iteration of the inner while loop, we have $\lt_t(P)=-\frac{\param_2(\param_2+1)(\param_3-4)}{t_2t_3}$ and $\delta(P,Q,B)=t_2^2t_3$.
We can iterate the inner loop once more.
Again, only $r_1$ can be used for reduction and we obtain
\begin{align*}
 P &= -\param_1\param_3(\param_3-3)\frac{t_2}{t_3^2}+\frac{(\param_2+1)(\param_1^2-4+8\param_3-2\param_3^2)}{t_3^2}+\text{lower terms}\\
 Q &= (\param_3-3)(\param_3-4)t_2-\frac{(\param_2+1)(\param_3-4)}{t_3}-\param_3(\param_3-3)\frac{t_2}{t_3^2}+\frac{\param_1(\param_2+1)}{t_3^2}+\frac{\param_2(\param_2+1)}{t_2t_3^2},
\end{align*}
which cannot be reduced further by any element of $S$.
So, we apply Algorithm~\ref{alg:CItoRR} to $(P,Q,B)$, whereby we obtain only one new reduction rule $r_6$.
It is given by
\begin{align*}
 P_6 = {}&(\param_2+1)(\param_1^2+4-2\param_3^2)+\frac{2\param_1\param_2(\param_2+1)}{t_2}+\frac{(\param_2-1)\param_2(\param_2+1)}{t_2^2}\\
  &+\param_3\left(-(\param_3-1)(\param_3+2)\frac{t_2}{t_3}+\frac{\param_1(\param_2+1)}{t_3}+\frac{\param_2(\param_2+1)}{t_2t_3}\right)\\
  Q_6 = {}&(\param_3-1)(\param_3-2)t_2t_3^2-(\param_2+1)(\param_3-2)t_3-(\param_3-1)(\param_3+2)t_2+\param_1(\param_2+1)+\frac{\param_2(\param_2+1)}{t_2}\\
  B_6 = {}&(\param_3=0 \wedge \param_1=0)
\end{align*}
and does not give rise to new critical pairs in $S=\{r_1,r_4,r_5,r_6\}$.
Note that, due to condition $B_6$ implying $\alpha_1=\alpha_3=0$ for all $\alpha\in\mathbb{N}^3$ satisfying $B_6|_{\param=\alpha}$, only the first and third term of $P_6$ can contribute to $P_6(\alpha,t)$ and, likewise, the coefficient of the second-last term of $Q_6$ vanishes in $Q_6(\alpha,t)$.

In the third and final iteration of the main loop, the only choice left is $i=1$ and $j=4$ satisfying $\lm_t(Q_i)=\frac{1}{t_3}<t_3=\lm_t(Q_j)$.
We remove $r_4$ from $S$ and we check that $B_4 \wedge \neg B_1$ has no solution in $\mathbb{N}^3$.
Reduction by $r_i$ yields
\begin{align*}
 P &= (\param_3-3)P_4-(\param_3+1)P_1 = -\frac{\param_1(\param_3+1)}{t_3}-\frac{\param_2(\param_3+1)}{t_2t_3}-\frac{(\param_3-1)(\param_3+1)}{t_3^2}\\
 Q &= (\param_3-3)Q_4-(\param_3+1)Q_1 = (\param_3-3)t_3-\frac{\param_3+1}{t_3}
\end{align*}
and simplified condition $B=(\param_3=1 \wedge \param_1=0 \wedge \param_2=0)$.
The rule $r_1 \in S$ is the only one with offset smaller than $\delta(P,Q,B)=t_3^2$, but cannot be used for reduction of $(P,Q,B)$.
The rule $r_6 \in S$ is the only one that could be used for reduction here, but $\delta(r_6)=\lm_t(Q_6)=t_2t_3^2$ is too large.
So the inner while loop cannot be entered and we proceed with creating new rules from $(P,Q,B)$.
Since condition $B$ implies $\alpha=(0,0,1)$, we have $P(\alpha,t)=0$ for all $\alpha\in\mathbb{N}^3$ satisfying $B|_{\param=\alpha}$.
Hence, no new rule is created by Algorithm~\ref{alg:CItoRR} and $Q(\alpha,t)t^\alpha=-2t_3^2-2$ is a nontrivial element of $\ker(L)$.
Since $S=\{r_1,r_5,r_6\}$ does not give rise to any critical pairs, our refined completion process stops and returns the complete reduction system $S$.

\section{Infinite reduction systems}
\label{sec:InfiniteSystems}

As explained in Section~\ref{sec:RedSys}, complete reduction systems may be infinite.
However, we can still find regular patterns of the reduction rules for certain examples.
So in this section, we present infinite complete reduction systems for two differential fields.
In each system shown below, the number of monomials in the $P$ and $Q$ components of rules is unbounded overall, so no finite reduction system can induce the same reduction relation.
To keep things simple, we only look for integrals with denominator $v=1$.
In other words, we look at the integration problem in the polynomial ring with derivation $\tilde{\Der}$.
The semigroup orders of monomials will be specified by matrices as follows.
Any matrix $M \in \mathrm{GL}_n(\mathbb{R})$ induces a semigroup order $<$ on (Laurent) monomials by letting $t^\alpha<t^\beta$ if and only if the first nonzero entry of $M\cdot(\beta-\alpha) \in \mathbb{R}^n$ is positive.

\subsection{Airy functions}
\label{sec:Airy}

As in Example~\ref{ex:AiryBoettner}, we consider the differential field generated by constants and the Airy function $\mathrm{Ai}(x)$ with the usual derivation $\frac{d}{dx}$.
This differential field is modelled as $C(t_1,t_2,t_3)$ with derivation $\Der$ such that
\[\partial t_1 = 1, \quad \partial t_2 = t_3, \quad \text{and} \quad \partial t_3 = t_1 t_2.\]
Since the denominator of the derivation is $1$, we have $\tilde \partial =\partial$ and $(C[t_1, t_2, t_3],\Der)$ forms a differential ring.
Using Norman's setting of reduction rules similar to Example~\ref{ex:NormanMain}, the authors presented two complete reduction systems for this situation w.r.t.\ different monomial orders in \cite{Airy}.
Below, we reformulate one of them using the new formalism introduced in Section~\ref{sec:RedSys} for further use in Section~\ref{sec:degboundsEx}.
In particular, we use the order induced by $\left(\begin{smallmatrix}0&1&1\\2&0&1\\0&0&1\end{smallmatrix}\right)$.

Since $\Der t^\alpha = \alpha_1t^{\alpha-(1,0,0)}+\alpha_2t^{\alpha+(0,-1,1)}+\alpha_3t^{\alpha+(1,1,-1)}$ for all $\alpha \in \mathbb{N}^3$, we have $p:=\param_1t_1^{-1}+\param_2t_2^{-1}t_3+\param_3t_1t_2t_3^{-1}$, cf.\ \eqref{eq:defP}, and $(p,1,true)$ encodes a conditional identity, which is not a reduction rule yet since $\lm_t(p)\neq1$.
Then, applying Algorithm~\ref{alg:CItoRR} to $(p,1,true)$ yields three basic rules
\begin{align}
&\big((\param_2+1)+(\param_3-1)t_1t_2^2t_3^{-2}+\param_1t_1^{-1}t_2t_3^{-1},&& t_2t_3^{-1},&& \param_3\ge1\big)\label{eq:AiryBasic1}\\
&((\param_3+1)+(\param_1-1)t_1^{-2}t_2^{-1}t_3,&& t_1^{-1}t_2^{-1}t_3,&& \param_2=1 \wedge \param_1\ge1)\label{eq:AiryBasic2}\\
&(\param_1+1,&& t_1,&& \param_2=0 \wedge \param_3=0),\label{eq:AiryBasic3}
\end{align}
where we only show the simplified and equivalent version of conditions $B$.

When applying Procedure~\ref{proc:NormanCompletion} and Procedure~\ref{proc:RefinedCompletion} to these basic rules, we find that neither of them terminates, which means the complete reduction system for Airy functions is infinite.
In particular, our refined completion process replaces the second basic rule by infinitely many new rules obtained via successive reduction by the generic rule.
However, due to the specific monomial order above and by computing a large number of reduction rules, we observe that monomials arise in the rules following a pattern and that their coefficients also satisfy some kind of pattern with respect to $(\param_1,\param_2,\param_3)$.
Based on that, we can present a complete reduction system as follows.

\begin{theorem}\label{TH:Airyreductionrule}
 The following are reduction rules for $L=\Der$ w.r.t.\ $<$ defined by $\left(\begin{smallmatrix}0&1&1\\2&0&1\\0&0&1\end{smallmatrix}\right)$.
 \begin{enumerate}
  \item[(i)] The generic rule is the first basic rule \eqref{eq:AiryBasic1}.
  \item[(ii)] For every odd integer $j\ge1$, a reduction rule is given by
  \begin{align*}
    P_j&=1+\sum_{m=1}^{\frac{j+1}{2}}c_{j,m}(\param_1-m)t_1^{-m-1}t_2^{-2m+1}t_3^{2m-1}\\
    Q_j&=\sum_{m=1}^{\frac{j+1}{2}}c_{j,m}t_1^{-m}t_2^{-2m+1}t_3^{2m-1}\\
    B_j&=\left(\param_3=0 \wedge \param_2=j \wedge \param_1\ge\frac{j+1}{2}\right),
  \end{align*}
  where \[c_{j, m} = (-1)^{m+1}\frac{(j-1)!!}{(j-2m+1)!! (2m-1)!!}.\]
  \item[(iii)] For every even integer $j\ge0$, a reduction rule is given by
  \begin{align*}
    P_j&=b_{j,0}+\sum_{m=1}^{\frac{j}{2}}b_{j,m+1}(\param_1-m)t_1^{-m-1}t_2^{-2m+1}t_3^{2m-1}\\
  Q_j&=b_{j,1}\sum_{m=0}^{\frac{j}{2}}(-1)^m \binom{\frac{j}{2}}{m} t_1^{-m+1} t_2^{-2m} t_3^{2m}+ \sum_{m=1}^{\frac{j}{2}} b_{j, m+1} t_1^{-m} t_2^{-2m+1} t_3^{2m-1}\\
    B_j&=\left(\param_3=0 \wedge \param_2=j \wedge \param_1\ge\frac{j}{2}-1\right),
  \end{align*}
   where $b_{j,0} = \param_1+1-\frac{j}{4}$ and
   \[
  \left(\begin{matrix}
     (\param_1+1)\binom{j/2}{0} & 1 & & & &\\[1ex]
     -\param_1\binom{j/2}{1} & j-1 & 3 & & &\\[1ex]
     (\param_1-1)\binom{j/2}{2} & & j-3 & 5 & &\\[1ex]
       \vdots   &     &           &  \ddots  &  \ddots  & \\[1ex]
     (-1)^{\frac{j}{2}-1}(\param_1+2-\frac{j}{2})\binom{j/2}{j/2-1} & & & & 3 & j-1\\[1ex]
     (-1)^{\frac{j}{2}}(\param_1+1-\frac{j}{2})\binom{j/2}{j/2} & & & & & 1
  \end{matrix}\right) \cdot \left(\begin{matrix}b_{j,1} \\[1ex] b_{j,2} \\[1ex] b_{j,3} \\[1ex] \vdots \\[1ex] b_{j, \frac{j}{2}} \\[1ex] b_{j, \frac{j}{2}+1}\end{matrix}\right) = \left(\begin{matrix}b_{j,0} \\[1ex] 0 \\[1ex] 0 \\[1ex] \vdots \\[1ex] 0 \\[1ex] 0\end{matrix}\right).\]
 \end{enumerate}
\end{theorem}

\begin{proof}
 For all $(P,Q,B)$ in the statement, it is straightforward to verify using the product rule that $L(Q(\alpha,t)t^\alpha)=P(\alpha,t)t^\alpha$ for all $\alpha\in\mathbb{N}^3$ with $B|_{\param=\alpha}$.
 We further prove that the determinant of above coefficient matrix in (iii) is equal to $b_{j,0}\cdot j!!$, so that each $b_{j,i}$ belongs to $C[\param_1,\param_2,\param_3]$ by Cramer's rule for $i \in \{1,2,\ldots,\frac{j}{2}+1\}$.
 We denote the above matrix by $A=(a_{m,n})_{0\le m\le \frac{j}{2},0\le n\le \frac{j}{2}}$ with
 \[\begin{cases}
  a_{m,0}=(-1)^m(\param_1+1-m)\binom{\frac{j}{2}}{m} & 0 \le m \le \frac{j}{2}\\[1ex]
  a_{m,m}=j-2m+1 & 1 \le m \le \frac{j}{2}\\[1ex]
  a_{m-1,m}=2m-1 & 1 \le m \le \frac{j}{2}\\[1ex]
  a_{m,n} = 0 & \mbox{otherwise}.
 \end{cases}\]
 Then by the Laplace expansion with respect to the first column of $A$ we get 
 \[\det(A)=\sum_{m=0}^{\frac{j}{2}}(\param_1+1-m)\binom{\frac{j}{2}}{m}(2m-1)!!(j-2m-1)!!.\]
 In order to show $\det(A)=b_{j,0} \cdot j!!$, we only need to prove the following identities
 \begin{align*}
 \sum_{m=0}^n \binom{n}{m}(2m-1)!!(2n-2m-1)!!&=(2n)!!\\
 \sum_{m=0}^n m \binom{n}{m}(2m-1)!!(2n-2m-1)!!&=\frac{n}{2} (2n)!!,
 \end{align*}
 where $n=\frac{j}{2}$ is a nonnegative integer.
 These two identities can be shown by the Wilf--Zeilberger method as described in \cite[Ch.~7]{AeqB}, for example.
 Consequently, we obtain $\det(A)=b_{j,0}\cdot j!!$ and that $b_{j,1},\ldots,b_{j, \frac{j}{2}+1} \in C[\param_1,\param_2,\param_3]$.
 To see that $P_j(\alpha,t)t^\alpha$ and $Q_j(\alpha,t)t^\alpha$ with even $j$ only involve monomials with nonnegative exponents for $\alpha\in\mathbb{N}^n$ satisfying $B_j$, we note that $b_{j,\frac{j}{2}+1}(\alpha)=(-1)^{\frac{j}{2}+1}(\alpha_1+1-\frac{j}{2})b_{j,1}(\alpha)$ is zero if $\alpha_1=\frac{j}{2}-1$.
\end{proof}

\begin{theorem}\label{TH:Airyunreduced}
 The reduction system given in Theorem~\ref{TH:Airyreductionrule} is complete.
 That is, no monomial of the form $t_1^it_2^j$ can be the leading monomial of any derivative in $C[t_1, t_2, t_3]$ if
 \begin{enumerate}
  \item $i\le\frac{j}{2}-2$ and $j$ is even or
  \item $i\le\frac{j-1}{2}$ and $j$ is odd.
 \end{enumerate}
\end{theorem}
\begin{proof}
 Let $S$ be the reduction system given by Theorem~\ref{TH:Airyreductionrule}.
 We first prove that every non-constant monomial occurs as leading monomial of $Q(\alpha,t)t^\alpha$ for some $(P,Q,B) \in S$ and $\alpha \in \mathbb{N}^3$ with $B|_{\param=\alpha}$.
 Let $(i,j,k) \in \mathbb{N}^3$ be nonzero.
 If $j \ge 1$, then $\lm(Q(\alpha,t)t^\alpha)$ arising from the generic rule with $\alpha=(i,j-1,k+1)$ is equal to $t_1^i t_2^j t_3^k$.
 Otherwise, $t_1^i t_3^k$ is the leading monomial of $Q_k(\alpha,t)t^\alpha$ for $\alpha=(i+\frac{k+1}{2},k,0)$, if $k$ is odd, or of $Q_k(\alpha,t)t^\alpha$ for $\alpha=(i+\frac{k}{2}-1,k,0)$, if $k$ is even.
 Note that $(i,j,k)$ is nonzero, so $i+\frac{k}{2}-1\ge 0$ if $j=0$.
 Altogether, $S$ is precomplete by \eqref{eq:weakcompleteness}.
 On the other hand, no two reduction rules in $S$ form a critical pair, because any two conditions among $\param_3\ge1$ and all $B_j$, $j \in \mathbb{N}$, are inconsistent.
 Then, by Lemma~\ref{lem:NoRedundancy}, the reduction system shown in Theorem~\ref{TH:Airyreductionrule} is complete, that is, a monomial is the leading monomial of a derivative in $C[t_1,t_2,t_3]$ if and only if it is reducible by $S$.
 So, a monomial whose exponent vector does not satisfy the condition $B$ of any $(P, Q, B) \in S$ is not the leading monomial of any derivative in $C[t_1,t_2,t_3]$.
\end{proof}

\begin{example}\label{ex:AiryBoettnerReduction}
Revisiting Example~\ref{ex:AiryBoettner}, we now apply the complete reduction system given by Theorem~\ref{TH:Airyreductionrule} to compute $u \in C[t_1,t_2,t_3]$ such that $\Der{u}=f$ for $f=t_3^2$.
First, the exponent vector $\alpha=(0,0,2)$ of $t_3^2$ satisfies the condition $\param_3\ge1$ of the generic rule.
So, instantiating \eqref{eq:rewriterule} as $t_3^2 = L(t_2t_3) - t_1t_2^2$, we reduce $f$ to the remainder $-t_1t_2^2$ and we obtain $t_2t_3$ as contribution to $u$.
Next, the exponent vector $\alpha=(1,2,0)$ of the leading monomial of the remainder satisfies the condition $B_2=(\param_3=0\wedge\param_2=2\wedge\param_1\ge0)$ of the reduction rule $(P_2,Q_2,B_2)$ obtained by setting $j=2$ in Theorem~\ref{TH:Airyreductionrule}~(iii).
Explicitly, we have $P_2=(\param_1+\tfrac{1}{2})+\frac{\param_1}{2}(\param_1-1)t_1^{-2}t_2^{-1}t_3$ as well as $Q_2=\tfrac{1}{2}(t_1-t_2^{-2}t_3^2)+\frac{\param_1}{2}t_1^{-1}t_2^{-1}t_3$.
This yields $P_2(\alpha,t)t^\alpha=\tfrac{3}{2}t_1t_2^2$ and $Q_2(\alpha,t)t^\alpha=\tfrac{1}{2}(t_1^2t_2^2-t_1t_3^2+t_2t_3)$.
Therefore, the remainder is reduced to $-t_1t_2^2-(-\tfrac{2}{3})P_2(\alpha,t)t^\alpha$, which equals zero.
Altogether, we obtain the solution $u=t_2t_3-\tfrac{2}{3}Q_2(\alpha,t)t^\alpha=-\tfrac{1}{3}t_1^2t_2^2+\tfrac{1}{3}t_1t_3^2+\tfrac{2}{3}t_2t_3$.
\end{example}

\subsection{Complete elliptic integrals}
\label{sec:CEI}

Let $K(x)$ and $E(x)$ be the complete elliptic integrals of first and second kind, respectively, where
\[K(x) = \int_{0}^{\pi/2} \frac{dy}{\sqrt{1-x^2 \sin(y)^2}} \quad \mbox{and} \quad E(x) = \int_{0}^{\pi/2} \sqrt{1-x^2 \sin(y)^2} \, dy.\]
In this subsection, we present a complete reduction system for the differential ring generated over constants by $x$ and above two complete elliptic integrals.
Modelling $x,K(x),E(x)$ by $t_1,t_2,t_3$, respectively, we first generate the differential field $(C(t_1,t_2,t_3),\Der)$ with
\begin{equation}\label{eq:CEIderivatives}
 \Der t_1=1, \quad \Der t_2 = \frac{t_3-(1-t_1^2)t_2}{t_1(1-t_1^2)}, \quad \mbox{and} \quad \Der t_3 = \frac{t_3-t_2}{t_1}.
\end{equation}
Multiplying $\Der$ with the denominator $t_1(1-t_1^2)$, we further get a differential ring $(C[t_1,t_2,t_3],\tilde{\Der})$ with
\begin{equation}\label{eq:CEIderivatives2}
 \tilde{\Der} t_1 = t_1(1-t_1^2), \quad \tilde{\Der} t_2 = t_3-(1-t_1^2)t_2, \quad \text{and} \quad \tilde{\Der} t_3 = (1-t_1^2)(t_3-t_2).
\end{equation}
Similar to the previous example involving Airy functions, by \eqref{eq:defP}, it is easy to find
\[
 p = (\param_1-\param_2+\param_3)(1-t_1^2)+\param_2t_2^{-1}t_3+\param_3(t_1^2-1)t_2t_3^{-1},
\]
such that $(p,1,true)$ encodes a conditional identity for $L=\tilde{\Der}$.
We are going to use two different monomial orders to construct a complete reduction system.

First, we use the block order induced by $\left(\begin{smallmatrix}0&1&1\\0&0&1\\1&0&0\end{smallmatrix}\right)$.
Then, applying Algorithm~\ref{alg:CItoRR} to $(p,1,true)$, we get basic reduction rules as follows:
\begin{align}
 &\left(1+\param_2-\frac{(\param_1-\param_2+\param_3-2)(t_1^2-1)t_2}{t_3}+\frac{(\param_3-1)(t_1^2-1)t_2^2}{t_3^{2}}, t_2t_3^{-1},\param_3\ge1\right)\label{eq:CEIgenericrule}\\
 &\left((2-\param_1+\param_2-\param_3)(1-t_1^{-2})+\frac{\param_3(1-t_1^{-2})t_2}{t_3}, t_1^{-2}, \param_2=0 \wedge \param_1\ge2 \wedge \param_1+\param_3\neq2\right).\label{eq:CEIbasicrule2}
\end{align}
According to Procedure~\ref{proc:RefinedCompletion}, the latter rule is replaced by infinitely many new ones, which can be equivalently described as follows.
\begin{theorem}\label{TH:CEIreductionrule}
 The following are reduction rules for $L=t_1(1-t_1^2)\Der$ w.r.t.\ the block order induced by $\left(\begin{smallmatrix}0&1&1\\0&0&1\\1&0&0\end{smallmatrix}\right)$.
 \begin{enumerate}
  \item[(i)] The generic rule is the first basic reduction rule \eqref{eq:CEIgenericrule}.
  \item[(ii)] For all $j \in \mathbb{N}$, we have reduction rules with
      \begin{align*}
  P_j&=\sum_{n=0}^{j+1}a_{j, n} t_1^{2n-2j-2}\\
  Q_j&=\sum_{m=0}^{j} \sum_{n=0}^m b_{j, m,n} t_1^{2n-2j-2} t_2^{m-j} t_3^{j-m}\\
    B_j&=(\param_3=0 \wedge \param_2=j \wedge \param_1\ge 2j+2 \wedge \param_1\neq 2),
     \end{align*}
   where $b_{j, 0,0}=1$ and $b_{j, m,n}$ with $1 \le m \le j$ and $0\le n \le m$ satisfies the recursion
  \begin{align*}
    b_{j, m,n}={}&\frac{-\param_1+j+2(m-n)}{m}b_{j, m-1,n}+\frac{\param_1-j-2(m-n+1)}{m}b_{j, m-1,n-1}\\
    &+\frac{j+2-m}{m}b_{j, m-2,n}+\frac{m-j-2}{m}b_{j, m-2,n-1},
  \end{align*}
  with the assumption that $b_{j,m,n}=0$ when $n>m$ or $n<0$, and where
   \[a_{j,n}=(\param_1-3j-2+2n)b_{j,j,n}-(\param_1-3j-4+2n)b_{j,j,n-1}-b_{j,j-1,n}+b_{j,j-1,n-1}.\]
 \end{enumerate}
\end{theorem}

\begin{proof}
 For all $(P,Q,B)$ in the statement, it is a straightforward computation in $C[t_1,t_2,t_3]$ to verify that $L(Q(\alpha,t)t^\alpha)=P(\alpha,t)t^\alpha$ for all $\alpha\in\mathbb{N}^3$ with $B|_{\param=\alpha}$.
 It remains to show that $a_{j,j+1}|_{\param=\alpha}\neq0$ whenever $B_j|_{\param=\alpha}$.
 By the assumption that $b_{j,m,n}=0$ for $n>m$, we have $a_{j,j+1}=(j+2-\param_1)b_{j,j,j}$ and $b_{j,m,m} = \frac{\param_1-j-2}{m}b_{j,m-1,m-1}$ for all $m$ with $1 \le m \le j$.
 Since $b_{j,0,0}=1$ for all $j \in \mathbb{N}$, it is easy to check that $a_{j,j+1} = -\frac{(\param_1-j-2)^{j+1}}{j!}$, hence $a_{j,j+1}|_{\param=\alpha}$ is nonzero for all $\alpha\in\mathbb{N}^3$ satisfying $B_j|_{\param=\alpha}$.
\end{proof}

\begin{theorem}\label{TH:CEIcomplete}
 The reduction system given in Theorem~\ref{TH:CEIreductionrule} is complete.
 In particular, no monomial of the form $t_1^i t_2^j$ with $i<\max(2j+2,3)$ is the leading monomial of any element in the image of $C[t_1,t_2,t_3]$ under $\tilde{\Der}$.
\end{theorem}
\begin{proof}
 Let $S$ be the set of reduction rules given in Theorem~\ref{TH:CEIreductionrule}.
 For showing completeness, we verify that
 \begin{enumerate}
  \item[(i)] every non-constant monomial occurs as leading monomial of $Q(\alpha, t) t^\alpha$ for some $(P, Q, B) \in S$ and $\alpha \in \mathbb{N}^3$ with $B|_{\param=\alpha}$ and
  \item[(ii)] no two elements of $S$ form a critical pair.
 \end{enumerate}
 For proving (i), let $(i,j,k) \in \mathbb{N}^3$ be nonzero.
 If $j \ge 1$, we have $\lm(Q(\alpha,t)t^\alpha)=t_1^i t_2^j t_3^k$ from the generic rule with $\alpha=(i,j-1,k+1)$ satisfying $\alpha_3\ge1$.
 If $j=0$, choosing $\alpha=(i+2k+2,k,0)$ yields $\lm(Q_k(\alpha,t)t^\alpha)=t_1^i t_3^k$ and $B_k|_{\param=\alpha}$.
 Thus, (i) holds and $S$ is precomplete.
 It is easy to see (ii), since any two conditions among $\param_3\ge1$ and all $B_j$ with $j \in \mathbb{N}$ in Theorem~\ref{TH:CEIreductionrule} are inconsistent.
 Then, by Lemma~\ref{lem:NoRedundancy}, we obtain completeness, i.e.\ the monomials reducible by $S$ are precisely those that arise as leading monomials of elements in the image of $C[t_1,t_2,t_3]$ under $\tilde{\Der}$.
\end{proof}

\begin{example}\label{EX:CEI1}
 The reduction system given by Theorem~\ref{TH:CEIreductionrule} can be used to find the following integrals, for instance.
 \begin{align}
  \int \frac{x K(x) E(x)}{1-x^2} dx &= \tfrac{1}{2}x^2 K(x)^2\label{eq:CEIexample1}\\
  \int \frac{2x^3 K(x) E(x)-(x^3-x)K(x)^2}{1-x^2}dx &= -\tfrac{1}{2}E(x)^2-(x^2-1)K(x)E(x)+\left(\tfrac{3}{2}x^2-\tfrac{1}{2}\right)K(x)^2\label{eq:CEIexample2}
 \end{align}
 We view the integrands as elements of the differential field $(C(t_1,t_2,t_3), \Der)$ defined by \eqref{eq:CEIderivatives} and aim to find integrals $\frac{u}{v}$ with $u \in C[t_1,t_2,t_3]$ and $v=1$.
 After multiplication by $\den(\Der)$, we need to find polynomials $u \in C[t_1,t_2,t_3]$ such that $\tilde{\Der}u=f_i$ with $f_1=t_1^2t_2t_3$ and $f_2 = 2t_1^4t_2t_3-t_1^4t_2^2+t_1^2t_2^2$, where $\tilde{\Der}$ is defined by \eqref{eq:CEIderivatives2}.
 So, we apply the reduction rules in Theorem~\ref{TH:CEIreductionrule} to $f_1$ and $f_2$.
\par
 Since the degree of $f_1$ in $t_3$ equals $1$, we first apply the generic rule in Theorem~\ref{TH:CEIreductionrule}~(i) to $f_1$ and obtain that $f_1=L\left(\frac{1}{2}t_1^2 t_2^2\right)$.
 Similarly, since the leading monomial of $f_2$ has exponent vector $(4,1,1)$, we also apply the generic rule to $f_2$ so that $f_2 = L\left(t_1^4t_2^2\right) + 2t_1^6t_2^2-3t_1^4t_2^2+t_1^2t_2^2$.
 This means, $f_2$ has been reduced to a remainder, which has leading term $2t_1^6t_2^2$ with exponent vector $\alpha=(6,2,0)$.
 Further reduction of this term requires use of the reduction rule $(P_2,Q_2,B_2)$ obtained by setting $j=2$ in Theorem~\ref{TH:CEIreductionrule}~(ii).
 Explicitly, $P_2(\alpha,t)t^\alpha = -4t_1^6t_2^2+6t_1^4t_2^2-2t_1^2t_2^2$ and $Q_2(\alpha,t)t^\alpha = t_3^2+(2t_1^2-2)t_2t_3+(2t_1^4-3t_1^2+1)t_2^2$ yield $2t_1^6t_2^2-3t_1^4t_2^2+t_1^2t_2^2 = L\left(-\frac{1}{2}Q_2(\alpha,t)t^\alpha\right)$, i.e. the remainder is reduced to zero.
 Altogether, we obtain $f_2=L\left(-\frac{1}{2}t_3^2+(-t_1^2+1)t_2t_3+(\frac{3}{2}t_1^2-\frac{1}{2})t_2^2\right)$ and hence \eqref{eq:CEIexample2}.
\end{example}

On the other hand, we are going to show another complete reduction system for the same operator $L$ with respect to a different order by swapping the two blocks of the above block order of monomials, i.e.\ we use the order induced by $\left(\begin{smallmatrix}1&0&0\\0&1&1\\0&0&1\end{smallmatrix}\right)$.
Then, we see that the reduction rules below are much easier than before with respect to this order.
They not only contain fewer monomials but even admit a simple fully explicit representation.
\begin{theorem}\label{TH:CEIreductionrule2}
 The following are reduction rules for $L=t_1(1-t_1^2)\Der$ w.r.t.\ the block order $<$ with $t_2<t_3<t_1$ defined by $\left(\begin{smallmatrix}1&0&0\\0&1&1\\0&0&1\end{smallmatrix}\right)$.
\begin{enumerate}
  \item[(i)] The generic rule is given in terms of
\begin{equation}\label{eq:CEInewgenericrule}
\begin{aligned}
     P_0={}&-(\param_1-\param_2+\param_3-2)+\param_3 t_2t_3^{-1}+\param_2t_1^{-2}t_2^{-1}t_3\\
     &+(\param_1-\param_2+\param_3-2)t_1^{-2}-\param_3t_1^{-2}t_2t_3^{-1}\\
     Q_0={}&t_1^{-2}\\
     B_0={}&(\param_1\ge2 \wedge \param_1-\param_2+\param_3\neq2)
\end{aligned}
\end{equation}
  \item[(ii)] For each $k \in \mathbb{N}^+$, we additionally have a reduction rule given by
\begin{equation}\label{eq:CEInewrule}
\begin{aligned}
     P_k&=\left(t_3-\frac{t_2}{2}\right)^{k-2}\left((\param_1+k-1)t_3-(\frac{\param_1}{2}+k-1)t_2\right)t_3^{-k+1}\\
     Q_k&=t_2 \left(t_3-\frac{t_2}{2}\right)^{k-1}t_3^{-k}\\
     B_k&=(\param_3= k\wedge \param_1+\param_3\neq1 \wedge \param_2= \param_1+k-2)
\end{aligned}
\end{equation}
\end{enumerate}
\end{theorem}

\begin{proof}
 Verifying that these are indeed reduction rules for $L$ w.r.t. $<$ can be done by easy computations in the differential ring.
\end{proof}
Again, for completeness of this reduction system, it is easy to verify that every non-constant monomial $t_1^i t_2^j t_3^k$, with $i, j, k$ not all zero, occurs as a leading monomial of either $Q_0 t_1^{i+2}t_2^jt_3^k$ or $Q_{k+1} t_1^i t_1^{i+k-1} t_3^{k+1}$, and that no critical pairs are formed.
So, monomials that cannot be reduced by these rules are indeed not reducible w.r.t.\ the chosen monomial order.
Nonetheless, we also present another proof of completeness as follows.
\begin{theorem}\label{th:CEInewcompleteness}
 If a monomial is neither reducible by \eqref{eq:CEInewgenericrule} nor by \eqref{eq:CEInewrule}, then it is not the leading monomial of any element in the image of $C[t_1,t_2,t_3]$ under $\tilde{\Der}$.
\end{theorem}
\begin{proof}
 Since $\tilde{\Der}t^\alpha$ is homogeneous w.r.t.\ $(0,1,1)$ with the same degree as $t^\alpha$, monomials with different $\deg_{(0,1,1)}$ can be considered independently of each other.
 For another monomial order, it has already been shown in Theorem~\ref{TH:CEIcomplete} that any monomial is the leading monomial of an element in the image of $C[t_1,t_2,t_3]$ under $\tilde{\Der}$, unless it is of the form $t_1^it_2^d$ with $i \le \max(2,2d+1)$.
 Therefore, it is sufficient to show that for every $d \in \mathbb{N}$ there are exactly $\max(3,2d+2)$ monomials with $\deg_{(0,1,1)}(t^\alpha)=d$ that can be reduced by neither \eqref{eq:CEInewgenericrule} nor \eqref{eq:CEInewrule} w.r.t.\ the present monomial order.\par
 If $d=0$, then only \eqref{eq:CEInewgenericrule} can be applied and all $t_1^i$ with $i\ge3$ can be reduced leaving exactly the same irreducible monomials $1,t_1,t_1^2$ as for the previous monomial order.
 If $d\ge1$, then a monomial $t_1^it_2^jt_3^k$, with $j+k=d$ and $i\ge2$, can be reduced by \eqref{eq:CEInewgenericrule}, if $j\neq i+k-2$, and by \eqref{eq:CEInewrule}, if $j=i+k-2$ and $k\ge1$.
 Hence, $t_1^{d+2}t_2^d$ is the only irreducible monomial with $i\ge2$.
 A monomial $t_1^it_2^jt_3^k$, with $j+k=d$ and $i\le1$, can only be reduced by \eqref{eq:CEInewrule} and only if $j=i+k-2$, which is equivalent to $j=\frac{d+i}{2}-1$ and $k=\frac{d-i}{2}+1\ge1$.
 Hence, among the monomials with $i\le1$, $t_1^{(d\bmod2)}t_2^{\lfloor\frac{d-1}{2}\rfloor}t_3^{\lfloor\frac{d}{2}\rfloor+1}$ is the only reducible one and the other $2d+1$ are irreducible.
 Altogether, this gives $2d+2$ irreducible monomials with $j+k=d$, which is precisely the number determined w.r.t.\ the previous monomial order.
\end{proof}

\begin{example}\label{EX:CEI2}
 Revisiting Example~\ref{EX:CEI1}, we now apply the reduction rules given in Theorem~\ref{TH:CEIreductionrule2} to $f_1=t_1^2t_2t_3$ and $f_2 = 2t_1^4t_2t_3-t_1^4t_2^2+t_1^2t_2^2$.
 Since the exponent vector $\alpha=(2,1,1)$ of $f_1$ satisfies the condition $B_1$ of the new rule given by setting $k=1$ in \eqref{eq:CEInewrule}, we apply this reduction rule $(P_1,Q_1,B_1)$ that has $P_1=\param_1$ and $Q_1=t_2t_3^{-1}$.
 Then, we have $f_1=L(\frac{1}{2}Q_1(\alpha,t)t^\alpha)$ and $f_1$ is reduced to zero.
\par
 For reducing $f_2$, we first apply the generic rule \eqref{eq:CEInewgenericrule} twice to reduce monomials $t^\alpha$ with $\alpha=(4,1,1)$ and $\alpha=(2,0,2)$ successively.
 Thereby, we obtain the contribution $-t_1^2t_2t_3-\frac{1}{2}t_3^2$ to the integral and $f_2$ is reduced to $3t_1^2t_2t_3+t_3^2-t_2t_3$, which has the same leading monomial as $f_1$.
 So, we apply $(P_1,Q_1,B_1)$ again to get remainder $t_3^2-t_2t_3$ by $L(\frac{3}{2}t_1^2t_2^2)$.
 Reduction of the leading monomial with exponent vector $\alpha=(0,0,2)$ requires to set $k=2$ in \eqref{eq:CEInewrule}, which yields the rule $(P_2,Q_2,B_2)$ with $P_2=(\param_1+1)-(\frac{\param_1}{2}+1)t_2t_3^{-1}$ and $Q_2=t_2t_3^{-1}-\frac{1}{2}t_2^2t_3^{-2}$.
 This gives $t_3^2-t_2t_3=L(Q_2(\alpha,t)t^\alpha)$ and reduces the remainder to zero.
 Altogether, we compute the same integrals as in Example~\ref{EX:CEI1}.
\end{example}

\section{Rigorous degree bounds}
\label{sec:degbound}
Solving \eqref{eq:PolynomialAnsatz} by ansatz usually relies on determining a finite candidate set for the monomials in $\supp(u)$ via degree bounds like \eqref{eq:degboundElem}--\eqref{eq:degboundparrisch}, which typically are only heuristic for given $f$ and $v$.
For solving our main problem \eqref{eq:MainEquation}, we now investigate degree bounds for $L$ given by \eqref{eq:defL} and \eqref{eq:defP} that are rigorous in the sense of the following definition.
Throughout this section, $C[t]$ is the ring of polynomials with coefficients in a field $C$ of characteristic zero in the indeterminates $t_1,\dots,t_n$ and $L$ denotes a $C$-linear map from $C[t]$ to itself.
\begin{definition}\label{def:DegreeBounds}
 Let $w \in \mathbb{R}^n$.
 A function $\varphi:\mathbb{R}\to[-\infty,\infty]$ is called a \emph{degree bound} for $L$ w.r.t.\ the weight vector $w$, if for every nonzero $f \in \im(L)$ there exists $g \in C[t]$ with $f=L(g)$ s.t.
 \begin{equation}\label{eq:DegreeBoundDef}
  \deg_w(g) \le \varphi(\deg_w(f)).
 \end{equation}
\end{definition}

\begin{example}\label{ex:DegreeBounds}
Revisiting Example~\ref{ex:NormanMain}, we use the denominator $v=t_2^2+1$ for finding integrals in the differential field $(C(t_1,t_2),\Der)$ with $\Der{t_1}=1$ and $\Der{t_2}=t_2^2+1$.
Given \eqref{eq:defP} and using results from the literature, this concrete derivation allows to obtain degree bounds for $L$ w.r.t.\ all partial degrees, i.e., let $w$ be $(1,0)$ and $(0,1)$, respectively.
Explicitly, we have $\varphi(x)=x+1$ as a degree bound for $L$ w.r.t.\ $(1,0)$ by Lemma~5.1.2 in \cite{Bronstein}, and for $w=(0,1)$, we get $\varphi(x)=\max(x-1,2)$ by Thm.~4.4.4 in \cite{Bronstein}.
In particular, $f=t_1$ and $g=\frac{1}{4}t_1^2t_2^2+\frac{1}{2}t_1t_2+\frac{1}{4}t_1^2+\frac{1}{4}$ from Example~\ref{ex:NormanMain} satisfy $\deg_{t_1}(g)=\deg_{t_1}(f)+1$ and $\deg_{t_2}(g)=2$, which illustrates tightness of both degree bounds.
\end{example}

Despite often being only heuristic, all standard bounds used in Risch--Norman integration could essentially be represented by functions of the form $\varphi(x)=\max(x+b,c)$ with $b,c \in \mathbb{Z}$ w.r.t.\ some $w=e_i$ or $w=(1,\dots,1)$.
By considering concrete examples, we will see later that the situation for rigorous degree bounds is different.
For given $L$ induced by \eqref{eq:defP}, there might be weight vectors $w$ that do not admit nontrivial degree bounds $\varphi$ with $\varphi(\deg_w(f))<\sup\{\deg_w(t^\alpha)\ |\ \alpha\in\mathbb{N}^n\}$ for some nonzero $f\in\im(L)$, see Theorem~\ref{thm:Airyinfinite} in the case of Airy functions.
Even if nontrivial degree bounds $\varphi$ exist for fixed $L$ and $w$, it might be that none of them is asymptotically equivalent to $x$ as $x\to\infty$, see Example~\ref{ex:AiryTightness} for $w=(1,\dots,1)$.
Abstractly, by considering the whole set of solutions $L^{-1}(f):=\{g\in{C[t]}\ |\ L(g)=f\}$ for any $f \in \im(L)$, we can think of degree bounds that are optimal as follows.

\begin{lemma}\label{lem:optimal}
 Let $w \in \mathbb{R}^n$ be any weight vector such that the function $\varphi_w^*:\mathbb{R}\to[-\infty,\infty]$ can be defined by
 \begin{equation}\label{eq:phiopt}
  \varphi_w^*(x) := \sup\left\{\min\{\deg_w(g)\ |\ g\in{L^{-1}(f)}\}\ \middle|\ f\in\im(L) \wedge \deg_w(f)=x\right\},
 \end{equation}
 i.e.\ for every nonzero $f \in \im(L)$ the set $\{\deg_w(g)\ |\ g\in{L^{-1}(f)}\}$ has a minimal element.
 Let $D:=\{\deg_w(f)\ |\ f \in \im(L)\setminus\{0\}\}$.
 Then, the following hold:
 \begin{enumerate}
  \item\label{item:phioptbound} For all $f \in \im(L)\setminus\{0\}$ there exists $g \in L^{-1}(f)$ with $\deg_w(g) \le \varphi_w^*(\deg_w(f))$.
  \item\label{item:phioptfinite} For all $x \in D$ with $\varphi_w^*(x)<\infty$ and all $\varepsilon>0$, there exists $f \in \im(L)$ with $\deg_w(f)=x$ such that for all $g \in L^{-1}(f)$ we have $\deg_w(g) > \varphi_w^*(x)-\varepsilon$.
  \item\label{item:phioptinfinite} For all $x \in D$ with $\varphi_w^*(x)=\infty$ and all $\varepsilon>0$, there exists $f \in \im(L)$ with $\deg_w(f)=x$ such that for all $g \in L^{-1}(f)$ we have $\deg_w(g) > \frac{1}{\varepsilon}$.
  \item\label{item:phioptmonotonic} On $D$, the restriction $\varphi_w^*|_D$ is (weakly) monotonically increasing.
 \end{enumerate}
\end{lemma}
\begin{proof}
 The first property immediately follows form the definition \eqref{eq:phiopt}, if we choose $g \in L^{-1}(f)$ with minimal $w$-degree, which exists by assumption.
 Similarly, properties \ref{item:phioptfinite} and \ref{item:phioptinfinite} are direct consequences of \eqref{eq:phiopt} as well.
\par
 Finally, to show monotonicity, we take $x_1,x_2 \in D$ with $x_1 \le x_2$ and we take $\varepsilon>0$ arbitrary.
 By virtue of properties \ref{item:phioptfinite} and \ref{item:phioptinfinite}, there exists $f_1 \in \im(L)$ with $\deg_w(f_1)=x_1$ such that for all $g_1 \in L^{-1}(f_1)$ we have $\deg_w(g_1) > \min(\varphi_w^*(x_1)-\varepsilon,\frac{1}{\varepsilon})$.
 We also fix some $f_2 \in \im(L)$ with $\deg_w(f_2)=x_2$.
 By assumption on $w$, we can choose $g_1,g_2 \in C[t]$ with $L(g_i)=f_i$ and $\deg_w(g_i) = \min\{\deg_w(g)\ |\ g\in{L^{-1}(f_i)}\}$.
\par
 Next, we show that $\deg_w(g_1) \le \varphi_w^*(x_2)$.
 If $\deg_w(g_1) \le \deg_w(g_2)$, then we immediately get $\deg_w(g_1) \le \deg_w(g_2) \le \varphi_w^*(x_2)$.
 Otherwise, if $\deg_w(g_1) > \deg_w(g_2)$, then there exists $c \in C$ such that $f:=f_1+cf_2$ satisfies $\deg_w(f)=x_2$ and $g:=g_1+cg_2$ satisfies $\deg_w(g)=\deg_w(g_1)$.
 If some $\tilde{g} \in C[t]$ would satisfy $L(\tilde{g})=f$ and $\deg_w(\tilde{g})<\deg_w(g)$, then $L(\tilde{g}-cg_2)=f_1$ and $\deg_w(\tilde{g}-cg_2) \le \max(\deg_w(\tilde{g}),\deg_w(g_2)) <  \max(\deg_w(g),\deg_w(g_1)) = \deg_w(g_1)$ would follow in contradiction to minimality of $\deg_w(g_1)$.
 Hence, $\deg_w(g)=\min\{\deg_w(\tilde{g})\ |\ \tilde{g}\in{L^{-1}(f)}\}$ holds.
 Together with $\deg_w(f)=x_2$, this implies $\deg_w(g_1) = \deg_w(g) \le \varphi_w^*(x_2)$ again.
\par
 Altogether, we obtain $\min(\varphi_w^*(x_1)-\varepsilon,\frac{1}{\varepsilon}) \le \varphi_w^*(x_2)$ independent of $g_1,g_2$.
 Since $\varepsilon>0$ was arbitrary, it follows from $\varphi_w^*(x_1)=\sup\{\min(\varphi_w^*(x_1)-\varepsilon,\frac{1}{\varepsilon})\ |\ \varepsilon>0\}$ that $\varphi_w^*(x_1) \le \varphi_w^*(x_2)$ as claimed.
\end{proof}

Monotonicity is not only a natural property of optimal degree bounds as shown by property~\ref{item:phioptmonotonic} in Lemma~\ref{lem:optimal}.
In general, monotonicity allows to construct degree bounds for new weight vectors from existing ones as shown by the following lemma.

\begin{lemma}\label{lem:conversion}
 Let $f \in C[t]$, $w_1,\dots,w_k \in \mathbb{R}^n$ and let $\varphi_1,\dots,\varphi_k$ be (weakly) monotonically increasing maps from $\mathbb{R}$ to $[-\infty,\infty]$.
 Let $v \in \mathbb{R}^n$, $c_1,\dots,c_k \in \mathbb{R}$, and $\lambda_1,\dots,\lambda_k>0$ such that $v_j \le \sum_{i=1}^k\lambda_iw_{i,j}$ and $\deg_{w_i}(f) \le c_i\deg_v(f)$. Then,
 \begin{equation}
  \varphi(x) := \sum_{i=1}^k\lambda_i\varphi_i(c_ix)
 \end{equation}
 satisfies $\deg_v(g) \le \varphi(\deg_v(f))$ for all $g \in C[t]$ with $\deg_{w_i}(g) \le \varphi_i(\deg_{w_i}(f))$ for every $i\in\{1,\dots,k\}$.
\end{lemma}
\begin{proof}
 Let $g \in C[t]$ such that $\deg_{w_i}(g) \le \varphi_i(\deg_{w_i}(f))$ holds for every $i\in\{1,\dots,k\}$.
 The case $g=0$ is trivial.
 For $g\neq0$, let $t^\alpha \in \supp(g)$ with $\deg_v(t^\alpha)=\deg_v(g)$.
 Then, by $v_j \le \sum_{i=1}^k\lambda_iw_{i,j}$, we have $\deg_v(g) = \sum_{j=1}^nv_j\alpha_j \le \sum_{j=1}^n\sum_{i=1}^k\lambda_iw_{i,j}\alpha_j = \sum_{i=1}^k\lambda_i\deg_{w_i}(t^\alpha)$ since $\alpha_j \ge 0$.
 Using $t^\alpha \in \supp(g)$ and the assumptions on $\varphi_i$ and $c_i$, it follows that $\deg_{w_i}(t^\alpha) \le \deg_{w_i}(g) \le \varphi_i(\deg_{w_i}(f)) \le \varphi_i(c_i\deg_v(f))$.
 Altogether, with $\lambda_i>0$, we obtain $\deg_v(g) \le \sum_{i=1}^k\lambda_i\deg_{w_i}(t^\alpha) \le \sum_{i=1}^k\lambda_i\varphi_i(c_i\deg_v(f)) = \varphi(\deg_v(f))$.
\end{proof}

Note that, in the following theorem, the value $+\infty$ does not appear in the degree bounds and we restrict to nonnegative weights from $\mathbb{R}_0^+=[0,\infty[$.
So, we can use the construction of Lemma~\ref{lem:conversion} to obtain a new degree bound that excludes the value $+\infty$ and is monotonically increasing as well.

\begin{theorem}\label{TH:degreeboundcombin}
 Let $w_1,\dots,w_k \in (\mathbb{R}_0^+)^n$ and let $\varphi_1,\dots,\varphi_k$ be (weakly) monotonically increasing maps from $\mathbb{R}$ to $[-\infty,\infty[$.
 Furthermore, let $v \in (\mathbb{R}_0^+)^n$ such that for all $j\in\{1,\dots,n\}$ with $v_j>0$ there exists $i\in\{1,\dots,k\}$ such that $w_{i,j}>0$ and $\forall\,{m\in\{1,\dots,n\}}: v_m=0 \Rightarrow w_{i,m}=0$.
 Then, there is a (weakly) monotonically increasing map $\varphi:\mathbb{R}\to[-\infty,\infty[$ such that $\deg_v(g) \le \varphi(\deg_v(f))$ holds for all $f,g \in C[t]$ that satisfy $\deg_{w_i}(g) \le \varphi_i(\deg_{w_i}(f))$ for every $i\in\{1,\dots,k\}$.
\end{theorem}
\begin{proof}
 If $v=0$, then the statement trivially holds with $\varphi(x):=0$.
 So, we assume $v\neq0$ now.
 By assumption on $v$, there are $\lambda_1,\dots,\lambda_k\ge0$ with $v_j \le \sum_{i=1}^k\lambda_iw_{i,j}$ for all $j\in\{1,\dots,n\}$ and such that for all $i$ with $\lambda_i>0$ we have $\forall\,{m\in\{1,\dots,n\}}: v_m=0 \Rightarrow w_{i,m}=0$.
 Without loss of generality, we assume that $\lambda_1,\dots,\lambda_k>0$, since we can remove some of the $w_i$ and their corresponding $\varphi_i$.
 Then, $w_{1,j}=\ldots=w_{k,j}=0$ holds for all $j\in\{1,\dots,n\}$ with $v_j=0$.
 Hence, with
 \[
  c_i:=\max\left\{\tfrac{w_{i,j}}{v_j}\ \middle|\ j\in\{1,\dots,n\} \wedge v_j>0\right\}\ge0,
 \]
 we have $\deg_{w_i}(t^\alpha) = \sum_{j=1}^nw_{i,j}\alpha_j \le \sum_{j=1}^nc_iv_j\alpha_j = c_i\deg_v(t^\alpha)$ for all $\alpha\in\mathbb{N}^n$ and therefore $\deg_{w_i}(f) \le c_i\deg_v(f)$ for all $f \in C[t]$.
 Now, Lemma~\ref{lem:conversion} implies that $\varphi(x) := \sum_{i=1}^k\lambda_i\varphi_i(c_ix)$ satisfies $\deg_v(g) \le \varphi(\deg_v(f))$ whenever $\deg_{w_i}(g) \le \varphi_i(\deg_{w_i}(f))$ holds for every $i\in\{1,\dots,k\}$.
 Since all $\lambda_i,c_i$ are nonnegative and all $\varphi_i$ are monotonically increasing and do not take the value $+\infty$, $\varphi$ too is monotonically increasing and does not take the value $+\infty$.
\end{proof}

In particular, the following two special cases are covered by Theorem~\ref{TH:degreeboundcombin}. By choosing the standard unit vectors $w_i=e_i$, we see that a collection of finite degree bounds $\varphi_1,\dots,\varphi_n$ for all partial degrees gives rise to a finite bound $\varphi(x):=\sum_{i=1}^n\varphi_i(x)$ for the total degree.
From a finite degree bound $\varphi$ w.r.t.\ a single nonzero weight vector $w \in (\mathbb{R}_0^+)^n$, we obtain finite bounds $\psi(x):=\lambda\varphi(cx)$ w.r.t.\ any weight vectors $v \in (\mathbb{R}_0^+)^n$ that have zeros in exactly the same positions as $w$ by letting $\lambda:=\max\{\frac{v_i}{w_i}\ |\ w_i\neq0\}$ and $c=\max\{\frac{w_i}{v_i}\ |\ w_i\neq0\}$.

\subsection{Bounds based on homogeneity}

For a given Laurent polynomial $p$ in $t_1,\dots,t_n$ with coefficients being polynomials in $\param_1,\dots,\param_n$, let us now assume that $w \in \mathbb{R}^n$ is such that $p$ is $w$-homogeneous.
Based on \eqref{eq:defL}, it is straightforward to see that the following is true with $d:=\deg_w(p)$.
For any $f \in \im(L)$ there exists $g \in C[t]$ such that $L(g)=f$ and $\deg_w(g) \le \deg_w(f)-d$.
In other words,
\begin{equation}\label{eq:homogeneousbounds}
 \varphi(x) := x-d
\end{equation}
is a degree bound for $L$ w.r.t.\ the weight vector $w$.
In fact, the bound results from the following stronger statement that allows to split the problem \eqref{eq:MainEquation}, if \eqref{eq:defP} is $w$-homogeneous.
\begin{lemma} \label{LM:homogeneousbound}
 Let $w \in \mathbb{R}^n$ such that $p \in C[\param][t,t^{-1}]$ is $w$-homogeneous of $w$-degree $d\in\mathbb{R}$ as a Laurent polynomial in $t_1,\dots,t_n$.
 Let $f \in C[t]$ and let $f_1,\dots,f_k$ be its nonzero $w$-homogeneous components.
 Then, $f \in \im(L)$ if and only if there are $w$-homogeneous $g_1,\dots,g_k \in C[t]$ such that $L(g_i)=f_i$ and $\deg_w(g_i)=\deg_w(f_i)-d$ for all $i\in\{1,\dots,k\}$.
\end{lemma}
\begin{proof}
 Trivially, $L(g_i)=f_i$ for all $i\in\{1,\dots,k\}$ yields $L(\sum_{i=1}^kg_i)=\sum_{i=1}^kf_i=f$.
 Conversely, let $g \in C[t]$ such that $L(g)=f$ and let $g_1,\dots,g_m$ be the $w$-homogeneous components of $g$.
 Without loss of generality, we assume that $L(g_i)\neq0$ for all $i\in\{1,\dots,m\}$.
 Since $L(g_i)$ is nonzero and $w$-homogeneous for all $i\in\{1,\dots,m\}$, $\sum_{i=1}^mL(g_i)=\sum_{i=1}^kf_i$ implies $m=k$ and (after possibly permuting the $g_i$) $L(g_i)=f_i$ for all $i\in\{1,\dots,k\}$.
 By \eqref{eq:defL} and definition of $d$, this also yields $\deg_w(g_i)=\deg_w(f_i)-d$.
\end{proof}

In general, one cannot expect that a nonzero weight vector with this property exists for a given Laurent polynomial $p$.
However, it is fairly easy to determine the set of all $w \in \mathbb{R}^n$ such that $p$ is $w$-homogeneous using linear algebra with exponent vectors of the monomials $t^\alpha\in\supp(p)$.
So, in practice, it can still be worthwhile to look for nonzero weight vectors that make $p$ homogeneous, since they provide an easy way of predicting part of the shape of solutions of $L(g)=f$.
Indeed, all the examples involving non-elementary functions considered in Section~\ref{sec:InfiniteSystems} are homogeneous w.r.t.\ $(0,1,1)$.

\subsection{Bounds based on complete reduction systems}
\label{sec:boundsfromsystems}

In order to prove a rigorous degree bound for $L$ using a reduction system $S$, a slightly stronger condition than completeness is needed for the given reduction system.
We require that every element of $\im(L)$ can be reduced to zero by $S$ in finitely many steps.
For shorter notation in what follows, we associate to a reduction system $S$ its set of instances
\begin{equation}\label{eq:instances}
 \Sigma:=\{(P(\alpha,t)t^\alpha,Q(\alpha,t)t^\alpha)\ |\ (P,Q,B)\in{S},\alpha\in\mathbb{N}^n,B|_{\param=\alpha}\}.
\end{equation}

\begin{theorem}\label{thm:tightbound}
 Let $S$ be a complete reduction system for $L$ w.r.t.\ $<$ that induces a normalizing reduction relation on $\im(L)$ and let $\Sigma$ be its set of instances.
 Let $w\in\mathbb{R}^n$ be an arbitrary weight vector that satisfies
 \begin{equation}\label{eq:compatibility}
  \forall\,(P,Q,B)\in{S}:\deg_w(P)=0.
 \end{equation}
 Then, the function $\varphi:\mathbb{R}\to[-\infty,\infty]$ given by
 \begin{equation}\label{eq:phi}
  \varphi(x) := \sup\{\deg_w(g)\ |\ (f,g)\in\Sigma \wedge \deg_w(f)\le{x}\}
 \end{equation}
 is a degree bound for $L$ w.r.t.\ $w$.
 Moreover, this bound is everywhere tight in the sense that, for all $x \in \{\deg_w(f)\ |\ f \in \im(L)\setminus\{0\}\}$ and all $\varepsilon>0$, there exist $f \in \im(L)$ and $g \in L^{-1}(f)$ with $\deg_w(f)=x$ and $\deg_w(g) > \varphi(\deg_w(f))-\varepsilon$, if $\varphi(x)<\infty$, or $\deg_w(g) > \frac{1}{\varepsilon}$, if $\varphi(x)=\infty$.
\end{theorem}
\begin{proof}
 First, let $f \in \im(L)$ be nonzero.
 Since the reduction system $S$ induces a normalizing reduction relation on $\im(L)$, there are $(f_1,g_1),\dots,(f_m,g_m) \in \Sigma$ and nonzero $c_1,\dots,c_m \in C$ such that $f-\sum_{i=1}^mc_if_i \in \im(L)$ cannot be reduced further and $\lm(f_j)\in\supp\left(f-\sum_{i=1}^{j-1}c_if_i\right)$ for all $j \in \{1,\dots,m\}$.
 From completeness of $S$ it follows that $f=\sum_{i=1}^mc_if_i$.
 Then, with \eqref{eq:compatibility}, we obtain $\deg_w(f_j) = \deg_w(\lm(f_j)) \le \deg_w\left(f-\sum_{i=1}^{j-1}c_if_i\right)$ for all $j \in \{1,\dots,m\}$, which implies $\deg_w(f_j)\le\deg_w(f)$ by induction.
 Consequently, with $g:=\sum_{i=1}^mc_ig_i$, we have $L(g)=f$ and $\deg_w(g) \le \max_{i=1}^m\deg_w(g_i) \le \varphi(\deg_w(f))$ by \eqref{eq:phi}.
 This proves that $\varphi$ is a degree bound for $L$ w.r.t.\ $w$.
\par
 Second, to show tightness, let $x \in \{\deg_w(f)\ |\ f \in \im(L)\setminus\{0\}\}$ and $\varepsilon>0$.
 By definition of $x$, there is $(f_1,g_1) \in \Sigma$ with $\deg_w(f_1)=x$.
 By \eqref{eq:phi}, there is $(f_2,g_2) \in \Sigma$ with $\deg_w(f_2)\le\deg_w(f_1)$ and $\deg_w(g_2) > \varphi(\deg_w(f_1))-\varepsilon$ resp.\ $\deg_w(g_2) > \frac{1}{\varepsilon}$, if $\varphi(x)<\infty$ resp.\ $\varphi(x)=\infty$.
 Furthermore, there exist $c_1,c_2 \in C$ not both zero such that $f:=c_1f_1+c_2f_2$ satisfies $\deg_w(f)=\deg_w(f_1)$ and $g:=c_1g_1+c_2g_2$ satisfies $\deg_w(g)=\max(\deg_w(g_1),\deg_w(g_2))$.
 Altogether, we have that $\deg_w(g) \ge \deg_w(g_2) > \varphi(\deg_w(f_1))-\varepsilon = \varphi(\deg_w(f))-\varepsilon$ resp.\ $\deg_w(g) \ge \deg_w(g_2) > \frac{1}{\varepsilon}$.
 This concludes the proof since $L(g)=f$.
\end{proof}

\begin{remark}
\label{rem:phi}
 We emphasize a few immediate properties of the degree bound given by Theorem~\ref{thm:tightbound}.
 \begin{enumerate}
  \item Note that the function $\varphi$ defined by \eqref{eq:phi} is (weakly) monotonically increasing.
   In particular, it is minimal among all monotonically increasing functions which satisfy on all of $\mathbb{R}$ that\label{rem:phiminimal}
   \[
    \varphi(x) \ge \sup\{\deg_w(g)\ |\ (f,g)\in\Sigma \wedge \deg_w(f)=x\}.
   \]
  \item It may happen that the bound $\varphi$ gives a trivial value, i.e.\ $\varphi(x)=\sup\{\deg_w(t^\alpha)\ |\ \alpha\in\mathbb{N}^n\}$, for some $x$.
  \item Even though Theorem~\ref{thm:tightbound} shows tightness of the bound in a certain sense, the values determined by \eqref{eq:phi} may still not be the lowest possible choice.
  Note the difference to the properties \ref{item:phioptfinite} and \ref{item:phioptinfinite} of Lemma~\ref{lem:optimal}.
  However, if for all $(f,g) \in \Sigma$ and all $\tilde{g} \in L^{-1}(f)$ we have $\deg_w(\tilde{g})\ge\deg_w(g)$, then \eqref{eq:phi} yields the optimal value \eqref{eq:phiopt} for every $x \in \{\deg_w(f)\ |\ f \in \im(L)\setminus\{0\}\}$.\qed
 \end{enumerate}
\end{remark}

Note that, if $<$ is Noetherian, then any reduction system w.r.t.\ $<$ trivially induces a normalizing reduction relation.
If the order $<$ is induced by $M \in \mathrm{GL}_n(\mathbb{R})$, then choosing $w$ equal to the first row of $M$ trivially yields \eqref{eq:compatibility}.
Clearly, any function that is pointwise larger or equal to $\varphi(x)$ defined in \eqref{eq:phi} will yield another degree bound w.r.t.\ $w$.
For instance, the following degree bound can be computed easily without looking at the conditions $B$ at all.

\begin{corollary}
\label{cor:simplebound}
 With the assumptions of Theorem~\ref{thm:tightbound}, we also have
 \begin{equation}\label{eq:phi2}
  \varphi(x):=x+\sup\{\deg_w(Q)\ |\ (P,Q,B) \in S\}
 \end{equation}
 as a degree bound for $L$ w.r.t.\ $w$.
\end{corollary}
\begin{proof}
 For $(f,g)=(P(\alpha,t)t^\alpha,Q(\alpha,t)t^\alpha) \in \Sigma$, we have $\deg_w(g) \le \deg_w(Q)+\deg_w(t^\alpha)$.
 By \eqref{eq:compatibility}, we have $\deg_w(t^\alpha) = \deg_w(f)$.
 Altogether, we obtain $\sup\{\deg_w(g)\ |\ (f,g)\in\Sigma \wedge \deg_w(f)\le{x}\} \le \varphi(x)$ by \eqref{eq:phi2}.
\end{proof}

If the underlying reduction system is finite, then \eqref{eq:phi2} trivially yields $\varphi$ of the form $\varphi(x)=x+b$ with $b\in\mathbb{R}$.
Consequently, $\varphi$ defined by \eqref{eq:phi} satisfies $\limsup_{x\to\infty}\frac{\varphi(x)}{x}\le1$ if the reduction system is finite, since \eqref{eq:phi} is pointwise less or equal \eqref{eq:phi2}.
If the reduction system is infinite, however, we may instead obtain $\varphi(x)=\infty$ by \eqref{eq:phi2}, even in cases when \eqref{eq:phi} satisfies $\varphi(x)<\infty$.
In particular, this necessarily happens when \eqref{eq:phi} satisfies $\limsup_{x\to\infty}\frac{\varphi(x)}{x}>1$.

\subsection{Examples}
\label{sec:degboundsEx}

According to the analysis in Section~\ref{sec:boundsfromsystems}, we can easily find weighted degree bounds as soon as a complete reduction system is available.
Since we have complete reduction systems for Airy functions and complete elliptic integrals in Section~\ref{sec:InfiniteSystems} w.r.t.\ Noetherian orders, we can find several weighted degree bounds for the two classes of functions as follows.
 
\subsubsection{Airy functions}
\label{sec:degboundsAiry}

For the elements of the differential ring $(C[t_1,t_2,t_3],\Der)$ used in Section~\ref{sec:Airy} for Airy functions, it is straightforward to obtain degree bounds for $L$ w.r.t.\ $(0,1,1)$ and $(2, 0, 1)$ because the order that we consider to find the complete reduction system is induced by $\left(\begin{smallmatrix}0&1&1\\2&0&1\\0&0&1\end{smallmatrix}\right)$.
Note that the derivative of each generator is a monomial and $\deg_{(0,1,1)}(\Der t_i)=\deg_{(0,1,1)}(t_i)$ for $i \in \{1,2,3\}$.
Then, by Lemma~\ref{LM:homogeneousbound}, a tight degree bound for $L$ w.r.t.\ $(0,1,1)$ is given by $\varphi(x)=x$.
 
Then, for the degree bound w.r.t.\ $(2, 0, 1)$, we need to apply Theorem~\ref{thm:tightbound} to the complete reduction system shown in Theorem~\ref{TH:Airyreductionrule} and, thanks to homogeneity, we can obtain an even more general result as follows.
While \eqref{eq:phi2} easily yields an explicit degree bound w.r.t.\ many weight vectors, direct application of \eqref{eq:phi} allows for refined degree bounds w.r.t.\ particular weights.

\begin{theorem}\label{TH:Airydegreebound}
 Let $w=(2, w_2, w_2+1)$ with $w_2 \in \mathbb{R}$.
 Then, a degree bound for $L$ w.r.t.\ $w$ is given by $\varphi(x)=x+2$.
 Moreover, if $w_2=-1$, another degree bound for $L$ w.r.t.\ $(2,-1,0)$ is given by
 \[
  \varphi(x)=\begin{cases}\lfloor{x}\rfloor-1&x<-2\\2\left\lfloor\tfrac{x}{2}\right\rfloor+2&x\ge-2\end{cases}.
 \]
\end{theorem}
\begin{proof}
 Let $S$ be the reduction system given in Theorem~\ref{TH:Airyreductionrule}, which is complete by Theorem~\ref{TH:Airyunreduced}.
 For each $(P,Q,B) \in S$, the leading monomial of $P$ is always equal to $1$ by definition and other monomials appearing in $P$ are of $w$-degree at most $0$, that is, $\deg_w(P)=0$.
 Since $<$ is Noetherian, the assumptions of Theorem~\ref{thm:tightbound} are satisfied.
 While we have $\deg_w(Q)=2$ for all $(P,Q,B)$ arising from Theorem~\ref{TH:Airyreductionrule}~(iii), all other rules arising from Theorem~\ref{TH:Airyreductionrule} have $\deg_w(Q)=-1$.
 Hence, by Corollary~\ref{cor:simplebound}, we immediately obtain $\varphi(x)=x+2$ as degree bound for $L$ w.r.t.\ $w$.
\par
 Moreover, if $w_2=-1$, we evaluate \eqref{eq:phi} noting that $(f,g)=(P(\alpha,t)t^\alpha,Q(\alpha,t)t^\alpha) \in \Sigma$ has $\deg_w(f)=\deg_w(t^\alpha)=2\alpha_1-\alpha_2$ and $\deg_w(g)=\deg_w(Q(\alpha,t))+\deg_w(f)$.
 For $(f,g)$ originating from Theorem~\ref{TH:Airyreductionrule}~(i), we see that $\deg_w(f)$ can assume any integer value and we have $\deg_w(g)=\deg_w(f)-1$.
 For $(f,g)$ originating from Theorem~\ref{TH:Airyreductionrule}~(iii), we have $\alpha_1\ge\frac{\alpha_2}{2}-1$ and $\alpha_2$ is even, hence $\deg_w(f)$ ranges over all even integers $\ge-2$, and we have $\deg_w(g)=\deg_w(f)+2$, since $b_{\alpha_2,1}=\frac{(\alpha_2-1)!!}{\alpha_2!!}\neq0$.
 Altogether, using Remark~\ref{rem:phi}.\ref{rem:phiminimal}, we obtain $\varphi(x)\ge{x+2}$ if $x$ is an even integer $\ge-2$, and $\varphi(x)\ge{x-1}$ if $x$ is any other integer.
 The minimal (weakly) monotonically increasing map satisfying these conditions is given by $\varphi(x)=\lfloor{x}\rfloor-1$ for $x<-2$, and by $\varphi(x)=2\lfloor\frac{x}{2}\rfloor+2$ for $x\ge-2$.
\end{proof}
Since the Airy differential equation has no nonzero Liouvillian solutions, it follows that $\const_\Der(F)=C$, see \cite{Airy}, hence $\ker(L)=C$.
For $w_2\ge0$, Theorem~\ref{TH:Airydegreebound} together with $\ker(L)=C$ implies that, for any nonzero $f,g \in C[t_1,t_2,t_3]$ with $f=\Der{g}$, we have
  \begin{equation}\label{EQ:Airydegreebound}
 \deg_{(2, w_2, w_2+1)}(g) \le \deg_{(2, w_2, w_2+1)}(f)+2.
  \end{equation}

By Theorems \ref{TH:degreeboundcombin} and \ref{TH:Airydegreebound}, we know that a non-trivial degree bound w.r.t.\ the total degree exists.
In particular, one can show the following.

\begin{corollary}\label{COR:Airytotaldegreebound}
For any nonzero $f, g \in C[t_1,t_2,t_3]$ such that $f = \partial g$, we have the total degree bound
\[
  \deg_{(1,1,1)}(g)\le\left\lfloor\tfrac{3}{2}\deg_{(1,1,1)}(f)\right\rfloor+1.
\]
\end{corollary}
\begin{proof}
 Choosing $w_2=2$, $\lambda=\frac{1}{2}$, and $c=3$, we obtain the total degree bound $\varphi(x)=\frac{3}{2}x+1$ from \eqref{EQ:Airydegreebound} using Lemma~\ref{lem:conversion}.
 Then, the statement follows, since the total degree of $g$ is an integer.
\end{proof}

\begin{example}
\label{ex:AiryTightness}
Actually, the degree bound w.r.t.\ $(2,w_2,w_2+1)$ given in Theorem~\ref{TH:Airydegreebound} and the total degree bound given in Corollary~\ref{COR:Airytotaldegreebound} are tight.
For example, the integrand $f=t_3^2$ and its integral $g=\frac{1}{3}t_1t_3^2+\frac{2}{3}t_2t_3-\frac{1}{3}t_1^2t_2^2$ considered in Example~\ref{ex:AiryBoettner} satisfy $\deg_{(2,w_2,w_2+1)}(f)=2w_2+2$ and $\deg_{(2,w_2,w_2+1)}(g)=2w_2+4$, so the degree bound~\eqref{EQ:Airydegreebound} is tight.
In addition, the total degrees of $f$ and $g$ are equal to $2$ and $4$, respectively. 
So $\deg_{(1,1,1)}(g)=\left\lfloor\tfrac{3}{2}\deg_{(1,1,1)}(f)\right\rfloor+1$, that is, the total degree bound given in Corollary~\ref{COR:Airytotaldegreebound} is also tight.
Furthermore, equality holds in \eqref{EQ:Airydegreebound} also for all instances arising from Theorem~\ref{TH:Airyreductionrule}~(iii) and tightness of Corollary~\ref{COR:Airytotaldegreebound} is exhibited by the following integrands.
\begin{enumerate}
 \item The integrand $f=t_3^4$ gives rise to the necessary condition $\deg_{(1,1,1)}(g)\le7$ on its integral, if it exists in $C[t_1,t_2,t_3]$.
  To find $g$, by Lemma~\ref{LM:homogeneousbound} and by \eqref{EQ:Airydegreebound}, only those $14$ monomials need to be included in the ansatz that have $(0,1,1)$-degree equal to $4$ and $(2,0,1)$-degree $\le6$.
  Plugging this into $f=L(g)$ and comparing coefficients, we obtain the solution
  \[
   g = \tfrac{3}{16}t_1t_3^4+\tfrac{13}{16}t_2t_3^3-\tfrac{3}{8}t_1^2t_2^2t_3^2-\tfrac{9}{16}t_1t_2^3t_3+\left(\tfrac{3}{16}t_1^3+\tfrac{9}{64}\right)t_2^4,
  \]
  which involves the monomial $t_1^3t_2^4$ of total degree $7$.
 \item The integrand $f=4t_1t_3^4-7t_1t_2^3t_3$ requires its integral $g \in C[t_1,t_2,t_3]$ to be of total degree $\le8$.
  Being homogeneous of $(0,1,1)$-degree $4$ in addition to having $(2,0,1)$-degree $\le8$, the ansatz for $g$ only needs $19$ monomials.
  Solving $f=L(g)$ by coefficient comparison reveals that the monomial $t_1^4t_2^4$ of total degree equal to $8$ actually appears in the solution
  \[
   g = \tfrac{1}{2}t_1^2t_3^4+3t_1t_2t_3^3-\left(t_1^3+\tfrac{3}{2}\right)t_2^2t_3^2-2t_1^2t_2^3t_3+\tfrac{1}{2}t_1^4t_2^4.
  \]
 \item The integrand $f=24t_3^6-77t_2^3t_3^3$ gives rise to an ansatz for its integral $g \in C[t_1,t_2,t_3]$ that consists of the $23$ monomials of $(0,1,1)$-degree equal to $6$ with $(2,0,1)$-degree $\le8$.
  Indeed, we find that the monomial $t_1^4t_2^6$ of total degree $10$ occurs in the solution
  \[
   g = 3t_1t_3^6+21t_2t_3^5-9t_1^2t_2^2t_3^4-29t_1t_2^3t_3^3+(9t_1^3-12)t_2^4t_3^2+12t_1^2t_2^5t_3-3t_1^4t_2^6.
  \]
 \item Similarly, for the integrand $f=1092t_1t_3^6-6449t_1t_2^3t_3^3$ of total degree $7$, we use an ansatz for $g \in C[t_1,t_2,t_3]$ that consists of the $30$ monomials with $(0,1,1)$-degree equal to $6$ having $(2,0,1)$-degree $\le10$.
  Thereby, we obtain the following integral of total degree $11$.
  \begin{multline*}
   g = \tfrac{195}{2}t_1^2t_3^6+897t_1t_2t_3^5-\left(\tfrac{585}{2}t_1^3+\tfrac{897}{2}\right)t_2^2t_3^4-\tfrac{2405}{2}t_1^2t_2^3t_3^3\\+\left(\tfrac{585}{2}t_1^4-\tfrac{1125}{2}t_1\right)t_2^4t_3^2+\left(\tfrac{975}{2}t_1^3+\tfrac{225}{2}\right)t_2^5t_3-\left(\tfrac{195}{2}t_1^5+\tfrac{225}{4}t_1^2\right)t_2^6
  \end{multline*}
 \item Analogously, for the integrand $f=8t_3^8-49t_2^6t_3^2$ of total degree $8$, we find an integral $g \in C[t_1,t_2,t_3]$ of total degree equal to $13$.
  Explicitly, via an ansatz consisting of the $34$ monomials that are of $(0,1,1)$-degree equal to $8$ with $(2,0,1)$-degree $\le10$, we obtain
  \begin{multline*}
   g = \tfrac{35}{48}t_1t_3^8+\tfrac{349}{48}t_2t_3^7-\tfrac{35}{12}t_1^2t_2^2t_3^6-\tfrac{721}{48}t_1t_2^3t_3^5+\left(\tfrac{35}{8}t_1^3+\tfrac{721}{192}\right)t_2^4t_3^4\\+\tfrac{595}{48}t_1^2t_2^5t_3^3-\left(\tfrac{35}{12}t_1^4+\tfrac{637}{96}t_1\right)t_2^6t_3^2-\left(\tfrac{175}{48}t_1^3+\tfrac{581}{96}\right)t_2^7t_3+\left(\tfrac{35}{48}t_1^5+\tfrac{581}{192}t_1^2\right)t_2^8.
  \end{multline*}
\end{enumerate}
In fact, for any even $n \in \mathbb{N}$, one can construct $f \in C[t_1,t_2,t_3]$, homogeneous of degree $n$ and free of $t_1$, such that there exists an integral $g \in C[t_1,t_2,t_3]$ satisfying
\[
 f = \partial g \quad\text{and}\quad g = t_1^{n/2+1}t_2^n + \text{terms of lower total degree},
\]
which implies that $\deg_{(1,1,1)}(g)=\tfrac{3}{2}\deg_{(1,1,1)}(f)+1$.
 The construction is similar to the one used in the proof of Theorem~\ref{thm:Airyinfinite} below.
\end{example}

From the weighted degree bound \eqref{EQ:Airydegreebound} w.r.t.\ the weight vector $w=(2,0,1)$, we trivially obtain a bound $\deg_{t_1}(g) \le \frac{1}{2}\deg_w(g) \le \frac{1}{2}\deg_w(f)+1$ on the partial degree w.r.t.\ $t_1$, which depends on the weighted degree of the integrand, where $f=\Der{g}$.
However, the only degree bound for $L$ w.r.t.\ $(1,0,0)$ is the trivial one $\varphi(x)=\infty$, as shown by Theorem~\ref{thm:Airyinfinite} below.
In fact, the same is true also for degree bounds w.r.t.\ partial degrees in $t_2$ or $t_3$.
For $t_3$, this can be seen by Thm.~2~(ii) in \cite{Airy}.

\begin{theorem}
\label{thm:Airyinfinite}
For any $m,n,d \in \mathbb{N}$ with $d\ge2m+1$ there exist $f,g \in C[t_1,t_2,t_3]$ homogeneous of degree $d$ w.r.t.\ $(0,1,1)$ such that $\Der{g}=f$, $\deg_{t_1}(f)=n$, and $\deg_{t_1}(g)=n+m$.
\end{theorem}
\begin{proof}
 Let $z$ be a new indeterminate and consider the differential ring $(\mathbb{Q}[z],')$ with derivation $'=\frac{d}{dz}$.
 We define $g_{-1}(z):=0$ and $g_0(z):=1$ and we recursively fix $g_1,\dots,g_{m+1} \in \mathbb{Q}[z]$ such that
 \[
  g_l^\prime(z) = z^2g_{l-1}^\prime(z)-d{\cdot}z{\cdot}g_{l-1}(z)-(m+n-l+2)g_{l-2}(z)
 \]
 for $l=1,\dots,m+1$.
 The choice of such $g_1,\dots,g_{m+1}$ is not unique, but always satisfies $\deg(g_l) \le 2l$ and $\coeff(g_l,z^{2l})=(-\frac{d}{2})_l/l!$, which follows by induction on $l$.
 Then, we let
 \[
  f:=-g_{m+1}^\prime\!\left(\frac{t_3}{t_2}\right)t_1^nt_2^d+ng_m\!\left(\frac{t_3}{t_2}\right)t_1^{n-1}t_2^d \quad\text{and}\quad g:=\sum_{l=0}^mg_l\!\left(\frac{t_3}{t_2}\right)t_1^{m+n-l}t_2^d.
 \]
 Evidently, $f,g \in C[t_1,t_2,t_3]$ are homogeneous of degree $d$ w.r.t.\ $(0,1,1)$.
 Applying $\Der$ to $g_l(\frac{t_3}{t_2})t_1^{m+n-l}t_2^d$ yields 
 $$\left(g_l^\prime\!\left(\frac{t_3}{t_2}\right)\left(t_1-\frac{t_3^2}{t_2^2}\right)+(m+n-l)g_l\!\left(\frac{t_3}{t_2}\right)\frac{1}{t_1}+dg_l\!\left(\frac{t_3}{t_2}\right)\frac{t_3}{t_2}\right)t_1^{m+n-l}t_2^d.$$
 So, collecting terms in $\Der{g}$ with same powers of $t_1$, we obtain
 \begin{align*}
  \Der{g} &= \underbrace{g_0^\prime\!\left(\frac{t_3}{t_2}\right)}_{=0}t_1^{m+n+1}t_2^d\\
  &\quad+\sum_{l=0}^{m-1}\underbrace{\left(g_{l+1}^\prime\!\left(\frac{t_3}{t_2}\right)-g_l^\prime\!\left(\frac{t_3}{t_2}\right)\frac{t_3^2}{t_2^2}+(m+n-l+1)g_{l-1}\!\left(\frac{t_3}{t_2}\right)+dg_l\!\left(\frac{t_3}{t_2}\right)\frac{t_3}{t_2}\right)}_{=0}t_1^{m+n-l}t_2^d\\
  &\quad+\underbrace{\left(-g_m^\prime\!\left(\frac{t_3}{t_2}\right)\frac{t_3^2}{t_2^2}+(n+1)g_{m-1}\!\left(\frac{t_3}{t_2}\right)+dg_m\!\left(\frac{t_3}{t_2}\right)\frac{t_3}{t_2}\right)}_{=-g_{m+1}^\prime(\frac{t_3}{t_2})}t_1^nt_2^d+ng_m\!\left(\frac{t_3}{t_2}\right)t_1^{n-1}t_2^d.
 \end{align*}
 Hence, we have $\Der{g}=f$.
 Since $d\ge2m+1$, $\coeff(g_{m+1},z^{2m+2})=(-\frac{d}{2})_{m+1}/(m+1)!$ implies $g_{m+1}^\prime(z)\neq0$ and therefore $\deg_{t_1}(f)=n$.
 Finally, $g_0(z)=1$ implies $\deg_{t_1}(g)=n+m$.
\end{proof}

\subsubsection{Complete elliptic integrals}
\label{sec:degboundsCEI}

Let $(C[t_1,t_2,t_3],\tilde{\Der})$ be the differential ring used in Section~\ref{sec:CEI} for the complete elliptic integrals of the first kind and of the second kind, which has
\[
 \tilde{\Der} t_1 = t_1(1-t_1^2), \quad \tilde{\Der} t_2 = t_3-(1-t_1^2)t_2, \quad \text{and} \quad \tilde{\Der} t_3 = (1-t_1^2)(t_3-t_2).
\]
Since $\tilde{\Der} t_i$ is homogeneous w.r.t.\ $(0,1,1)$ with degree equal to $\deg_{(0,1,1)}(t_i)$ for each $i \in \{1,2,3\}$, a tight degree bound for $L$ w.r.t.\ $(0,1,1)$ is given by $\varphi(x)=x$.
Then, different from the Airy case, we look at the degree bound w.r.t.\ $w = (1, w_2, w_2+2)$, where $w_2 \in \mathbb{R}$.
Note that $w$ is chosen such that the monomials along the slope of the triangle $\supp(Q_j)$ for all $j \in \mathbb{N}$ arising from the reduction rules shown in Theorem~\ref{TH:CEIreductionrule} (ii) have the same degree.
\begin{theorem}\label{TH:CEIdegreebound}
  Let $w=(1, w_2, w_2+2)$ with $w_2 \in \mathbb{R}$.
  Then, $\varphi(x)=x-2$ is a degree bound for $L$ w.r.t.\ $w$.
\end{theorem}
\begin{proof}
 Let $S$ be the reduction system given in Theorem~\ref{TH:CEIreductionrule}, which is complete by Theorem~\ref{TH:CEIcomplete}.
 For each $(P,Q,B) \in S$, we have $\deg_w(P)=0$ and $\deg_w(Q)=-2$.
 Since the order used in Theorem~\ref{TH:CEIreductionrule} is Noetherian, the statement then follows from Corollary~\ref{cor:simplebound}.
\end{proof}

\begin{theorem}
\label{thm:CEIdegreebound2}
 Let $w=(1,w_2,w_2)$ with $w_2 \in \mathbb{R}$.
 Then, $\varphi(x)=x$ is a degree bound for $L$ w.r.t.\ $w$.
 Moreover, if $w_2=-1$, a smaller degree bound for $L$ w.r.t.\ $(1,-1,-1)$ is given by
 \[
  \varphi(x)=\begin{cases}\lfloor{x}\rfloor&x<0\\
  0&0\le{x}<3\\
 \lfloor{x}\rfloor-2&x\ge3\end{cases}
 \]
 and, if $w_2=1$, another degree bound for $L$ w.r.t.\ the total degree is given by
 \[
  \varphi(x)=\begin{cases}-\infty&x<2\\
  2\left\lfloor\tfrac{x}{2}\right\rfloor&x\ge2\end{cases}.
 \]
\end{theorem}
\begin{proof}
 Let $S$ be the reduction system given in Theorem~\ref{TH:CEIreductionrule2}, which is complete by Theorem~\ref{th:CEInewcompleteness}.
 For each $(P,Q,B) \in S$, we have $\deg_w(P)=0$.
 Since the order we use for the reduction system $S$ is Noetherian, the assumptions of Theorem~\ref{thm:tightbound} are satisfied.
 We have $\deg_w(Q)=-2$ for the rule $(P,Q,B)$ given by \eqref{eq:CEInewgenericrule}, and $\deg_w(Q)=0$ for all rules $(P,Q,B)$ given by \eqref{eq:CEInewrule}.
 This implies that $\varphi(x)=x$ is a degree bound for $L$ w.r.t.~$w$ by Corollary~\ref{cor:simplebound}.
\par
 In the following, note that $(f,g)=(P(\alpha,t)t^\alpha,Q(\alpha,t)t^\alpha) \in \Sigma$ has $\deg_w(f)=\deg_w(t^\alpha)$ and $\deg_w(g)=\deg_w(Q(\alpha,t))+\deg_w(f)$.
 First, let $w_2=-1$.
 For $(f,g)$ originating from \eqref{eq:CEInewgenericrule}, $\alpha_1\ge2 \wedge \alpha_1-\alpha_2+\alpha_3\neq2$ implies that $\deg_w(f)=\alpha_1-\alpha_2-\alpha_3$ can assume any integer value and we have $\deg_w(g)=\deg_w(f)-2$.
 For $(f,g)$ originating from \eqref{eq:CEInewrule}, $\alpha_3\ge1 \wedge \alpha_2=\alpha_1+\alpha_3-2 \wedge \alpha_1+\alpha_3\neq1$ implies that $\deg_w(f)=-2(\alpha_3-1)$ ranges over all even integers $\le0$ and we have $\deg_w(g)=\deg_w(f)$.
 Consequently, $\varphi(x)$ defined by \eqref{eq:phi} satisfies $\varphi(x) \ge x$, if $x$ is an even integer less than or equal to zero, and $\varphi(x) \ge x-2$, if $x$ is any other integer.
 Theorem~\ref{thm:tightbound} together with Remark~\ref{rem:phi}.\ref{rem:phiminimal} then proves the claim.
\par
 Now, assume $w_2=1$.
 For $(f,g) \in \Sigma$ originating from \eqref{eq:CEInewgenericrule}, $\alpha_1\ge2 \wedge \alpha_1-\alpha_2+\alpha_3\neq2$ implies that $\deg_w(f)$ ranges over all integer values $\ge3$ and we have $\deg_w(g)=\deg_w(f)-2$.
 For $(f,g)$ originating from \eqref{eq:CEInewrule}, $\alpha_3\ge1 \wedge \alpha_2=\alpha_1+\alpha_3-2 \wedge \alpha_1+\alpha_3\neq1$ implies that $\deg_w(f)=2(\alpha_1+\alpha_3-1)$ ranges over all even values $\ge2$ and we have $\deg_w(g)=\deg_w(f)$.
 Evidently, $\varphi(x)$ defined by \eqref{eq:phi} evaluates to $-\infty$ for $x<2$.
 For $x\ge2$, we conclude that $\varphi(x) \ge x$ if $x$ is an even integer and $\varphi(x) \ge x-2$ if $x$ is an odd integer.
 Altogether, since $\varphi(x)$ is (weakly) monotonically increasing and minimal with these constraints, we obtain $\varphi(x)=2\lfloor\tfrac{x}{2}\rfloor$ for $x\ge2$.
\end{proof}

For $w_2=1$, Theorem~\ref{thm:CEIdegreebound2} gives a degree bound w.r.t.\ the total degree.
It can be shown that the first-order linear differential system with coefficients in $\mathbb{Q}(x)$ satisfied by $K(x)$ and $E(x)$ does not have any nonzero Liouvillian solutions.
So, similar to Section~\ref{sec:degboundsAiry}, we conclude $\const_\Der(F)=C$, and therefore $\ker(L)=C$.
Hence, we get the following statement that is analogous to Corollary~\ref{COR:Airytotaldegreebound}:
For any nonzero $f,g \in C[t_1,t_2,t_3]$ such that $f = \tilde{\Der}g$, we have $\deg_{(1,1,1)}(f)\ge2$ and
\[
 \deg_{(1,1,1)}(g)\le2\left\lfloor\tfrac{1}{2}\deg_{(1,1,1)}(f)\right\rfloor.
\]
Tightness is illustrated by $(f,g)\in\Sigma$ obtained from \eqref{eq:CEInewrule}, which are given by $f=t_1t_3$ and $g=t_1t_2$ as well as by $f=(k-1)t_2^{k-2}t_3(t_3-t_2)(t_3-\frac{1}{2}t_2)^{k-2}$ and $g=t_2^{k-1}(t_3-\frac{1}{2}t_2)^{k-1}$ for $k\ge2$, for example.

\section{Discussion} \label{sec:discussion}

Any reduction system or rigorous degree bound computed by Norman's or our methods depends on the map $L$ acting on polynomials.
Since this map is induced by \eqref{eq:defP}, it depends not only on the derivation, but also on the choice of $v$.
So, using a particular reduction system (or resulting degree bounds), one can solve Problem~\ref{prob:MainProblem} for different choices of $f_i$, but only for particular $v$.
Consequently, it allows finding elementary integrals at most of those integrands for which the integral \eqref{eq:ParallelAnsatz} can be written with the chosen denominator $v$.
If $(F,\Der)$ is such that the denominator $v$ and the logarithmic part of elementary integrals can be determined correctly, e.g.\ by Thm.~4 in \cite{BronsteinParallel}, then successful computation of complete reduction systems allows to decide elementary integrability over $(F,\Der)$.
Moreover, we can apply a given reduction system with fixed $v$ to the polynomial part of $\frac{v^2\den(\Der)f}{\gcd(v,\tilde{\Der}v)}$ determined by the matryoshka decomposition presented in \cite{DuGuoLiWong}.
Since the matryoshka decomposition provides a direct complement of $C[t]$ in $F$, this yields, for any $f \in F$, an additive decomposition $f=\Der(\frac{u}{v})+r$ with $u \in C[t]$ and $r \in F$.

\begin{example}\label{ex:LogIntegralDenominator}
 With the logarithmic integral $\mathrm{li}(x)$, we have $\frac{d}{dx}\mathrm{li}(\tfrac{1}{x})=\frac{1}{x^2\ln(x)}$ away from the branch cut.
 Based on this identity, we consider the following integral involving $\mathrm{li}(\tfrac{1}{x})$.
 \begin{multline}\label{eq:LogIntegralDenominator}
  \int(2\ln(x)^2+3\ln(x)-1)\mathrm{li}(\tfrac{1}{x})^3\,dx =\\ x\ln(x)(2\ln(x)-1)\mathrm{li}(\tfrac{1}{x})^3-3\ln(x)(\ln(x)-1)\mathrm{li}(\tfrac{1}{x})^2-\frac{6\ln(x)\mathrm{li}(\tfrac{1}{x})}{x}-\frac{3}{x^2}
 \end{multline}
 Modelling $x,\ln(x),\mathrm{li}(\tfrac{1}{x})$ by $t_1,t_2,t_3$, respectively, we arrive at $(F,\Der)=(C(t_1,t_2,t_3),\Der)$ with $\Der{t_1}=1$, $\Der{t_2}=\frac{1}{t_1}$, and $\Der{t_3}=\frac{1}{t_1^2t_2}$.
 The integrand $f=(2t_2^2+3t_2-1)t_3^3$ has trivial denominator.
 So, to compute an elementary integral of $f$ over $(F,\Der)$ via \eqref{eq:ParallelAnsatz}, one might expect to use $v=1$.
 However, any attempt to satisfy \eqref{eq:ParallelAnsatz} necessarily will fail unless $t_1^2|v$, as can be seen from \eqref{eq:LogIntegralDenominator}.
 Using $v=1$ and $v=t_1^2$ in \eqref{eq:defP}, respectively, the two sets of basic reduction rules created by Algorithm~\ref{alg:CItoRR} w.r.t.\ the lexicographic monomial order with $t_1<t_2<t_3$ are even complete.
 Indeed, the latter reduction system succeeds in finding the integral by reducing $t_1^4t_2f$ to zero, while the former can only decompose $t_1^2t_2f=\tilde{\Der}\big(t_1t_2(2t_2-1)t_3^3-3t_2(t_2-1)t_3^2\big)+6t_2(t_2-1)t_3$ to find the polynomial part of the integral.
 In fact, the integral of $f$ can also be found by the reduction algorithm presented in \cite{DuGuoLiWong}, since this $(F,\Der)$ is an S-primitive tower as defined there.
\end{example}

\begin{example}\label{ex:TanSpecial}
 Independent of the choice of denominator $v$ in \eqref{eq:ParallelAnsatz}, there are also other issues that can prevent successful computation of an elementary integral over $(F,\Der)$.
 To exemplify these, we again use the differential field $(C(t_1,t_2),\Der)$ with $\Der{t_1}=1$ and $\Der{t_2}=t_2^2+1$, introduced in Example~\ref{ex:NormanMain}.
 Consider the integrand $f=t_2$, which implies $m=0$ in \eqref{eq:ParallelAnsatz} by $\den(f)=1$.
 If we cannot find polynomials $s$ with $s|\Der{s}$ that are necessary to satisfy \eqref{eq:ParallelAnsatz}, then there is no way to compute the elementary integral just in terms of $f=\Der(u/v)$.
 This is because no such integral exists and the elementary integral involves a nontrivial logarithmic part.
 However, by choosing $s=t_2^2+1$, which satisfies $\Der{s}=2t_2s$, we can obtain zero in the right hand side of \eqref{eq:PolynomialAnsatz} and thereby find $f = \frac{1}{2}\frac{\Der s}{s}$, as discussed in \cite{DavenportParallel3}. 
 This implies that $\int\tan(x)\,dx=\frac{1}{2}\ln\!\big(\tan(x)^2+1\big)$.
 Moreover, it is evident that, for all integrands $f$ that do not have any elementary integral over $(F,\Der)$, \eqref{eq:ParallelAnsatz} cannot be satisfied regardless of the choices made.
 For example, it can be shown that this is the case for $f=t_1t_2$, i.e.\ $\int x\tan(x)\,dx$ is not elementary.
\end{example}

For D-finite functions, an alternative approach that allows to find antiderivatives was given in \cite{AbramovHoeij}.
In short, it aims to find antiderivatives that are expressible as linear combination with rational function coefficients of the integrand's derivatives.
While this rules out finding even some simple integrals like $\int\frac{\ln(x)}{x}\,dx=\frac{1}{2}\ln(x)^2$ by design, the algorithm is applicable in principle to a very large class of integrands and is able to find the denominator, which is a polynomial in $x$ only, required for the solution.
Representing the integrand by a differential operator annihilating it, this approach, however, may also fail due to non-minimality of the order of the representation, as it can result from straightforward application of D-finite closure properties.
For instance, the integrands in \eqref{eq:CEIexample1} and \eqref{eq:CEIexample2} admit a representation of minimal order $3$, based on which the integrals can be found, while straightforward closure properties yield orders $4$ and $7$, respectively, which make this approach fail.

Our presentation focused on complete reduction systems, their construction, and their use for solving integration problems in the form \eqref{eq:MainEquation} via reduction or via degree bounds.
It should be noted that, in practice, even incomplete reduction systems can be used in a similar way.
Whether we use it for performing reduction or for obtaining degree bounds, completeness of the reduction system is only needed to ensure that a solution of \eqref{eq:MainEquation} is always found if one exists.
In particular, a precomplete reduction system consisting of basic reduction rules can be obtained with relatively little computational effort, as mentioned in Section~\ref{sec:RedSys}.
Given any reduction system for a fixed $L$, we evidently can try to solve \eqref{eq:MainEquation} by reducing the right hand side to zero.
If zero can be reached for a particular right hand side, we still obtain a valid solution by the reduction system, regardless of its completeness.
Alternatively, we can also apply \eqref{eq:phi}, or any variant like \eqref{eq:phi2}, to compute an ansatz for the solution $u$ via bounds on the degree of its monomials based on the reduction system.
For particular right hand sides, valid solutions may still be found via such heuristic degree bounds, just like with existing heuristic degree bounds like those mentioned in Section~\ref{sec:RischNorman}.

Moreover, one can also use Procedure~\ref{proc:RefinedCompletion} at the core of a semi-decision procedure that mixes the computation of a complete reduction system with computing a solution of  \eqref{eq:MainEquation}.
After a limited number of iterations of the main loop, or after a limited number of new reduction rules have been found in Procedure~\ref{proc:RefinedCompletion}, the procedure would try to solve \eqref{eq:MainEquation} via reduction or via degree bounds based on the intermediate reduction system obtained so far.
Unless a solution is found, we resume Procedure~\ref{proc:RefinedCompletion} and keep trying.
Using a Noetherian order and a fair selection that ensures that every critical pair is treated eventually and by enforcing termination of the inner loop in Procedure~\ref{proc:RefinedCompletion}, we indeed obtain a semi-decision procedure for solving \eqref{eq:MainEquation} even in cases when Procedure~\ref{proc:RefinedCompletion} does not terminate.
Since we do not have a criterion to decide at a given point if a given monomial will become reducible during one of the infinitely many remaining iterations, the outlined semi-decision procedure obviously is not a decision procedure in cases when Procedure~\ref{proc:RefinedCompletion} does not terminate.

Even in cases when the completion process terminates, there are two aspects that can cause considerable computational effort in practice.
First, conditional identities and reduction rules computed during completion can exhibit considerable expression swell.
Reducing a conditional identity by a reduction rule according to Definition~\ref{defn:reducible} can lead to a conditional identity where the degree of $P,Q$ in the variables $\param$ is increased.
The degrees in $\param$ of intermediate reduction rules can be higher than the degrees needed to express the coefficients in the corresponding complete reduction system.
For example, computing the reduction system of Section~\ref{sec:Airy} by completion requires intermediate reduction rules with necessarily unbounded $\param_3$-degree even though the infinite complete reduction system can be expressed with coefficients that are linear in the variables $\param$.
Second, conditions of reduction rules can involve nonlinear equations and inequations.
Nonlinear conditions can be introduced during completion in steps \ref{line:newrule} and \ref{line:updateB} of Algorithm~\ref{alg:CItoRR} if the leading coefficient of $P$ at this stage is not linear in the variables $\param$.
This is also the case for the reduction system of Section~\ref{sec:Airy} before simplification of conditions, for example.
While the exact form of coefficients appearing in reduction rules is essential for performing reduction, knowledge of these coefficients is not relevant for constructing degree bounds based on \eqref{eq:phi}, since the degree depends only on which monomials appear.
In future research, we plan to investigate how the completion process can be simplified to obtain degree bounds more efficiently by keeping only partial information about reduction rules during completion.
The resulting degree bounds may no longer be rigorous, and so it will be important to find a good balance between accuracy and efficiency.

Another open problem is the choice of the order of monomials used.
As Norman already pointed out, it can influence termination of the completion process.
This applies to our refined completion process as well.
In addition, when computing a degree bound based on a reduction system using Theorem~\ref{thm:tightbound}, the order restricts the possible weight vectors by \eqref{eq:compatibility}.
For example, the total degree $w=(1,1,1)$ satisfies this condition only for the second but not for the first reduction system given for the same $L$ in Section~\ref{sec:CEI}.
When a solution of $L(u)=f$ is computed via ansatz based on degree bounds, the choice of weights can have a substantial influence on the size of the ansatz determined by the resulting bounds.
For example, both weight vectors $(1,0,0)$ and $(1,0,2)$ satisfy condition \eqref{eq:compatibility} for the reduction system of Theorem~\ref{TH:CEIreductionrule2}.
For fixed $k\ge2$, the same reduction system yields $\deg_{(1,0,0)}(u)\le0$ and $\deg_{(1,0,2)}(u)\le2k-2$ for $f=t_2^{k-2}t_3(t_3-t_2)(2t_3-t_2)^{k-2}$ that is reduced to zero by setting $(\param_1,\param_2,\param_3)=(0,k-2,k)$ in \eqref{eq:CEInewrule}.
Combining each of these bounds with homogeneity of $L$ w.r.t.\ $(0,1,1)$, the former bound yields an ansatz with $2k-1$ monomials for $u$ while the latter would allow $k^2$ monomials in the ansatz to find the solution $u=\frac{1}{2(k-1)}t_2^{k-1}(2t_3-t_2)^{k-1}$ that involves only those $k$ monomials that appear in both ansatzes.

Our motivation and focus is on solving first-order differential equations \eqref{eq:MainEquation} arising from the Risch--Norman approach in symbolic integration, but we made use of the concrete form \eqref{eq:defP} only when discussing examples.
Consequently, the methods presented are also applicable to other linear operators of the form \eqref{eq:defL}, like higher-order differential operators.
Throughout the paper, we used the monomial basis $\{t^\alpha\}_{\alpha\in\mathbb{N}^n}$ to work with polynomials.
In the linear algebra view, it is clear that the framework can also be adapted to other bases $\{b_\alpha\}_{\alpha\in\mathbb{N}^n}$ of $C[t]$, if applying the operator of interest can be expressed as finite sum $\sum_\beta c_\beta(\alpha)b_{\alpha+\beta}$ with $c_\beta(\param) \in C[\param]$.
This sum and coefficients $c_\beta(\param)$ generalize \eqref{eq:defL} and the finitely many coefficients $\coeff(p,t^\beta)$ of the Laurent polynomial $p$.

Finally, we note that our notion of complete reduction systems is closely related to the notion of \emph{staggered linear bases} introduced in \cite{GebauerMoellerStaggered}.
In particular, a reduction system $S$ for $L$ is complete according to Definition~\ref{DEF:complete} if and only if the set $\{P(\alpha,t)t^\alpha\ |\ (P,Q,B)\in S,\alpha\in\mathbb{N}^n,B|_{\param=\alpha}\}$ is a staggered generator of $\im(L)$ as defined in \cite{MoellerMoraTraverso}.
Moreover, that set is even a staggered linear basis of $\im(L)$ if and only if the reduction system is complete and no two rules in $S$ form a critical pair, which is the case for the output of Procedure~\ref{proc:RefinedCompletion} as well as for all reduction systems shown in Section~\ref{sec:InfiniteSystems}.
It will be interesting to explore if recent ideas for the computation of staggered linear bases of polynomial ideals \cite{HashemiJavanbakht,HashemiMoeller} can be adapted to improve Procedure~\ref{proc:RefinedCompletion}.

\subsubsection*{Acknowledgements}

The authors would like to thank the reviewers for their careful reading and for their comprehensive feedback, which helped to improve the presentation of the material and motivated to include more examples.

\end{document}